\documentclass[aps,onecolumn,preprint,superscriptaddress,nofootinbib,floats]{revtex4}
\usepackage{amsmath,amssymb,color,mathrsfs, graphicx,verbatim,epsfig, bbm, wasysym}
\usepackage[hyperfootnotes=false]{hyperref}
\usepackage{slashed}
\usepackage{subfigure}
\usepackage{booktabs}
\usepackage{multirow}
\usepackage{slashbox}
\usepackage{soul}
\allowdisplaybreaks

\usepackage{pifont}

\def\lsim{\mathrel{\rlap{\lower4pt\hbox{\hskip1pt$\sim$}}
    \raise1pt\hbox{$<$}}}
\def\gsim{\mathrel{\rlap{\lower4pt\hbox{\hskip1pt$\sim$}}
    \raise1pt\hbox{$>$}}} 
\newcommand{\vev}[1]{ \left\langle {#1} \right\rangle }

\newcommand{\be}{\begin{eqnarray}}
\newcommand{\ee}{\end{eqnarray}}

\def\addresses#1#2{\hbox to \hsize{\@tablebox{#1}\hfil\@tablebox{#2}}}
\def\@tablebox#1{\vtop{\hsize=5in \begin{flushleft} #1 \end{flushleft}}}

\def\beq{\begin{equation}}
\def\eeq{\end{equation}}
\def\bit{\begin{itemize}}
\def\eit{\end{itemize}}
\def\beqa{\begin{eqnarray}}
\def\eeqa{\end{eqnarray}}

\def\Pythia{{\tt Pythia}}
\def\Pythia8{{\tt Pythia8}}
\def\MadGraph{{\tt MadGraph}}
\def\MadGraph5{{\tt MadGraph5}}

\def\Lxy{{\rm L}_{\rm xy}}

\newcommand{\met}{\ensuremath{\displaystyle{\not}E_T}}

\begin{document}

\baselineskip 0.6cm

\begin{titlepage}

\thispagestyle{empty}

\begin{flushright}
PITT PACC 1503\\
FERMILAB-PUB-15-087-T
\end{flushright}

\begin{center}

\vskip 1cm

{\Large \bf The Fate of Long-Lived Superparticles with \\ \vspace{2mm} Hadronic Decays after LHC Run 1}

\vskip 1.0cm
{\large Zhen Liu$^{1,2}$ and Brock Tweedie$^1$}
\vskip 0.4cm
$^1${\it PITT PACC, Department of Physics and Astronomy, University of Pittsburgh, \\ Pittsburgh, PA 15260, USA}
\vskip 0.2cm
$^2${\it Theoretical Physics Department, Fermi National Accelerator Laboratory, \\ Batavia, IL 60510, USA}
\vskip 2.0cm

\end{center}

\noindent   Supersymmetry searches at the LHC are both highly varied and highly constraining, but the vast majority are focused on cases where the final-stage visible decays are prompt.  Scenarios featuring superparticles with detector-scale lifetimes have therefore remained a tantalizing possibility for sub-TeV SUSY, since explicit limits are relatively sparse.  Nonetheless, the extremely low backgrounds of the few existing searches for collider-stable and displaced new particles facilitates recastings into powerful long-lived superparticle searches, even for models for which those searches are highly non-optimized.  In this paper, we assess the status of such models in the context of baryonic R-parity violation, gauge mediation, and mini-split SUSY.  We explore a number of common simplified spectra where hadronic decays can be important, employing recasts of LHC searches that utilize different detector systems and final-state objects.  The LSP/NLSP possibilities considered here include generic colored superparticles such as the gluino and light-flavor squarks, as well as the lighter stop and the quasi-degenerate Higgsino multiplet motivated by naturalness.  We find that complementary coverage over large swaths of mass and lifetime is achievable by superimposing limits, particularly from CMS's tracker-based displaced dijet search and heavy stable charged particle searches.  Adding in prompt searches, we find many cases where a range of sparticle masses is now excluded from zero lifetime to infinite lifetime with no gaps.  In other cases, the displaced searches furnish the only extant limits at any lifetime.

\end{titlepage}

\setcounter{page}{1}

\section{Introduction}
\label{sec:intro}

In spite of progressively tightening limits~\cite{ATLASSUSY,CMSSUSY}, supersymmetry (SUSY) continues to serve as one of the most compelling scenarios for new physics at the Large Hadron Collider (LHC).  In the absence of traditional signals below about 1~TeV, there has been a growing interest in exploring models in which superparticles are light enough for the 8~TeV LHC to produce, but somehow manage to evade the existing searches.  Indeed, this situation remains a strong possibility, in large part because the observable signatures of SUSY depend extremely sensitively on the details of the spectrum and on the decays of the lightest superparticles.  In particular, a broad class of highly-motivated scenarios that has so far received relatively limited dedicated attention in LHC searches includes superparticles with macroscopic decay lengths spanning from sub-mm to tens of meters.  Such ``displaced'' particles can occur in models with R-parity violation (RPV)~\cite{Barbier:2004ez} or gauge mediated SUSY breaking~(GMSB)~\cite{Giudice:1998bp}, and also in mini-split spectra where all squarks are at roughly the 1000~TeV scale~\cite{Arvanitaki:2012ps,ArkaniHamed:2012gw} and the gluino lifetime becomes extended.

However, displaced decay signals occupy a subtle place in collider phenomenology.  At the high end of the above lifetime range, long-lived charged particle searches~\cite{ATLAS:2014fka,Chatrchyan:2013oca,CMS-EXO-13-006} may become appropriate if the displaced particle is charged or has a nontrivial chance to form charged hadrons, but the efficiency drops exponentially for lower lifetimes.  At the low end of the lifetime range, any number of prompt searches or searches involving bottom or charm hadrons might pick up the signals~\cite{ATLAS-2014-037}, but may subject them to unnecessarily large backgrounds.  Many different targeted strategies have been applied to search the broad range lifetimes in between~\cite{Aad:2014gfa,Chatrchyan:2012jwg,Aad:2013yna,ATLAS-2013-092,Khachatryan:2014mea,Aad:2013gva,CMS:2014wda,CMS:2014hka,ATLAS-2014-041,ATLAS:2012av,Aad:2012kw,Aad:2014yea,Aaij:2014nma,Khachatryan:2015jha}.  Because energetic particle production originating in the bulk of the detector is extremely rare, most of these searches benefit from tiny backgrounds, often $O$(1) event or smaller while maintaining good signal efficiency.  The non-observation of excesses in such clean searches then begs the question of their implications for more general classes of models, where even very low efficiency can still lead to significant limits.  Several recent phenomenological works have investigated the power of these searches, or proposed new searches for similarly striking signals~\cite{Strassler:2006im,Strassler:2006ri,Meade:2010ji,Graham:2012th,Jaiswal:2013xra,Csaki:2013jza,Covi:2014fba,Buckley:2014ika,Falkowski:2014sma,Cui:2014twa,Cahill-Rowley-Talk,Schwaller:2015gea,Jung:2015boa}.

In this paper, we attempt to develop a more refined understanding of the status of SUSY scenarios with displaced decays, in light of the small but powerful collection of existing LHC displaced particle searches.  Most of the models that we study have either never been searched for at nonzero lifetime, or explicit searches cover only one possible signature out of several.  In practice, it is typical for each highly specialized displaced decay search to phrase its results in terms of only one or a handful of highly specialized new physics models.  This is understandable, given the vast range of possible interesting models and the computational overhead required to fully simulate and interpret them.  However, inferring the implications of those searches for a different model, and in particular how they might interplay with each other in covering the parameter space of that different model, then requires careful recastings.  We present here, what we believe to be for the first time in the context of our chosen SUSY models, a comprehensive set of such recastings for multiple displaced particle searches simultaneously.  Our results highlight the strengths and weaknesses of the various searches, and give a clear indication of what regions may currently be lacking in sensitivity.  The process of undertaking these recastings also illustrates some of the difficulties and ambiguities that can arise when attempting to extrapolate the results of displaced decay searches beyond their original target models, especially given the unconventional approaches to event reconstruction.  A saving feature is that total rates near the boundaries of sensitivity are usually very strong functions of both mass and lifetime, such that even $O$(1) uncertainties in our estimation of experimental acceptance can still lead to only $O$(10\%) uncertainties in model reach.  Nonetheless, where possible we point out aspects of the searches that could be particularly prone to mismodeling by recasters.  We also make several suggestions for how some of these searches might be adapted to serve as more powerful probes of SUSY or other models beyond their original targets.

Of course, even restricting ourselves to simple variations on the particle content of the Minimal Supersymmetric Standard Model (MSSM), the variety of possible displaced final-state signatures is extremely rich.  To narrow down the possibilities to a manageable level, we first of all focus on simplified models where only one type of superparticle is produced, and undergoes a single-stage displaced decay back into SM particles and/or the LSP.  These simplified models can generally be embedded into a variety of more complete spectra with additional production channels, such that our results are both broad in applicability and conservative within any given model.  (For a more inclusive survey approach, see~\cite{Cahill-Rowley-Talk}.)  Within the still rather large set of possible simplified models, we focus on ones that have a sizable fraction of hadronic visible decays, either directly or due to subsequent decays of electroweak bosons.  Such hadronic signals are nominally the most challenging and the least constrained by explicit displaced searches, and in some cases unconstrained or only mildly constrained even in prompt decay searches.  Significantly, some of the simplified models that are most motivated by naturalness~\cite{Dimopoulos:1995mi,Cohen:1996vb,Brust:2011tb,Papucci:2011wy,Kats:2011qh} can fall into this category, including direct production of the lightest stop eigenstate or of a quasi-degenerate multiplet of Higgsinos.  All together, the models that we consider include:
\begin{itemize}
\item  $\tilde t \to \bar d_i \bar d_j $  via baryonic RPV, including $\tilde t \to \bar b \bar b $~\cite{Csaki:2013jza} (Figs.~\ref{fig:stop},~\ref{fig:stopDynamical})
\item  $\tilde g \to u_i d_j d_k$ via baryonic RPV (Fig.~\ref{fig:RPVgluino})
\item  $\tilde H \to u_id_jd_k \; (+ {\rm soft})$ via baryonic RPV (Fig.~\ref{fig:RPVHiggsino})
\item  $\tilde q \to q \: \tilde G$ in GMSB (Fig.~\ref{fig:GMSBsquarks})
\item  $\tilde g \to g \: \tilde G$ in GMSB (Fig.~\ref{fig:GMSBgluino})
\item  $\tilde t \to t^{(*)} \: \tilde G$  in GMSB (Fig.~\ref{fig:GMSBstop})
\item  $\tilde H \to h/Z \: \tilde G \; (+ {\rm soft})$ in GMSB (Fig.~\ref{fig:GMSBhiggsino})
\item  $\tilde g \to q\bar q \tilde B$ in mini-split SUSY (Fig.~\ref{fig:MiniSplitgluino})
\end{itemize}
Some major options missing from this list are sleptons, electroweak gauginos, simplified spectra with leptonic RPV, and mini-split SUSY with gluino decays dominated by heavy flavor.  As discussed in more detail below, some of these other possibilities are covered already by existing searches or recasts, and some we expect to have significant overlap with the above signals, but some would also be worth a closer look in future work.

The most powerful displaced decay limits within our selection of models typically come from the CMS tracker-based search for displaced dijets~\cite{CMS:2014wda}.  This search often remains sensitive to models with $c\tau$ much larger than the tracker radius, as well as to models with decay topologies different from the nominal dijets.  For models where the long-lived particle is colored, hadronization implies a sizable charged fraction that can also be picked up by stable charged particle searches in events where the decay takes place outside of the detector.  Similarly, these searches maintain some sensitivity for $c\tau$ much smaller than the 5--10~m outer detector radius.  The overlap of exclusions between displaced decay searches and stable charged particle searches can then be significant, sometimes more than three orders of magnitude in lifetime.  At the low end of the lifetime range, prompt searches also become sensitive.  While it is not possible for us to precisely map out the lifetime range over which these searches remain efficient, conservative guesses again allow for significant overlap.  This complementarity often allows for exclusions that span from prompt lifetimes to infinity with no gaps.

The main results of this paper consist of a series of exclusion plots over the mass-lifetime plane of each displaced particle, Figs.~\ref{fig:stop} through ~\ref{fig:MiniSplitgluino}.  For the colored production models, the mass reach in the $c\tau$ range of $O$(mm--m) is usually comparable to, and in some cases better than, the $\approx$~1~TeV reach from collider-stable charged particle searches.  In particular stops, which are expected to have mass less than about 1~TeV in a natural model, have very little viable model space surviving in this lifetime range under these decay scenarios.  For the electroweak Higgsino production, stable charged particle limits do not apply, and prompt searches are typically limited in sensitivity, but a large number of displaced searches yield powerful limits, especially in the GMSB case.  We find that for $c\tau \sim 10$~cm, masses below about 600--800~GeV are excluded, giving serious tension with naturalness at those lifetimes.  For natural masses near 100~GeV, the excluded lifetime ranges from $O$(10~microns) to $O$(10~m) in RPV, and up to $O$(100~m) in GMSB, dominated there by CMS's tracker-based displaced dilepton search~\cite{CMS:2014hka}.

For all models, the region of lifetimes around $c\tau \sim 10$~m could in principle benefit from searches in the hadronic calorimeters and muon chambers, such as those performed by ATLAS~\cite{ATLAS:2012av,ATLAS-2014-041}.  But the existing searches are highly limited in sensitivity by their focus on lower-mass models and by requiring very tight reconstruction cuts on both sides of the event.  For the Higgsino models, improvements in this direction might be the only option for extending the sensitivity to higher lifetimes, without ultimately appealing to more standard-style SUSY searches that assume that assume that both final-state Higgsinos escape the detector unseen.

The paper is organized as follows.  In the next section, we review the existing LHC collider-stable and displaced particle searches that we use in our limit-setting.  (This section may be bypassed by a reader who is not interested in the details of these analyses.)  Section~\ref{sec:models} specifies the motivations and features of the simplified SUSY models under investigation, and presents our derived limits.  We conclude and present some ideas for future searches in Section~\ref{sec:conclusions}.  An appendix discusses the details and calibrations of our detector simulations used for recasting.

\section{The LHC Searches Under Consideration}
\label{sec:searches}

Displaced decay searches at the LHC are currently limited to a handful of specific new physics scenarios.\footnote{Displaced decay searches have also previously been carried at the Tevatron~\cite{Aaltonen:2009kea,Aaltonen:2013har,Aaltonen:2011rja,CDF-7244,Abazov:2012ina,Abazov:2009ik,Abazov:2008zm,Abazov:2007ht,Abazov:2006as} and at LEP~\cite{Barate:2000qf,Barate:1998zp,Barate:1999gm}.  These searches have for the most part either been superseded by the LHC or do not have immediate relevance to the SUSY models we consider.  We do not attempt to recast any of them.  However, we practically assume that long-lived particles below 100~GeV should have been highly visible to some of these searches.}  Searches that target minimal SUSY include non-pointing photons in gauge mediation (assuming a mostly-bino LSP)~\cite{Aad:2014gfa,Chatrchyan:2012jwg}, the ``disappearing track'' signature of NLSP charginos in anomaly mediation~\cite{Aad:2013yna,CMS:2014gxa}, displaced leptons from neutralino or stop decays with leptonic RPV~\cite{ATLAS-2013-092,CMS:2014wda,Khachatryan:2014mea}, and late decays of gluino R-hadrons stopped in the calorimeters in mini-split SUSY~\cite{Aad:2013gva,Khachatryan:2015jha}.  Other searches focus on models such as Hidden Valleys~\cite{Strassler:2006im,Strassler:2006ri,CMS:2014wda,CMS:2014hka,ATLAS-2014-041,ATLAS:2012av,Aaij:2014nma} or light hidden-sector gauge bosons~\cite{Aad:2012kw,Aad:2014yea}.  Recently, ATLAS has also re-interpreted its prompt gluino limits, accounting for the effect of displacement on the signal acceptance~\cite{ATLAS-2014-037}, results that we put into broader context here.  CMS has re-interpreted its stable charged particle searches for a large ensemble of pMSSM models with long but finite lifetimes~\cite{CMS-EXO-13-006}, and we apply a similar strategy to our more focused set of models.

\begin{table}
\begin{center}
\begin{tabular}{ l|c|c|c|c|c }
                        & \ beam energy \ & lumi & refs & \ our analysis \ & \ our calibration   \\ \hline
CMS heavy stable charged & 7+8~TeV & \ 5.0+18.8~fb$^{-1}$  & \cite{Chatrchyan:2013oca} & \ref{sec:CMSHSCP_analysis} & Appendix~\ref{sec:CMSHSCP_calibration}  \\
CMS displaced dijets    & 8~TeV & 18.5~fb$^{-1}$ & \cite{CMS-EXO-12-038,CMS:2014wda} &  \ref{sec:CMSdisplacedDijets_analysis} & Appendix~\ref{sec:CMSdisplacedDijets_calibration}\\
CMS displaced dileptons & 8~TeV & 19.6/20.5~fb$^{-1}$ & \cite{CMS:2014hka} &  \ref{sec:CMSdisplacedDileptons_analysis} & Appendix~\ref{sec:CMSdisplacedDileptons_calibration} \\
CMS displaced $e$+$\mu$ & 8~TeV & 19.7~fb$^{-1}$ & \cite{Khachatryan:2014mea} &  \ref{sec:CMSemu_analysis} & Appendix~\ref{sec:CMSemu_calibration} \\
ATLAS muon spectrometer \ &  7~TeV & 1.94~fb$^{-1}$ & \cite{ATLAS:2012av} &  \ref{sec:ATLASmuonChamber_analysis} & Appendix~\ref{sec:ATLASmuonChamber_calibration}  \\
ATLAS low-EM jets & 8~TeV & 20.3~fb$^{-1}$ & \cite{ATLAS-2014-041} &  \ref{sec:ATLASlowEM_analysis} & Appendix~\ref{sec:ATLASlowEM_calibration}  \\
ATLAS $\mu$+tracks & 8~TeV & 20.3~fb$^{-1}$ & \cite{ATLAS-2013-092} &  \ref{sec:ATLASmuonTracks_analysis} & Appendix~\ref{sec:ATLASmuonTracks_calibration}  \\
\end{tabular}
\end{center}
\caption{A summary of the LHC searches recast in this paper.}
\label{table:searches}
\end{table}

From this modest but growing list of analyses, we select seven that appear to be of greatest relevance for the SUSY models studied in Section~\ref{sec:models}:  the CMS heavy stable charged particle search~\cite{Chatrchyan:2013oca}, the CMS displaced dijets search~\cite{CMS:2014wda}, the CMS displaced dileptons search~\cite{CMS:2014hka}, the CMS displaced electron and muon search~\cite{Khachatryan:2014mea}, the ATLAS muon chamber search\cite{ATLAS:2012av}, the ATLAS low-EM jet search~\cite{ATLAS-2014-041}, and the ATLAS displaced muon plus tracks search~\cite{ATLAS-2013-092}.  Except for the ATLAS muon chamber, all of these have been performed with the full 8~TeV dataset.  The following subsections summarize the relevant aspects of each analysis that we use for our recasts, as well as commentary on the reliability of these recasts where appropriate.  Our approximate reproduction of each of these analyses relies on simplified detector simulations.  Descriptions of these simulations and their calibration to known experimental results is provided in a corresponding set of subsections in the appendix.  Table~\ref{table:searches} provides a compact overview, including the associated references and subsections.

\subsection{CMS Heavy Stable Charged Particles}
\label{sec:CMSHSCP_analysis}

Both ATLAS~\cite{ATLAS:2014fka} and CMS~\cite{Chatrchyan:2013oca} have conducted searches for heavy stable charged particles~(HSCP) that traverse the entire detector, and appear as a ``heavy muon'' with anomalously small velocity or $dE/dx$ in the detector material.  We choose to focus on the CMS searches~\cite{Chatrchyan:2013oca}, though we expect very similar performance from the ATLAS searches.  Stable squarks and gluinos have already been explicitly considered by both experiments, and will simply generalize these results to cases with finite lifetimes.

The only major subtlety when dealing with meta-stable colored particles is that they are only seen bound into R-hadrons~\cite{Farrar:1978xj}.  The hadronization fractions can be estimated from simple models, and are likely fairly accurate for squarks given the extensive theoretical and experimental experience with heavy quarks.  Hadronization of the color-octet gluino is less certain, but we assume here the default behavior in \Pythia8~\cite{Sjostrand:2007gs}.  (This results in a charged hadronization fraction of approximately 46\%.)  A more subtle issue is how these R-hadrons interact with the detector material, especially the chance that a charged R-hadron will pass through the calorimeters without a net charge exchange, and thus manage to trigger in the muon system.  CMS considers two models: a nominal hadronic cloud interaction and a more extreme ``charge-stripped'' assumption where all R-hadrons emerge from the back of the calorimeter in a neutral state.  Thankfully, the complicated interplay with the detector has been accounted for by CMS, and to extract our own finite-lifetime limits we can concentrate on simpler, geometric considerations.

We consider two of their search strategies.  For the nominal hadronic interaction model, we take the tracker plus time-of-flight analysis.  For the pessimistic charge-stripped assumption, we take the tracker-only analysis.

The tracker plus time-of-flight analysis relies dominantly on the muon trigger, and searches for anomalous track candidates that are matched between the muon chamber and inner tracker.  This track must be reconstructed with $|\eta| < 2.1$ and $p_T > 70$~GeV, inverse-velocity above $1.225/c$ (measured using timing information), and a high $dE/dx$.  There are also additional requirements on the mass inferred from the momentum and velocity measurements, which are constructed to be highly efficient for signal.  In order to recast the cross section limit for a given model at finite lifetime, we form a conservative rescaling factor.  The numerator is the number of charged R-hadrons that would pass into the analysis given the above cuts, excepting the $dE/dx$ cut, which we cannot model but which should also be highly efficient for signal.  The denominator is the number of charged R-hadrons that pass these criteria and that also decay fully outside the detector.  We only consider decaying R-hadrons where {\it none} of the visible decay products re-intercept the detector volume, as this may cause additional activity in the muon chambers and could have an unpredictable effect on the acceptance.\footnote{CMS has also provided a full efficiency map of this analysis~\cite{CMS-EXO-13-006}, which can be extremely useful in general recasts.  However, we do not use this map since our physics models are identical to the ones that CMS studies, up to the finite lifetime.  There could in principle be some interplay between the variation in stable particle acceptance and non-decay probability versus kinematics, which we are not simulating, but our treatment should be conservative.  For example, slow particles would tend to decay earlier and become vetoed from the analysis, but slow stable particles (especially with $\beta \lsim 0.4$) are anyway less efficiently accepted.  Similarly, particles at higher $|\eta|$ must survive over a longer three-dimensional path length before exiting the detector, and again are more likely to be lost due to decay, but high-$|\eta|$ is also less efficient even for stable particles.  Therefore, our naive approach, which effectively assumes a flat acceptance within the fiducial region, misses the fact that the particles that survive undecayed also tend to be in kinematic regions with higher acceptance.  In any case, the turn-off of overall acceptance for this analysis due to decays at lower lifetimes is {\it exponential}, and we expect this behavior to dominate.}

In order to access the charge-suppressed scenario, the tracker-only analysis exploits a subtlety of the \met\ trigger.  A charged R-hadron may leave a track in the inner tracker, but if it leaves no track in the muon chamber and minimal calorimeter activity along its trajectory, the particle-flow algorithm used in triggering will assume that the track is spurious and not count it toward the \met\ calculation.  The R-hadron therefore adds to the apparent \met.  Because each R-hadron either leaves such a ``trigger-invisible'' track or is neutral to begin with, the apparent \met\ is the total recoil $p_T$ of the heavy particle pair.\footnote{For all SUSY pair production, processed through \Pythia8, we damp the ISR, which has been shown to better-reproduce matched results~\cite{Corke:2010zj}.}  Offline, events from this \met-triggered sample can be analyzed for inner tracks consistent with heavy stable particles.  The basic track $|\eta|$ and $p_T$ requirements are the same, but there is no velocity cut (as no timing information with respect to the muon chambers is available), and the $dE/dx$ requirement is tightened.  Again, we cannot model the $dE/dx$ cut, so we assign an ad hoc velocity ceiling of $0.7c$, which puts us on the steep section of the Bethe-Bloch stopping power curve~\cite{Agashe:2014kda}.  (Our final results are not very sensitive to the placement of this velocity cut.)  We again form a rescaling factor for the infinite-lifetime cross section limits presented by CMS.  For the numerator, we take the number of {\it events} where the recoil $p_T$ exceeds 150~GeV and at least one R-hadron is charged and passes the reconstruction cuts.  The denominator is the number of events that satisfy these criteria and where both R-hadrons decay fully outside the detector (including the non-intercept requirement on the visible daughters, as above).

\subsection{CMS Displaced Dijets}
\label{sec:CMSdisplacedDijets_analysis}

The CMS displaced dijet search~\cite{CMS-EXO-12-038,CMS:2014wda} uses a specialized trigger to capture events containing a pair of high-$p_T$ jets containing displaced tracks at the level of several hundred microns.  In the offline analysis, it counts the total number of jet-pairs that appear to be consistent with common vertices with a large number of such displaced tracks.  For our recasts, we focus on the ``High-$\Lxy$'' analysis, which has $1.14 \pm 0.54$ expected background vertices and one observed, placing an upper limit of $3.7$ signal vertices.  We have found that the High-$\Lxy$ works well for all of our models, even ones with short decay lengths ($\vev{\Lxy} < 20$~cm), and that the choice of High-$\Lxy$ versus the very similar Low-$\Lxy$ analyses has only minor impact on our results.

The jets used in the analysis have $p_T > 60$~GeV and $|\eta| < 2$, and the total event $H_T$ must exceed 300~GeV at trigger level.  (In practice we use a slightly tighter 320~GeV to account for the observed turn-on of the trigger with $H_T$ measured offline~\cite{Zuranski-talk}.)  Each pair of such jets is inspected for associated tracks with impact parameters larger than 500~$\mu$m, and this set of displaced tracks is checked for consistency with a common vertex.  At least one displaced track from each jet is required to fit that vertex, and there are a number of additional quality requirements on the vertex itself:  total track-mass greater than 4~GeV, total track-$p_T$ greater than 8~GeV, ``significant'' transverse distance $\Lxy$ from the primary vertex, and a multivariate likelihood-ratio discriminant cut.  The discriminant is formed from distributions over the vertex track multiplicity, the fraction of tracks with positive impact parameters (based on the sign of the dot product of the track's $p_T$ vector and transverse displacement vector at the transverse point of closest approach to the beamline), and two additional variables based on a special clustering of track crossing points along a line starting at the detector's center and oriented with the dijet $p_T$ direction.  The exact algorithm for this clustering is not given by CMS.  For our reproduction of the analysis, we create a sliding window of full-width $0.15\times\Lxy$, and adjust it to surround a maximal number of crossing points.  When multiple window locations would surround different crossing point collections with equal multiplicities, we choose the one with the smallest RMS.  For its multivariate discriminant, CMS includes both the cluster multiplicity and the RMS relative to the vertex $\Lxy$.  We determine the event-by-event value of the discriminant by using CMS's own distributions for the four variables.  (Our calibration distributions for these variables and the multivariate discriminant can be found in Appendix~\ref{sec:CMSdisplacedDijets_calibration}.)

The rest of the High-$\Lxy$ selection demands at most one prompt track per jet within the dijet candidate, less than 9\% of the energy of each jet associated to prompt tracks, and a multivariate discriminant value greater than 0.8.  In the analysis note~\cite{CMS-EXO-12-038} (which has identical results as the more recent preprint~\cite{CMS:2014wda}), CMS's new physics limit is phrased in terms of the number of dijet candidates that pass all of these requirements over the full 8~TeV run.

While this defines the basic search, we point out a few possible subtleties:
\begin{itemize}
\item  CMS has only performed full simulation on models with dijet masses up to 350~GeV.  We assume that there are no major obstacles to probing masses beyond 1~TeV.  These may experience more tracking confusions due to the greater multiplicity of hits, but on the other hand should also be capable of surviving the analysis cuts with a smaller fraction of successfully-reconstructed tracks.
\item  For all of the physics models that CMS has studied, the displaced particle is neutral and leaves no tracker hits before its decay.  It is unclear what would happen for charged displaced particles, such as R-hadrons, which would leave a signature sometimes called an ``exploding track.''  Presumably the extra hits would lead to additional confusions of the tracker algorithms.  We simply exclude such cases from the analysis, effectively assuming zero efficiency.  This is certainly an over-conservative treatment, especially, for decays that occur before crossing the first pixel layer.
\item  For some of our SUSY models, more than two jets can originate from the same displaced vertex.  In principle, in the note~\cite{CMS-EXO-12-038} CMS considers all possible displaced dijet pairs, allowing the same jet to appear multiple times.  Such a decay could therefore contribute much more than one dijet candidate.  However, since the exact procedure is not unambiguously described in the CMS note, when a dijet pair passes the basic selection cuts (before the High-$\Lxy$ selections), we remove its constituent jets from further consideration for constructing other dijet pairs.
\item  For most of our SUSY models, the displaced dijet candidate will be produced in a decay with additional activity, and will therefore by itself not reconstruct the decaying particle's momentum vector.  Since CMS's cluster discriminant variables are constructed under the assumption that the vertex displacement vector and dijet momentum vector are well-aligned, there is a question of whether the discriminant is particularly inefficient for models where this is no longer true.  We have found that any such effect is quite minor, and that the multivariate discriminant is mainly driven by the vertex track multiplicity variable.\footnote{The same conclusion was reached in~\cite{Cui:2014twa}.}
\item  When a decay contains bottom and/or charm quarks, it may generate multiple nearby displaced vertices rather than a single displaced vertex.  CMS gives some explicit indication of how the reconstruction rate differs for heavy flavor decays, and it appears that such small secondary displacements do not play a significant role, but only for models down to $O$(cm) lifetimes.  The behavior for shorter lifetimes is not specified.  Due to the ambiguity, we simply ignore decays that contain heavy flavor and have $\Lxy < 1$~cm.
\item  It is clarified in the more recent analysis preprint~\cite{CMS:2014wda} that no lepton identification is utilized, and that electrons and muons would function as jet constituents in this analysis.  However, the earlier note~\cite{CMS-EXO-12-038}, on which we base our analysis, is not explicit on this point.  We have therefore excluded isolated leptons (as per the definition of~\cite{CMS:2014hka}) from jet clustering.  This may lead to slightly over-conservative limits when leptonic decays are available.
\end{itemize}

\subsection{CMS Displaced Dileptons}
\label{sec:CMSdisplacedDileptons_analysis}

The CMS displaced dilepton search~\cite{CMS:2014hka} is in some ways a simpler version of the displaced dijet search above.\footnote{We do not consider the superceded version of this search at lower luminosity and beam energy~\cite{Chatrchyan:2012jna}.}  The analysis operates on pairs of isolated $e^+e^-$ or $\mu^+\mu^-$, demanding that a pair reconstruct to a common highly-displaced vertex.  The analysis is both background-free and has zero observed events, leading to a 95\% limit of approximately 3 signal events.

Electron-pair candidates in the event must consist of a leading (subleading) electron with $p_T > 40$~GeV (25~GeV).  Muon-pair candidates must consist of muons with $p_T > 26$~GeV.  Each lepton should be in the well-instrumented region of the tracker, at $|\eta| < 2.0$, and should have a ``significant'' transverse displacement of 12$\sigma$ relative to the resolution.  Since here we do do not simulate track-by-track resolutions, we simply replace this last criterion with a fixed cut of 250~$\mu$m.  This choice appears to give conservative results for very low lifetimes.  CMS further requires the leptons to be isolated from other tracks, excluding other identified leptons, using hollow cones of outer radius $0.3$ built around the lepton candidates.  Since our own lepton ID is ``perfect'' (up to the efficiency factors discussed in the appendix), we use solid cones.  This again should furnish a conservative approximation, and should roughly approximate the ID failures that would occur in reality, e.g. for leptons inside of $b$/$c$-jets.  Non-lepton tracks with $p_T > 1$~GeV count toward the isolation, and should tally to less than 10\% of the lepton $p_T$.  For muon-pairs, the two candidate tracks must be separated by $\Delta R > 0.2$, to operate in the region of high dimuon trigger efficiency.  For electron-pairs, we here additionally require $\Delta R > 0.1$, assuming that the ECAL patterns of closely-spaced electrons could start to become difficult to reconstruct.  A number of other basic quality cuts are applied: the dilepton pair should have invariant mass above 15~GeV to avoid hadronic resonances, the azimuthal angle between the dilepton momentum vector and its displaced vertex position vector should be less than $\pi/2$, and the 3D opening angle between a pair of candidate muon tracks should have a cosine greater than $-0.79$ to veto cosmics.

Given the simpler topology and cuts relative to the displaced dijet search, we should naively be more confident in the robustness in our recasting procedure for different models.  Nonetheless, as discussed in Appendix~\ref{sec:CMSdisplacedDileptons_calibration}, our modeled acceptances appear to be low for some of CMS's benchmark models, possibly because we err on the conservative side for the lepton track-finding efficiencies at large decay radii.  The ``exploding track'' question also persists in principle, but does not arise in any of the models that we consider.  (Displaced dileptons only occur for our GMSB Higgsino models, in which case the long-lived particle is always neutral.)  There is still some question about how this analysis would behave when additional tracks emerge from the same vertex, such as $h \to ZZ^* \to l^+l^-$+jets from Higgsino decay.  However, this model is the only case that we study where such a question would arise, and the limits from this search are regardless not the most powerful for displaced Higgs decays.

\subsection{CMS Displaced Electron and Muon}
\label{sec:CMSemu_analysis}

The CMS displaced electron and muon search~\cite{Khachatryan:2014mea} uses a highly minimalistic and inclusive analysis, simply demanding the presence of exactly one electron and one muon, each with significant 2D impact parameter up to 2~cm.  The analysis is broken down into three exclusive bins in joint impact parameters, with varying degrees of background.  Consequently, the statistical analysis is somewhat more involved, as discussed below.

The electron and muon must each have $p_T > 25$~GeV, $|\eta| < 2.5$, and transverse impact parameter between $0.02$~cm and 2~cm.  The last requirement allows the analysis to focus on the region of tracking parameters where the efficiencies are largely the same as for prompt tracks.  The two leptons must also be well-isolated.  In addition to a basic particle isolation requirement, which we form by tallying all hadrons (charged and neutral) within $\Delta R < 0.3$ and demanding a relative $p_T$ less than 10\%, each lepton must also be isolated from jets with a $p_T$ threshold of 15~GeV within $\Delta R < 0.5$.  The leptons themselves must also be separated by at least this distance and have opposite charges.

The three exclusive signal regions consist of a high-displacement/lower-background region SR3 where both leptons' impact parameters are above $0.1$~cm, an intermediate region SR2 where the event fails this criterion but still has impact parameters above $0.05$~cm, and a low-displacement/higher-background SR1 where the event fails both of these criteria but still has impact parameters above $0.02$~cm.  The observed (expected) event counts are, respectively, 0 (0.051), 0 (1.01), and 19 (18.0).  We have coded this three-bin statistical analysis as a toy Monte Carlo using a Poisson likelihood-ratio discriminator built from the central background predictions, and including the systematic uncertainties on the background (assumed Gaussian and uncorrelated) as perturbations on the simulated pseudo-experiments.  This allows us to map out the 95\% $CL_S$ boundaries in the three-dimensional space of signal bin counts.  Our statistical analysis is not exactly the same as that performed by CMS, but it should furnish an adequate approximation.  We have verified this by reproducing sections of the leptonic RPV stop limits presented in the analysis note.  Approximately speaking, high-lifetime signals that are concentrated in SR2 and/or SR3 have a limit $N$(SR2)+$N$(SR3) $ < 3$, whereas low-lifetime signals that are concentrated in SR1 have a limit $N$(SR1) $ < 13.5$.

Other than the nontrivial statistical analysis, this particular displaced search was one of the simplest for us to implement, since it is insensitive to efficiency degradations and nontrivial geometries that occur for decays in the body of the detector.  Also, due to the fact that the focus is on decays that occur before traversing the pixels, the issue of whether the displaced particle is charged does not appear.

\subsection{ATLAS Muon Spectrometer}
\label{sec:ATLASmuonChamber_analysis}

The ATLAS muon spectrometer search~\cite{ATLAS:2012av} (7~TeV, 2~fb$^{-1}$) is focused on models where particle decay lengths are several meters, and have high probability of decaying outside of the HCAL.  It uses a novel vertex-finding algorithm~\cite{Aad:2013ela} to identify the sprays of tracks from a displaced decay within the muon chambers.  Events with two successfully identified candidates are used in the analysis, with zero events expected and zero observed (again, setting an upper limit of about 3 signal events).

To pass the analysis cuts, first of all both decays must occur within fiducial regions of the muon spectrometer.  The identified decays must be well-isolated from tracks above 5~GeV out to $\Delta R < 0.4$, and from jets above 15~GeV out to $\Delta R < 0.7$.  (Again, we do not consider displaced decays from charged particles, since the particle's own track would veto it.)

This analysis appears to fill an important niche at lifetimes intermediate between the tracker radius and beyond the outer detector radius for colored long-lived particles, and uniquely probes out to the largest possible decay distances for long-lived particles that lack charged states.  Nonetheless, we will see that the search is ultimately not very powerful.  There are several reasons for this:
\begin{itemize}
\item  Unlike all of the other analyses that we study here, it has not (yet) been performed at the full beam energy and luminosity.  We will indicate how much this might help below by making a naive projection to 8~TeV, 20~fb$^{-1}$, assuming identical signal efficiencies and zero background.
\item  The ATLAS data acquisition becomes highly inefficient for particles traveling large distances at sub-light speeds, and loses the signal when a time delay of $\approx$~7~ns has accrued relative to a light-speed particle.  For example, a particle traveling at $0.7c$ and decaying at $r = 5$~m (at $\eta = 0$) accrues this much delay, and this velocity is already above the typical median for pair-produced heavy particles.  The timing requirement therefore usually has an $O(1)$ impact on our signal acceptance.
\item  The analysis requires coincident behavior for both decays, and therefore pays every possible inefficiency factor twice.  These include basic requirements such as neutral R-hadronization probabilities, the geometric constraints of the muon spectrometer, the vertexing efficiency, and the isolation requirements.  The geometry in particular becomes a major factor when considering either very long or very short lifetimes, where it respectively leads to either a power suppression or an exponential suppression.  Analyses that can rely on one candidate only need to be ``lucky'' once.
\end{itemize}

Regarding the last point, it would be very interesting to see if this analysis could be run with a single-candidate option.  On the one hand, this would result in much higher background rates (effectively $O$(10~pb)) with the original reconstructions and cuts.  On the other hand, many of the models that we consider here have quite appreciable cross sections, and because they have much higher masses than the baseline models that ATLAS studied, should lead to even more spectacular multi-track signatures in the muon spectrometer.  Relaxing the isolation requirements somewhat, to allow more of the decay particles to point back to the HCAL, could also be beneficial.  These HCAL signals would also be rather distinctive given that they would contribute mainly in the outermost layers.  However, it is unclear whether ATLAS's jet reconstruction requirements would anyway ignore such anomalous deposition patterns.

It should be also noted that the information on the performance of the displaced vertex reconstruction from ATLAS's papers is limited to a rather small set of new physics models, using the common benchmark scenario of a Higgs-like scalar decaying to a pair of displaced pseudoscalars.  Only four mass points with fairly similar kinematics are studied, the most energetic decays coming from a 140~GeV scalar decaying into a 40~GeV pseudoscalar.  It is therefore unclear how this very complex search would perform on, say, a 1~TeV RPV gluino decaying into three jets.  Moreso than most of our other recasts, this one must be then viewed cautiously and somewhat conditionally.  Still, because of the search's limitations discussed above, in practice it does not end up probing masses beyond a few hundred GeV.

\subsection{ATLAS Low-EM Jets}
\label{sec:ATLASlowEM_analysis}

The ATLAS low-EM jets search~\cite{ATLAS-2014-041} focuses on coincident jet-like signals confined entirely to the HCAL, with stringent cuts on nearby ECAL and tracker activity.  It is sensitive to pairs of displaced decays within the HCAL volume.  The analysis has a small but non-negligible background, leading to an upper limit of about 20 signal events.

For a pair of jets to pass into the analysis, the leading (subleading) jet must have $p_T > 60$~GeV (40~GeV).  Each jet should have no associated tracks with $p_T > 1$~GeV within $\Delta R < 0.2$, and should have at least approximately 16 ($10^{1.2}$) times more energy recorded in the HCAL than in the ECAL.  As discussed in Appendix~\ref{sec:ATLASlowEM_calibration}, our default model of these isolation requirements uses a combination of a flat efficiency factor and an overconservative veto on decays in the HCAL body that produce particles pointing back to the ECAL.  An alternative, looser version removes the latter requirement, and the two extremes define an approximate error band for our modeling of this analysis.  We also automatically veto events containing charged long-lived particles, which as usual would leave a track.  ATLAS further makes an explicit cut of 5~ns on the signal timing delay relative to what would be deposited by a particle moving at light-speed.  Given that the linear distances involved are about half as large as those for the muon spectrometer analysis, this ends up having a relatively less detrimental effect (though still potentially $O(1)$) on the signal acceptance.  Finally, there is a requirement that the event has $\met < 50$~GeV in order to reject non-collision events, though this can severely impact many of our models with invisible LSPs.

Similar to the ATLAS muon spectrometer search, the requirement of coincident decays within the same detector system significantly limits the model reach.  For our SUSY models, much of the candidate-by-candidate inefficiency comes from our somewhat ad hoc requirement that no decay particles can point back to the ECAL, and we take results with and without this requirement to define our uncertainty band.  Still, we conclude that this search is not very competitive, even to a luminosity-scaled muon spectrometer search.  It seems as if it is simply too constrained by geometry.

As in the previous subsection, it would potentially be useful if the search could be adapted for single-candidate acceptance instead of requiring decays with coincident properties.  The fact that this would increase the background could be offset in several ways that should maintain high signal acceptance.  First of all, most of the SUSY models that we consider would deposit far more energy than ATLAS's benchmark models.  A search that is broken down into different jet $E_T$ bins could already improve sensitivity.  Second, these very energetic and wide-angle multibody decays might leave quite unusual 3D spatial and timing deposition patterns in the HCAL, possibly so unusual that even the relative ECAL deposition requirement could be relaxed.  In fact, we are again already giving ATLAS the benefit of the doubt by assuming that such unusual jets would pass the reconstructions of their analysis.  But the limited mass reach of the search by itself limits the possible impact of this subtlety.

It would also be very advantageous if the $\met$ cut could be eliminated.  For GMSB and mini-split models in particular, the cut is a major handicap.  It can also contribute a subtle geometric problem in models without true $\met$.  Under the zeroth-order assumption that a particle decaying within the HCAL has all of its energy absorbed at one point, the particle's reconstructed momentum vector is effectively rescaled by its lab-frame energy.  For sparticle pair production that is not exactly back-to-back in 3D space, true transverse momentum balances, but the energy-scaled transverse momentum need not.  Again, this is mainly an issue for models with large masses, beyond the sensitivity of the original search.  But it could become problematic if the search were to be extended.

\subsection{ATLAS Displaced Muon Plus Tracks}
\label{sec:ATLASmuonTracks_analysis}

The ATLAS muon plus tracks search~\cite{ATLAS-2013-092} uses the inner tracker to reconstruct highly displaced vertices containing at least one muon.\footnote{We do not consider the superceded versions of this search at lower luminosity and beam energy~\cite{Aad:2012zx,Aad:2011zb}.}  The basic search, which counts the number of events containing at least one such vertex, is background-free.  However, a looser version of the search relaxes the demand that the muon is matched to the displaced vertex candidate.  This version is also background-free, and has improved signal acceptance, especially for models with a mixture of low-multiplicity leptonic decays and high-multiplicity hadronic decays (i.e., our GMSB Higgsino and stop models).  We utilize this version for our recasts.

Vertices are reconstructed from displaced tracks originating within the inner region of the tracker, $r < 18$~cm and $|z| < 30$~cm.  The tracks used toward the vertex reconstruction must have $p_T > 1$~GeV and transverse impact parameter greater than 2~mm.  A good candidate vertex must be reconstructed within the fiducial volume, have at least five associated tracks, a track-mass greater than 10~GeV.  The muon selection is $p_T > 55$~GeV and $|\eta| < 1.07$, and transverse impact parameter greater than 1.5~mm.

Unlike the other ATLAS searches, this one requires neither very tight activity cuts nor coincident behavior for the decays.  Consequently, it is much more powerful within its realm of applicability.  Again, we conservatively ignore displaced particles produced in the decays of charged R-hadrons. 

Because this is a tracker-based analysis, similar to the CMS displaced dijets, the possibility also exists for subtleties when heavy flavor secondary decays occur after the displaced decay.  However, ATLAS states explicitly that vertices less than 1~mm from one another are merged.  While the exact behavior for heavy flavor final-states is not given by ATLAS, we assume that this merging procedure effectively makes them insensitive to this issue.

\section{Models and Limits}
\label{sec:models}

We now describe three of the well-known scenarios that lead to displaced sparticle decays, and present our new limits on several simplified models within those scenarios.  In all three scenarios, we consider models with LSP or NLSP gluinos, which are common search targets due to their enormous pair production cross sections.  For gauge mediation, we also consider the closely-related case of light-squark NLSPs.  Otherwise, we focus on either stop pair production or Higgsino pair production, as both particles are expected to be below 1~TeV in a truly natural SUSY theory~\cite{Dimopoulos:1995mi,Cohen:1996vb,Brust:2011tb,Papucci:2011wy,Kats:2011qh}.\footnote{We save explicit investigation of a left-handed sbottom (N)LSP for future work.}  In most of what follows in this section, we restrict ourselves to presenting the basic search results, and reserve commentary on implications for Section~\ref{sec:conclusions}.

We generate event samples for most models with \Pythia8~\cite{Sjostrand:2007gs}, making extensive use of that program's R-hadronization capabilities.  Final-state particles from each lowest-level R-hadron or Higgsino decay are subsequently displaced before detector simulation and event reconstructions.  For some of the colored production models where multibody decay kinematics can be important, we have generated events in \MadGraph5~\cite{Alwall:2011uj}.  In this case, the \Pythia8 R-hadronization routine does not work because the R-odd colored particle has already decayed, and its daughters given color connections to other parts of the event.  There, we identify final-state hadrons as descendants of the long-lived colored particles in an approximate way:  each hadron is associated to the closest quark/gluon in $\Delta R$ as viewed the end of the parton shower, and the ancestry of that quark/gluon is traced by proxy.\footnote{ Using the simple 2-body $\tilde t \to \bar d_i \bar d_j$ as a cross-check, we find that the invariant mass of each stop is reconstructed to within about 10\% and without bias, and that most search acceptances are only mildly sensitive to to the ordering of decay and hadronization.  The largest effects are on the tracker-based searches, with the efficiencies dropping by 10--15\% for the sample decayed before hadronization, and therefore yielding conservative limits.}

All pair production cross sections are normalized to their NLO+NLL predictions, including colored production through pure QCD~\cite{Beenakker:2010nq} and electroweak Higgsino production~\cite{Fuks:2013vua}.  These predictions are all conveniently tabulated for 7~and 8~TeV by the LHC SUSY Cross Section Working Group~\cite{LHCSUSYCSWG}.  The Higgsino predictions assume nearly mass-degenerate Higgsino states with small mixings into the gauginos (assuming $M_{1,2} \sim 1$~TeV).  Those cross sections are only provided up to 410~GeV, but we assume a flat K-factor for higher masses.

\begin{figure}[tp!]
\centering
\subfigure{
\includegraphics[scale=0.55]{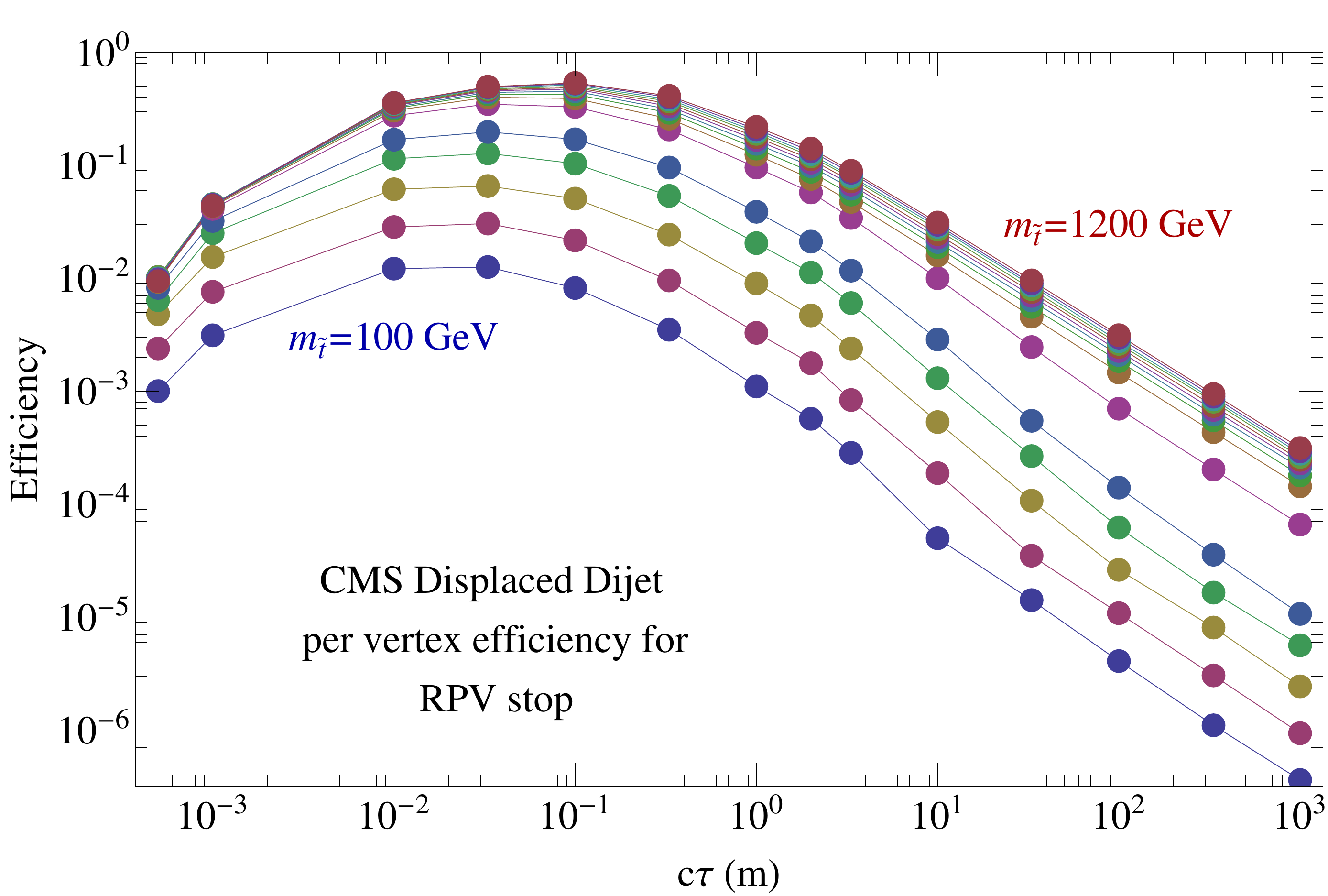}}
\caption[]{Our central predictions for the CMS per-decay displaced dijet acceptances for $\tilde t \to \bar d \bar s$, using the detector model described in Appendix~\ref{sec:CMSdisplacedDijets_calibration} and following the analysis described in Section~\ref{sec:CMSdisplacedDijets_analysis}.  The different colored lines are steps of 100~GeV in mass.}
\label{fig:stopeff}
\end{figure}

The generated particle-level events are passed through the detector simulations described in the Appendix and then subjected to the various analysis cuts described in Section~\ref{sec:searches}.  The main output is a set of experimental acceptances for each individual simplified model (either per-decay or per-event), scanned over mass and lifetime.  As an illustrative example, we provide our acceptances for the RPV $\tilde t \to \bar d \bar s$ model described below, passed through the CMS displaced dijet analysis.  As can be seen, this specific analysis is most efficient for masses greater than a few hundred GeV and lifetimes at the $O$(10~cm) scale.

When constructing limits, for most of our recast experimental searches we provide a very rough error band on our predicted exclusion regions by varying these signal acceptance estimates up and down by a factor of 1.5 (not allowing those rescaled acceptances to exceed unity).  This gives some indication of sensitivity to possible recasting errors.  There are only two specific searches where we do not follow this protocol.  The first such search is for stable charged particles, for which we do not explicitly include an error band.  Our modeling here is fairly basic and conservative, and the acceptance anyway turns off exponentially fast at low lifetimes.  We have also recast CMS's over-conservative ``charge stripped'' limits, estimated in a scenario where interactions in the calorimeters always strip off the R-hadron charges.  The second search is for low-EM jets at ATLAS, where we have opted to instead define over-conservative and under-conservative treatments of the isolation against EM calorimeter activity, which we cannot reliably model.  Here, we require either that no decay particles point back to the ECAL, or do not place any explicit isolation criterion (though in both cases we employ a flat $O(1)$ reconstruction efficiency factor given in Appendix~\ref{sec:ATLASlowEM_calibration}).

\subsection{Baryonic R-Parity Violation}

One of the simplest extensions to the MSSM is the introduction of R-parity-violating Yukawa superpotential couplings and/or a $\mu$-term between the lepton doublet and down-type Higgs doublet superfields~\cite{Barbier:2004ez}.  R-parity violation may also be introduced in the soft SUSY-breaking potential, or in the K\"ahler potential~\cite{Csaki:2013jza}.  These are all typically set with zero coefficients in order to enforce R-parity.  R-parity trivially prevents dimension-four proton decay and stabilizes the LSP, providing a possible dark matter candidate.  However, these virtues are hardly strict requirements of the MSSM.  Proton decay can be prevented by alternative stabilizing symmetries, which in any case may be required given the existence of potentially dangerous R-even operators at dimension-five~\cite{Brust:2011tb}.  Dark matter could easily arise from a different particle sector instead of the MSSM neutralino.

Violation of R-parity can lead to radical changes in collider phenomenology, depending sensitively on which operators are activated, on the magnitude and flavor structure of those operators, and on the identity of the LSP, which is no longer stable and no longer needs to be electromagnetically neutral.  Broad ranges of coupling strengths allow for the LSP to decay at displaced locations within the LHC detector volumes.  For example, for two-body sfermion decays into light SM fermions, mediated by one of the R-parity-violating Yukawas, a dimensionless coupling of $O(10^{-10}$--$10^{-6})$ would yield a substantial population of measurably-displaced decays.

Proton stabilization allows the active RPV operators to violate either lepton number or baryon number, but not both.  This partitions the RPV scenarios into two mutually-exclusive classes, which we can call leptonic RPV and baryonic RPV.  The only explicit RPV displaced searches so far at the LHC have assumed leptonic RPV~\cite{ATLAS-2013-092,CMS:2014wda,CMS:2014hka,Khachatryan:2014mea}, capitalizing on the presence of leptons in the final state to help with triggering and with controlling backgrounds.  Here we pursue the completely untested case of baryonic RPV.  (However, we will put the leptonic RPV searches to an alternative use in the next subsection, applying them to gauge mediation models.)

We begin with the case of stop LSP.  Minimalistically, the decay would be mediated by the usual Yukawa superpotential interaction of right-handed chiral quark/squark fields, $\lambda''_{ijk} U_i D_j D_k$ ($ijk$ are flavor indices).  This is the only baryon-number-violating interaction in the MSSM that respects SUSY at dimension-four.  The required antisymmetrization over color indices requires a commensurate antisymmetrization over down-type flavor indices, leading to allowed decays $\tilde t \to \bar d \bar s$, $\bar d \bar b$, and $\bar s \bar b$.   Recently, it has also been observed that stop decays may proceed through a different combination of chiral quark/squark fields, via dimension-five operator $Q_i Q_j D_k^\dagger$ in the K\"ahler potential~\cite{Csaki:2013jza}.  The resulting component-field operators allow for a decay $\tilde t \to \bar b \bar b$, and indeed this is generally preferred since the decay amplitudes are chirally-suppressed (analogous to pion decay).  Prompt decays $\tilde t \to \bar d_j \bar d_k$ have only just begun to be probed by the LHC, in the mass region 200--400~GeV~\cite{Khachatryan:2014lpa}.  It has been estimated that a search based on jet substructure could also push down into the lower-mass region currently not covered~\cite{Bai:2013xla}.  (For longer-term projections, also see~\cite{Bai:2013xla} as well as~\cite{Duggan:2013yna}.)  The only other available limits are when the stops are detector-stable, the strongest ($\approx$~900~GeV) coming from the CMS and ATLAS charged R-hadron searches~\cite{Chatrchyan:2013oca,ATLAS:2014fka}.

\begin{figure}[tp!]
\centering
\subfigure{
\includegraphics[scale=0.55]{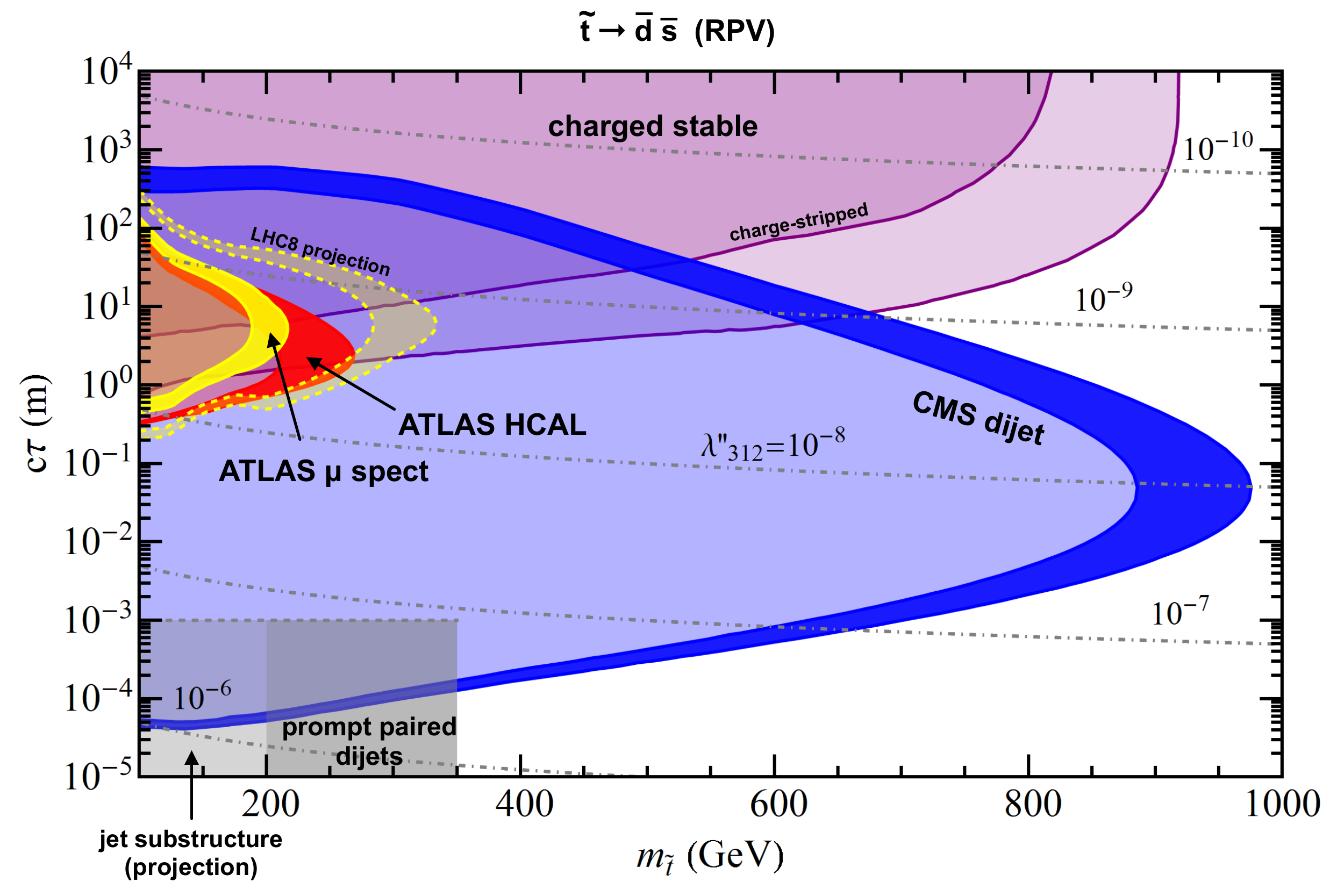}}
\caption[]{Recast constraints on displaced $\tilde t \to \bar d_j \bar d_k$ via baryonic RPV.  Colored bands indicate acceptance variations up/down by 1.5.  The dot-dashed lines indicate contours of $\lambda''_{312}$, assumed to be the only contributing RPV coupling.  Prompt limits (dark gray) are from~\cite{Khachatryan:2014lpa}, and low-mass search projections (light gray) are from~\cite{Bai:2013xla}.  They are conservatively cut off at 1~mm.}
\label{fig:stop}
\end{figure}

\begin{figure}[tp!]
\centering
\subfigure{
\includegraphics[scale=0.55]{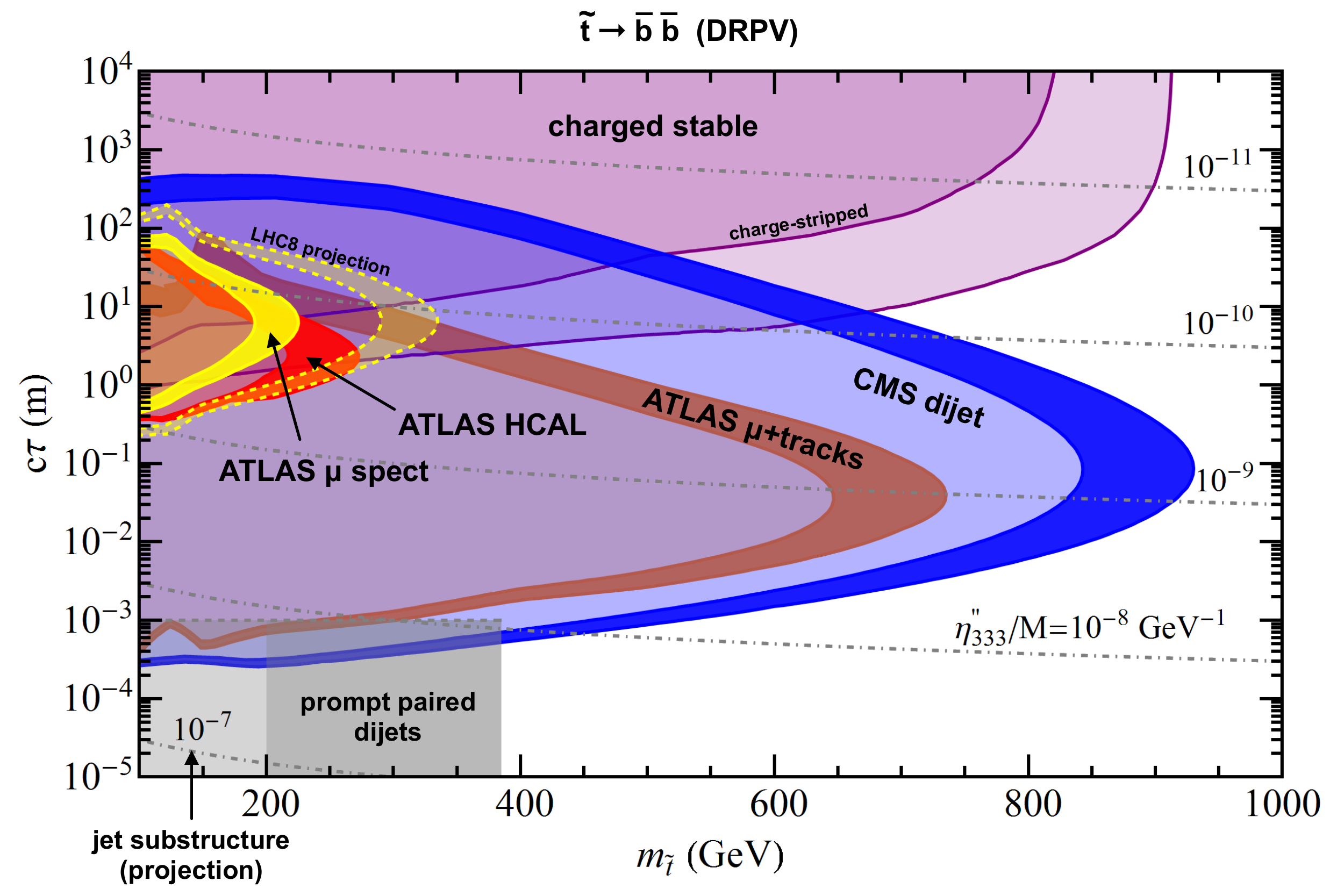}}
\caption[]{Recast constraints on displaced $\tilde t \to \bar b \bar b$ in the Dynamical RPV framework.  Colored bands indicate acceptance variations up/down by 1.5.  The dot-dashed lines indicate contours of $\eta''_{333}/M$, assumed to be the only contributing RPV coupling.  It arises from the K\"ahler potential operator $(\eta''_{333}/M)Q_3Q_3D_3^\dagger + {\rm (h.c.)}$.  Prompt limits (dark gray) are from~\cite{Khachatryan:2014lpa} (neglecting a possible improvement in the limits due to the higher b-jet multiplicity in the DRPV model), and low-mass search projections (light gray) are from~\cite{Bai:2013xla}.  They are conservatively cut off at 1~mm.}
\label{fig:stopDynamical}
\end{figure} 

Figs.~\ref{fig:stop} and~\ref{fig:stopDynamical} show the regions of mass and lifetime for $\tilde t \to \bar d_j \bar d_k$ that have now been excluded according to our recasts, taking the two extreme cases of only light-flavor decays and only $\bar b \bar b$ decays.  The sensitivity is dominated by the charged R-hadron and displaced dijet searches, a pattern that will recur often in our colored sparticle limits.  For both models there is nearly complete coverage out to almost 1~TeV, with a notable weak-spot at $c\tau \sim 10$~m and of course much weaker limits for displacements $\lsim$~mm.  This weakening at low lifetimes is more pronounced for the $\bar b \bar b$ decays, partially because the CMS dijet search is intrinsically less efficient for heavy flavor decays due to the somewhat lower particle track multiplicities, but also because of the conservative choice in our modeling of displaced vertex reconstruction for $b$-jets, discussed in Section~\ref{sec:CMSdisplacedDijets_analysis}.  At lower lifetimes, we have also indicated the existing and projected prompt limits, applying a conservative sensitivity cutoff at 1~mm.  (There should still be sensitivity from prompt searches for longer lifetimes, but we do not have enough information to reliably model this.)  Combining these three searches, unbroken coverage is achieved for {\it all} lifetimes for masses where the prompt searches are sensitive.  Indeed, for stop masses up to a few hundred GeV, the CMS dijet search alone spans 6--7 orders of magnitude in lifetime.  This amazing performance capitalizes heavily on the fact that millions of stop pairs would have been produced at such small masses, with sizable enough kinematic tails to pass the jet $H_T$ and $p_T$ cuts, and enough remaining rate to catch anomalously early or late decays from models at the edges of the exclusion region.  Two other displaced searches, ATLAS muon chamber (including our naive 8~TeV projection) and ATLAS low-EM jets, are much less competitive for the reasons discussed in Sections~\ref{sec:ATLASmuonChamber_analysis} and~\ref{sec:ATLASlowEM_analysis}, though they do offer useful complementarity in that their limits are derived from completely different detector systems.  Finally, we point out a sizable region in the $\bar b \bar b$ decay case that is also covered by the ATLAS $\mu$+tracks search, from a small population of events where one of the bottom decays produces a hard muon.\footnote{The muon in this search is triggered from the standalone muon spectrometer, and is not explicitly required to be isolated.}

The next model that we consider is gluino LSP.  Considering only traditional superpotential RPV, the gluino decays by first transitioning into a virtual squark and a corresponding real quark.  The virtual squark then splits to two quarks through the $UDD$ operator.  The full 3-body decay is $\tilde g \to jjj$.  There are again many options for flavor structure, which may be engineered both at the level of the $\lambda''_{ijk}$ couplings and the squark mass spectrum.  Here, we simply assume decays into light flavors, though decays involving $b$-quarks could again be subjected to weaker limits at low lifetimes, and decays involving $t$-quarks would also receive constraints from the displaced searches involving leptons.  Otherwise, we expect fairly similar coverage.  Of course, branching ratios into top also suffer additional phase space suppression.

\begin{figure}[tp!]
\centering
\subfigure{
\includegraphics[scale=0.5]{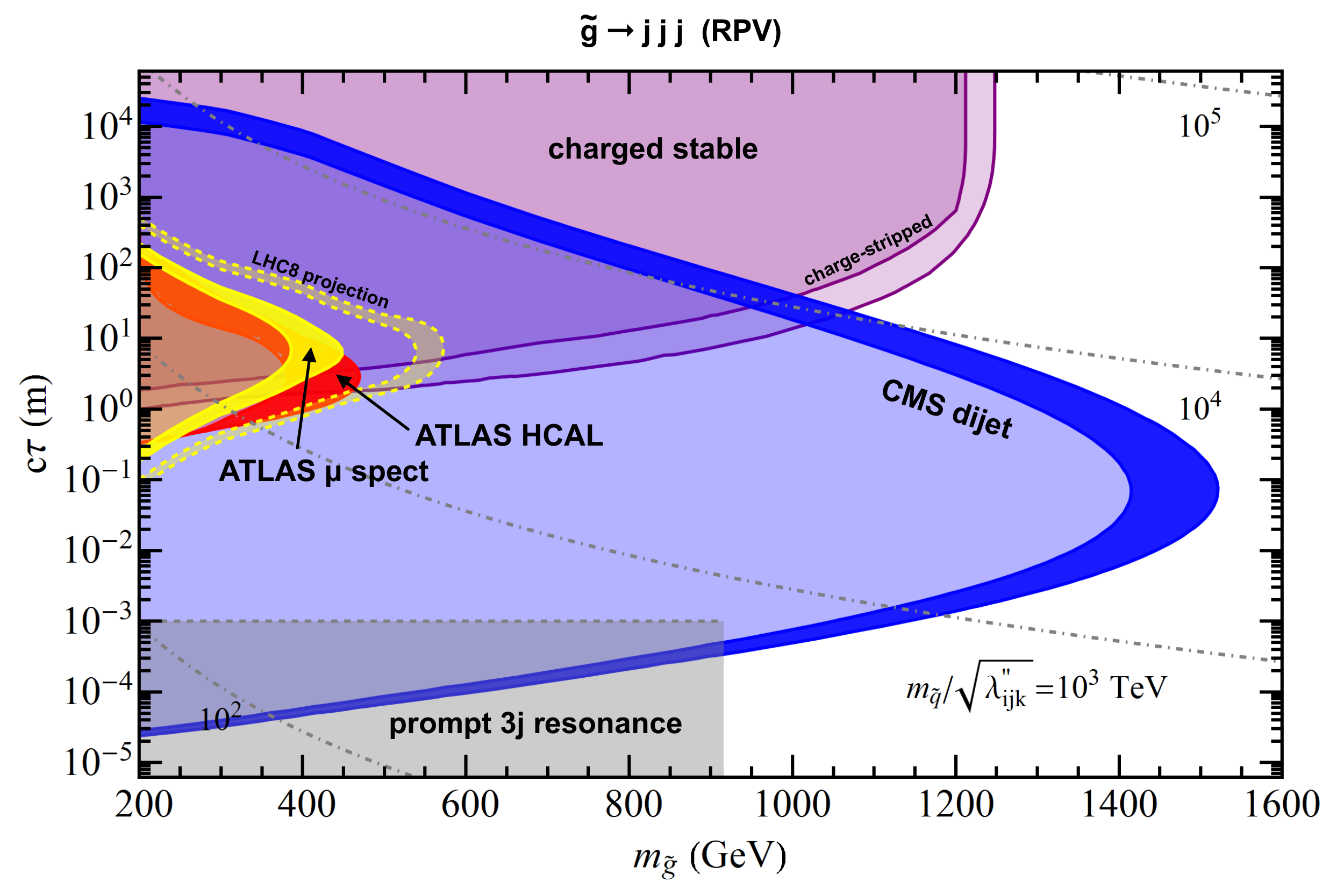}}
\caption[]{Recast constraints on displaced $\tilde g \to jjj$ via baryonic RPV.  Colored bands indicate acceptance variations up/down by 1.5.  The dot-dashed lines indicate contours of $m_{\tilde q}/\sqrt{\lambda''_{ijk}}$.  We have parametrized the decay assuming that one species of off-shell RH squark dominates, and splits into quarks via a single $\lambda''_{ijk}$ coupling.  All final-state quarks are also assumed to be from the first two generations.  Prompt limits (gray) are derived from~\cite{ATLAS-CONF-2013-091}.  They are conservatively cut off at 1~mm.}
\label{fig:RPVgluino}
\end{figure}

Fig.~\ref{fig:RPVgluino} shows our estimated exclusions for $\tilde g \to jjj$.  The qualitative features are quite similar to the RPV stop decays, though the much higher cross sections yield a significantly extended mass reach for all searches.  CMS dijets in particular reaches above 1.5~TeV, close to the production limit of $O$(1) event in the entire run, and exceeding the mass reach of the stable R-hadron search by several hundred GeV.  Notably, the displaced trijet configuration is very efficiently picked up by the CMS dijets search, which was designed for a very different signal.  The weak spot at 10~m is still apparent, but much less pronounced since the CMS dijet search nearly matches the HSCP search sensitivity at that lifetime.  It is also interesting to supplement with the limits from prompt searches~\cite{ATLAS-CONF-2013-091,Chatrchyan:2013gia}, which are similar for purely light-flavor decays and decays containing $b$-quarks.  Again applying an ad hoc 1~mm cutoff on the lifetime sensitivity of the prompt searches, there is currently unbroken coverage for {\it all} possible lifetimes for masses potentially as high as 900~GeV.

\begin{figure}[tp!]
\centering
\subfigure{
\includegraphics[scale=0.5]{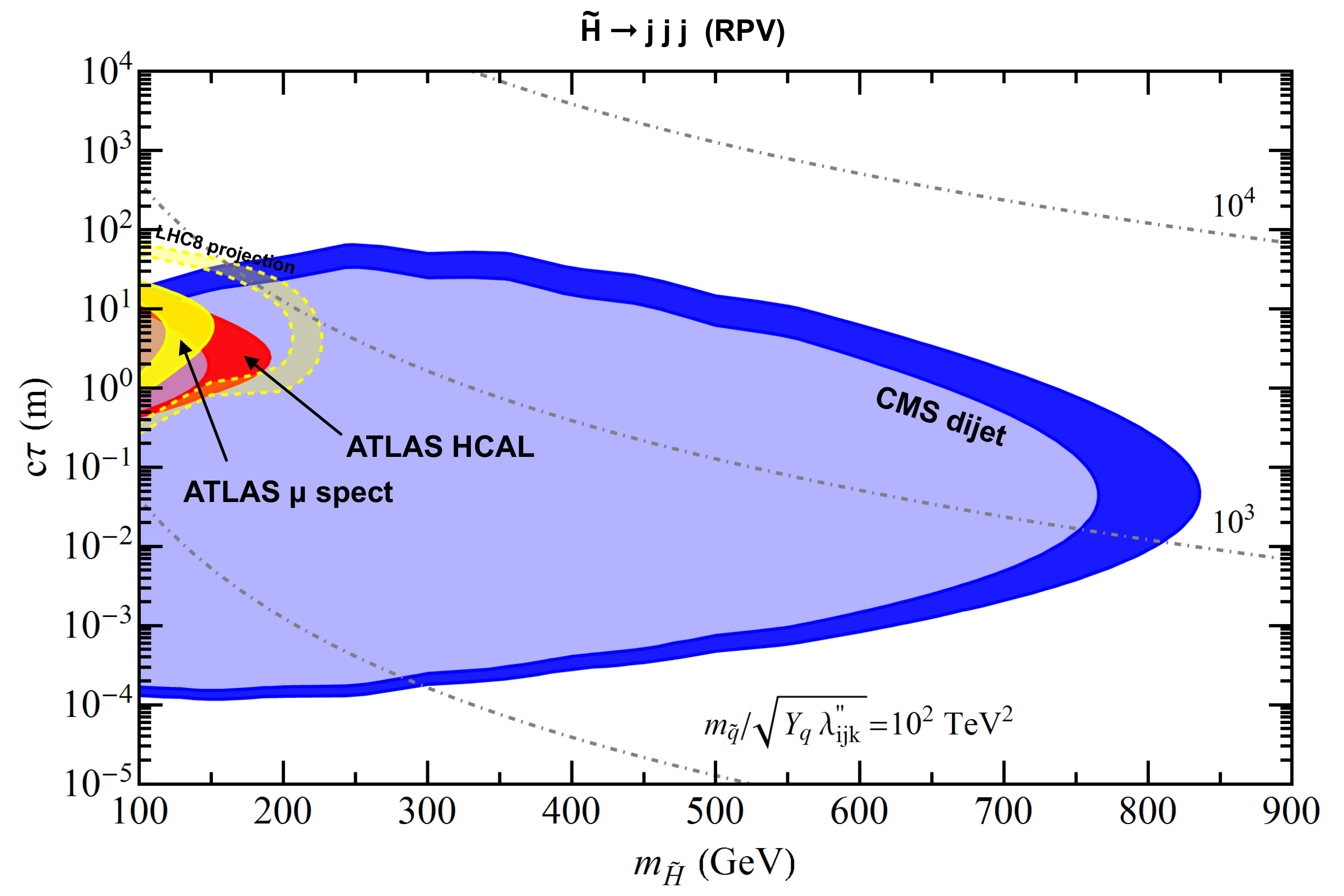}}
\caption[]{Recast constraints on displaced $\tilde H^0 \to jjj$ via baryonic RPV.  Colored bands indicate acceptance variations up/down by 1.5.  The dot-dashed lines indicate contours of $m_{\tilde q}/\sqrt{Y_q\lambda''_{ijk}}$.  We have parametrized the decay assuming that one species of off-shell RH squark dominates, coupling to the Higgsino according to its up-type or down-type Yukawa $Y_q \propto m_q/(v\sin\beta)$ or $m_q/(v\cos\beta)$, and splits into quarks via a single $\lambda''_{ijk}$ coupling.  All final-state quarks are also assumed to be from the first two generations.}
\label{fig:RPVHiggsino}
\end{figure}

The last baryonic RPV example model that we consider is a Higgsino multiplet ``co-LSP.''  The four Higgsino states are assumed to be only mildly mixed into heavier electroweak gauginos, and the multiplet split by $O$(10~GeV) or less.  The heavier Higgsinos undergo a soft but prompt cascade via virtual gauge boson emission into the lightest, neutral Higgsino.  The displaced decay of this lightest Higgsino proceeds in almost exact analogy with the gluino decay, though now the virtual squark is accessed via a super-Yukawa coupling instead of a super-QCD coupling.  Again, the flavor structure of the decay can be nontrivial, but as a first pass we simply assume that the final quarks are all light.  Decays involving tops deserve a dedicated investigation, especially in the context of a natural theory, though we anticipate fairly similar limits.

Fig.~\ref{fig:RPVHiggsino} shows our estimated exclusions for $\tilde H \to jjj$.  The qualitative picture from the displaced decay searches remains similar, though with reduced mass reach due to the smaller production cross sections.  Unlike the preceding stop and gluino examples, there are no explicit limits on $\tilde H \to jjj$ in either prompt or stable charged particle searches.  For the prompt case, we can compare to $\tilde g \to jjj$ searches~\cite{ATLAS-CONF-2013-091,Chatrchyan:2013gia}.  Using simple cross section scaling suggests that that promptly decaying RPV Higgsinos are genuinely unconstrained, since the Higgsino cross section is roughly 500 times smaller at a given mass.  (A more aggressive dedicated prompt search could be useful, though would be highly challenging.)  For the stable case, the LSP here is generically neutral, and hence does not leave a track. Therefore, our reported direct Higgsino production limits here are the first for {\it any} lifetime.

While we have only studied a small sample of possible spectra, these results clearly illustrate the power of the LHC in probing baryonic RPV in general via displaced decays.  An obvious extension of our observations would be an application to a broader class of flavor assumptions, though as indicated we do not expect radically different sensitivity.  The remaining extensions would be to consider different LSPs, and perhaps more model-dependent scenarios where the LSP is created in cascades in addition to prompt production.  An LSP squark could represent a rather trivial example, since the production and decay could be very similar to the LSP stop.  However, decays into the top quark could also open up, and the effective production cross section could also be highly enhanced by the multiplicity of nearby squark states (cascading promptly into one another) and/or by gluino exchanges.  Direct LSP slepton production represents a qualitatively different direction, wherein a 4-body decay to $ljjj$ or $\nu jjj$ (via virtual electroweakino and squark) might dominate, even for much larger values of the $\lambda''_{ijk}$.  Finally, production of different electroweakinos, such as a mostly-bino or mostly-wino, could be considered.  In fact, the latter has recently been investigated in~\cite{Cui:2014twa}, and similarly finds very high mass reach using the CMS displaced dijets search.  With the generality and power of the HSCP and dijets searches, the main missing pieces in covering the mass-lifetime plane for these varied models would be prompt and b-tagged searches (possibly recast from other models) and more aggressive muon-chamber and calorimeter searches, especially for the cases without long-lived charged states.  Additional studies within the framework of dynamical RPV could also be interesting, since this allows for additional flavor and chirality structures in the decays.

\subsection{General Gauge Mediation}
\label{sec:GMSB}

Traditional forms of gauge mediation assume fairly minimal messenger sector constructions, and consequently relatively predictive patterns of sparticle masses~\cite{Giudice:1998bp}.  For example, relationships like $M_1$:$M_2$:$M_3$~= $\alpha_1$:$\alpha_2$:$\alpha_3$ for the three gauginos favor a bino-like NLSP and a much heavier gluino.  A much more general perspective has been advocated in~\cite{Meade:2008wd}, acknowledging the full range of possible MSSM spectra derivable from arbitrary messenger sectors, perturbative or not.  Practically speaking, the individual sfermion and gaugino soft masses become almost freely-adjustable, up to two sfermion sum rules and flavor universality, as well as vanishing $A$-terms at the mediation scale.  Even more general model frameworks allow for the possibility of flavor nonuniversality effects, either by mixing into a supersymmetric composite sector~\cite{Sundrum:2009gv,Csaki:2012fh,Gherghetta:2011wc,Larsen:2012rq}, by gauging flavor symmetries~\cite{Craig:2012di}, or by introducing large $A$-terms through non-minimal interactions between MSSM and messenger fields~\cite{Evans:2013kxa}.

Such freedom of model-building in gauge mediation allows for a number of intriguing options for LHC phenomenology.  Practically any superparticle can be made the NLSP, which then decays into its SM-partner and a light gravitino (i.e., Goldstino) at a rate controlled only by the fundamental SUSY-breaking mass scale, $\sqrt{F}$.  Roughly speaking, when the SM partner is light,
\beq
c\tau \: \sim \:  0.3~{\rm mm} \, \left(\frac{100~{\rm GeV}}{\tilde m}\right)^5  \left(\frac{\sqrt{F}}{100~{\rm TeV}}\right)^4,
\label{eq:GMSBlt}
\eeq
implying displaced decays at the $O$(mm~--~10~m) scale for $\sqrt{F}$ in the range of a few-hundred to a few-thousand TeV.  With the traditional option of the bino-like neutralino as the NLSP, the decays are dominated by photon and gravitino.  Consequently, the experimental effort in displaced GMSB searches has focused on signals of displaced/delayed photons and missing energy~\cite{Aad:2014gfa,Chatrchyan:2012jwg}.\footnote{These are also necessarily the most model-dependent of the available signals, since a simplified model containing only the bino would have vanishing tree-level pair production cross section.  The most powerful existing search~\cite{Aad:2014gfa} mainly capitalizes on pair production of mostly-winos cascading down into a mostly-bino, relying on the relationship $M_2/M_1 \simeq \alpha_2/\alpha_1$.}  In~\cite{Meade:2009qv,Meade:2010ji}, it was pointed out that an NLSP neutralino with larger wino or Higgsino fraction would also yield displaced $W$, $Z$, and Higgs.  The possibility of displaced NLSP stops was emphasized in~\cite{Chou:1999zb,Kats:2011it,Covi:2014fba}, especially for stop masses near or below the top quark mass.  The simplest remaining displaced NLSP options would be slepton decaying to lepton, squarks decaying to (non-top) quarks, and gluino decaying to gluon.\footnote{An NLSP sneutrino from light $\tilde \ell_L$ doublets would decay fully invisibly, making the displacement irrelevant for its experimental signatures.  The distinctive phenomenology of such a scenario has been studied in~\cite{Katz:2009qx,Katz:2010xg,Covi:2007xj}.}  None of these other possibilities have been searched for in the case of displaced decays, though a number of searches have been performed for prompt decays and for the collider-stable cases.  However, most NLSP possibilities are actually already under tight constraint, as we will see, again sometimes with coverage over the full range of possible lifetimes.

A notable exception is any variation on the slepton NLSP, such as the standard mostly-$\tilde \tau_R$ when the sleptons are degenerate up to Yukawa effects, or possibly $\tilde e_R / \tilde\mu_R$ if the staus can be made heavier~\cite{Calibbi:2014pza}.  These would be largely unconstrained by the displaced vertex searches due to the low track multiplicities and vertex masses, and searches for displaced activity in the calorimeters or muon chambers would usually fail to pick up the signal, for example because the associated slepton track would cause isolation failure.  CMS's displaced $e$+$\mu$ search~\cite{Khachatryan:2014mea} should yield some sensitivity for leptonic tau decays.  Though we reserve for future work a more comprehensive study of the status of displaced slepton NLSPs in GMSB, we anticipate that planned searches for ``kinked track'' topologies will be important to more fully cover the parameter space, and that existing disappearing track searches might also provide some sensitivity.

The remaining NLSP options that yield a single SM final-state particle, without passing through an intermediate heavy SM decay (top or electroweak), are the non-top squarks and the gluino.  As in the baryonic RPV models, presently the only potentially applicable tracker-based search is the CMS displaced dijets, but the nominal number of partons in the decay is not two.  Here, in order for the displaced dijet search to be sensitive, the decay must undergo a hard enough final-state bremsstrahlung to create a second jet.  Given the large strong production cross sections, this is an affordable penalty: of order $(\alpha_s/\pi)\log(\tilde m/(60~{\rm GeV}))$, times color factors, with the sparticle mass and jet $p_T$ cut appearing inside the logarithm.  Quantitatively, the chance to radiate a second jet is roughly at the 1--10\% level.  Of course, a search explicitly geared toward the one-jet topology could be more efficient, and the displaced dijet trigger would already capture this signal in cases where both decays occur within the inner 60~cm radius of the tracker and with at least $O$(mm) displacement.  However, at the very high and very low lifetime ranges, the inefficiency induced by requiring an extra jet may be less than the inefficiency that would be induced by forcing both decays to occur at improbably short or long proper times.  A more fruitful option for future analyses could be to exploit the traditional jet, $H_T$, and \met\ triggers, and apply an offline search for individual displaced jets.\footnote{We thank Joshua Hardenbrook for emphasizing this possibility.}

Proper simulation of the decays for the existing displaced dijet search requires some level of matrix element matching.  This is performed automatically by the \Pythia8 shower in the case of $\tilde q \to q \tilde \chi^0$ with a massless neutralino LSP, while the desired decay $\tilde q \to q \tilde G$ is not matched.  However, we have observed essentially identical rates and kinematics for extra jet production in $\tilde q$ decays between explicit \MadGraph5 2- and 3-body decay simulations with neutralino and gravitino LSPs, and close agreement with \Pythia8's matched predictions for the first shower emission.  We therefore feel confident using the massless neutralino LSP as a proxy for the gravitino LSP for squark decays in \Pythia8.   For the gluino, such an analogous decay to neutralino does not exist at tree level, is not part of the \MadGraph5 MSSM model, and would not obviously be matched if forced to proceed in \Pythia8.  Instead, we compare the unmatched \Pythia8 predictions for its first shower emission to \MadGraph5, both with gravitino LSP.  We again find similar decay kinematic distributions, with \Pythia8 predicting a somewhat slower falloff out to $\Delta R(j,j) \sim \pi$.  But the major difference is in the total emission rate, which \Pythia8 over-estimates by a factor of about 1.8.  To approximately compensate for this, we rescale the individual vertex reconstruction efficiencies by 1/2.  It should be understood that $O(10\%)$ modeling uncertainties on the displaced dijet reconstruction efficiencies for GMSB gluinos should likely be applied, though we anyway effectively absorb this into our ad hoc systematic variations.

\begin{figure}[tp!]
\centering
\subfigure{
\includegraphics[scale=0.5]{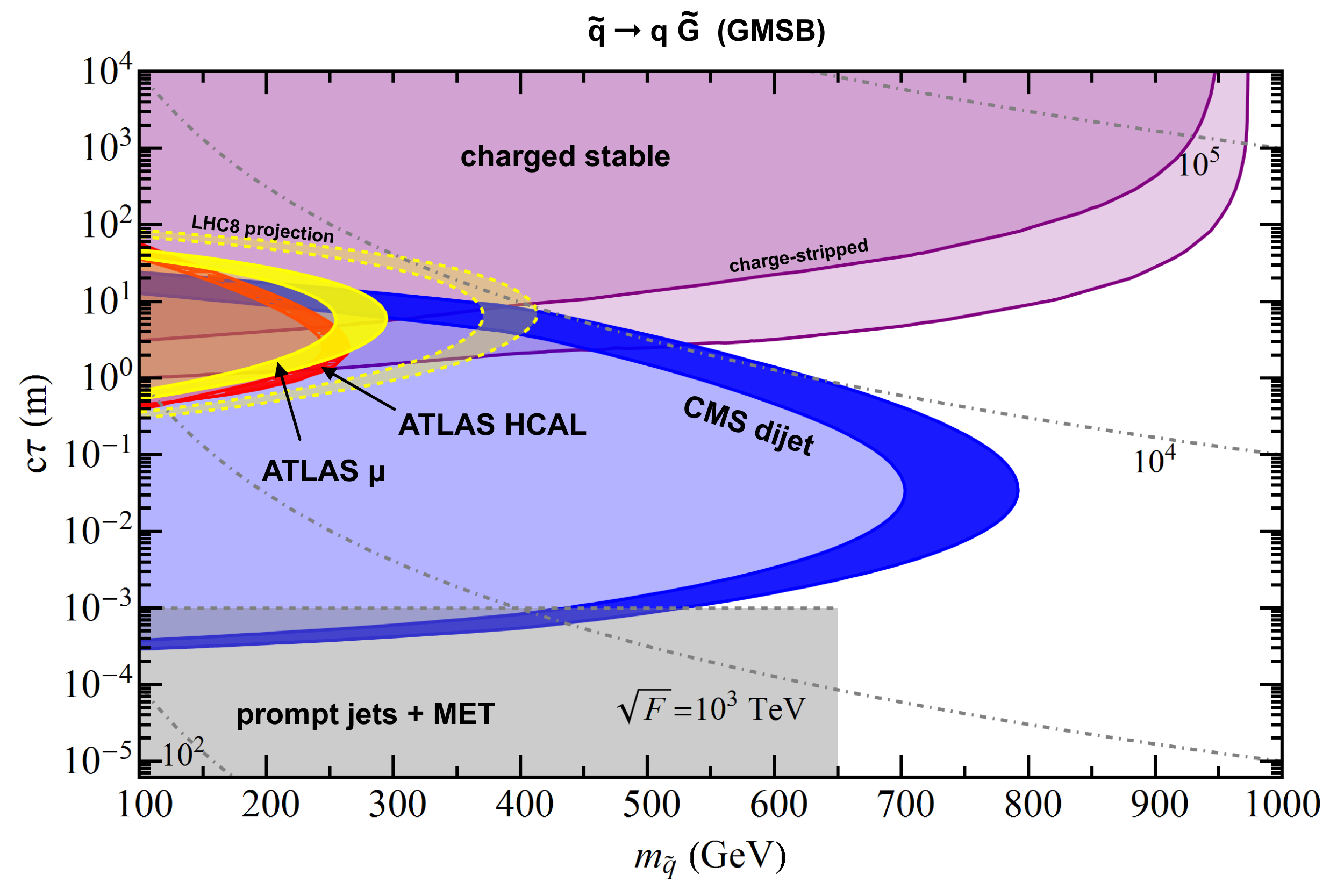}}
\caption[]{Recast constraints on displaced $\tilde q \to q \tilde G$ in general GMSB, conservatively assuming contributions from only $\tilde d_R$ and $\tilde s_R$.  Colored bands indicate acceptance variations up/down by 1.5.  The dot-dashed lines indicate contours of the SUSY-breaking scale $\sqrt{F}$.  Prompt limits (gray) are derived from~\cite{Aad:2014wea}.  They are conservatively cut off at 1~mm.}
\label{fig:GMSBsquarks}
\end{figure}

Starting with the squark NLSP, we display the results in Fig.~\ref{fig:GMSBsquarks}.  We conservatively assume just two degenerate species, $\tilde d_R$ and $\tilde s_R$.  This is a technical possibility if the $SU(3)$ contributions to the sfermion masses are small, the $SU(2)$ contributions are large, and the third-generation squarks receive additional mass contributions.  The exclusions are similar to those of the RPV stops (Figs.~\ref{fig:stop} and~\ref{fig:stopDynamical}), although now with much stronger prompt jets+$\met$ searches.  Unbroken coverage over lifetime is achieved up to about 450--550~GeV, limited by the crossover between the HSCP and displaced dijet searches.  

\begin{figure}[tp!]
\centering
\subfigure{
\includegraphics[scale=0.5]{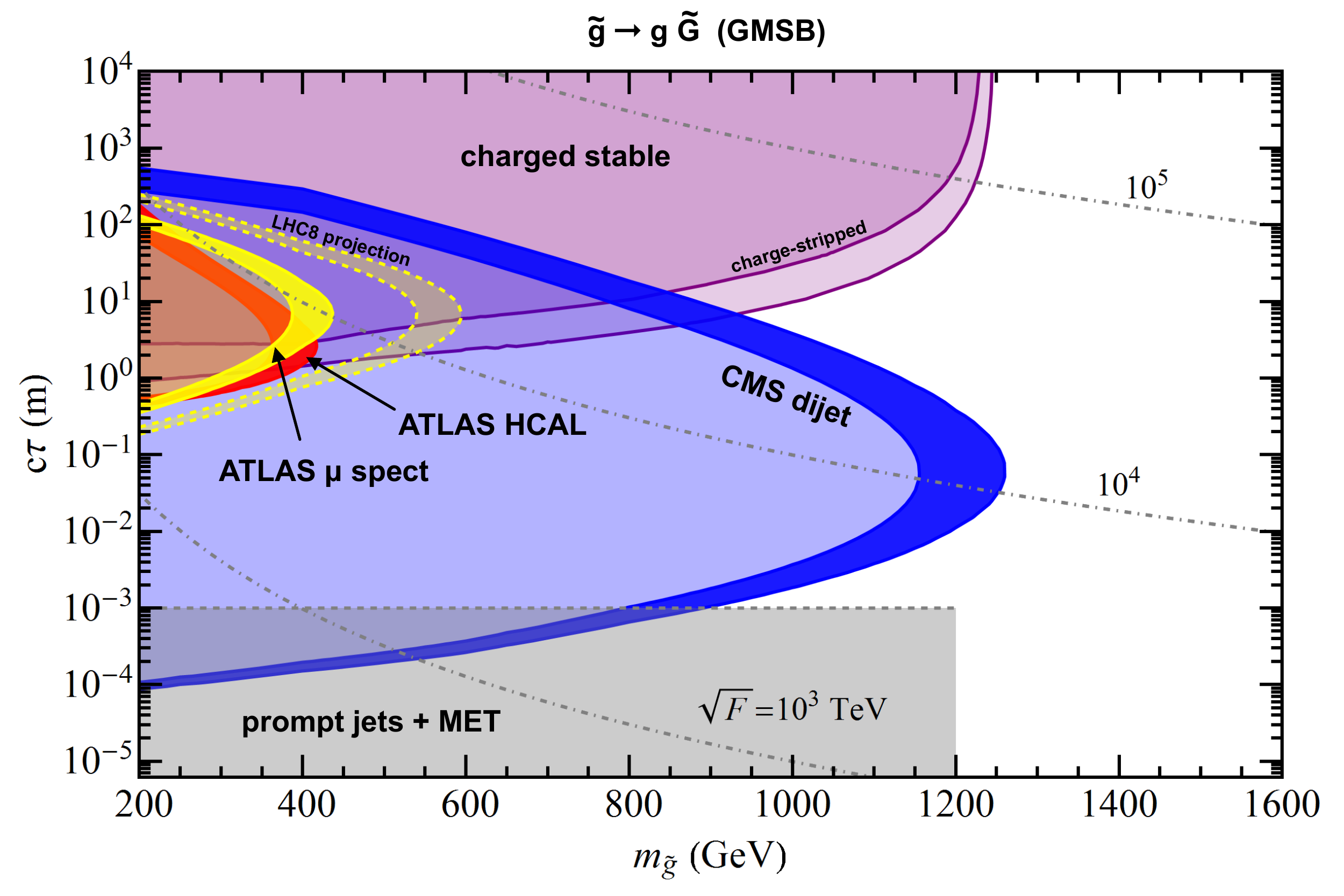}}
\caption[]{Recast constraints on displaced $\tilde g \to g \tilde G$ in general GMSB.  Colored bands indicate acceptance variations up/down by 1.5.  The dot-dashed lines indicate contours of the SUSY-breaking scale $\sqrt{F}$.  Prompt limits (gray) are derived from the $\tilde q \to q \tilde G$ of~\cite{Aad:2014wea}.  They are conservatively cut off at 1~mm.}
\label{fig:GMSBgluino}
\end{figure}

Next we consider the gluino NLSP in Fig.~\ref{fig:GMSBgluino}.  Comparing to the RPV results for $\tilde g \to jjj$ in Fig.~\ref{fig:RPVgluino}, we observe much weaker displaced decay limits and much stronger prompt decay limits.  The former is due to the requirement of additional hard radiation in the decay to pass the CMS displaced dijet reconstruction.  The latter is due to the much more distinctive jets+$\met$ signature.  Most of the model space below 1200~GeV is covered, with expected weak spots at $O$(mm) and $O$(1--10~m), though much of the surviving space at smaller lifetimes would likely be probed by a more detailed jets+$\met$ recast, as in~\cite{ATLAS-2014-037}.  Full coverage over all lifetimes is only achieved for masses below 800~GeV.

\begin{figure}[tp!]
\centering
\subfigure{
\includegraphics[scale=0.5]{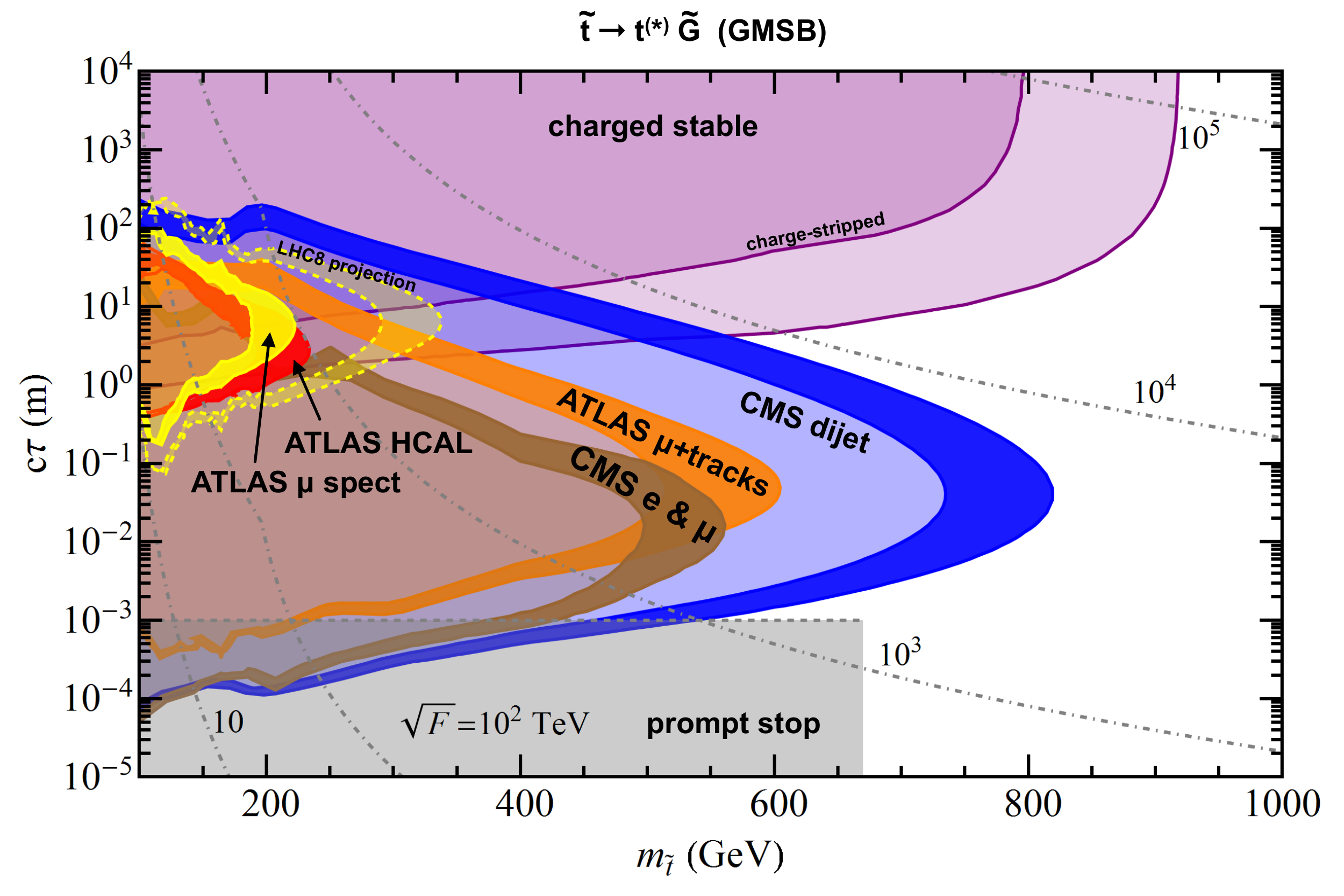}}
\caption[]{Recast constraints on displaced $\tilde t \to t^{(*)} \tilde G$ in general GMSB.  Colored bands indicate acceptance variations up/down by 1.5.  The dot-dashed lines indicate contours of the SUSY-breaking scale $\sqrt{F}$.  Prompt limits (gray) are derived from~\cite{Aad:2014kra,CMS-PAS-SUS-14-011}.  They are conservatively cut off at 1~mm.}
\label{fig:GMSBstop}
\end{figure}

We now move on to the naturalness-motivated options, starting with the NLSP stop in Fig.~\ref{fig:GMSBstop}.  We consider stops of any mass above 100~GeV, including a range of masses below $m_t$ and through the compressed region where $m_{\tilde t} = m_t$.\footnote{Because of the smallness of the $t\tilde t \tilde G$ coupling, top quark decay into a light stop and $\tilde G$ would be rare.}  In these regions, the decays are dominantly 3-body $\tilde t \to Wb\tilde G$, with a large fraction of energy going into the effectively derivatively-coupled gravitino/Goldstino.  Also, in addition to the by-now familiar searches that have appeared in all of our recasts above, the semileptonic decays of the stop open up sensitivity in the ATLAS $\mu$+tracks search and the CMS displaced $e$+$\mu$ search.  While the sensitivity regions for these searches are fully contained by CMS displaced dijets, corroborating coverage is provided by the leptonic searches over much of the excluded region.  Adding in the prompt searches~\cite{Aad:2014kra,CMS-PAS-SUS-14-011}, which likely give unbroken coverage between 100~GeV and 670~GeV,\footnote{The prompt searches face some subtleties.  On the one hand, for stop masses well above $m_t$, existing searches for $\tilde t \to t\tilde\chi^0$ with massless neutralino should offer identical coverage.  On the other hand, the decay kinematics near or below $m_t$ can be significantly different than the corresponding decays to neutralinos.  The expectation is that the GMSB limits there should be much stronger than the nominal limits, and not subject to the usual sensitivity gap with a compressed spectrum~\cite{Kilic:2012kw}. The major exception is a mostly-$\tilde t_L$ stop, for which spin effects would reshape the \met-sensitive distributions and weaken the limits in searches with semileptonic decays.  Without recasting the most recent searches, it is not possible to precisely delineate this gap, though the results of~\cite{Kilic:2012kw} suggest that it may be several 10's of GeV wide, centered in the vicinity of 200~GeV.  Since the displaced searches are not designed to cut on \met-sensitive tails, we do not expect such spin effects to be significant there.} we infer that GMSB stops of any lifetime are excluded below about 500~GeV.  For lifetimes at the cm-scale, exclusions extend beyond 700~GeV, and, as noted before, out to about 900~GeV for lifetimes longer than $O$(10~m).  (For other estimates of displaced stop exclusions in GMSB, see~\cite{Cahill-Rowley-Talk}.)

\begin{figure}[tp!]
\centering
\subfigure{
\includegraphics[scale=0.4]{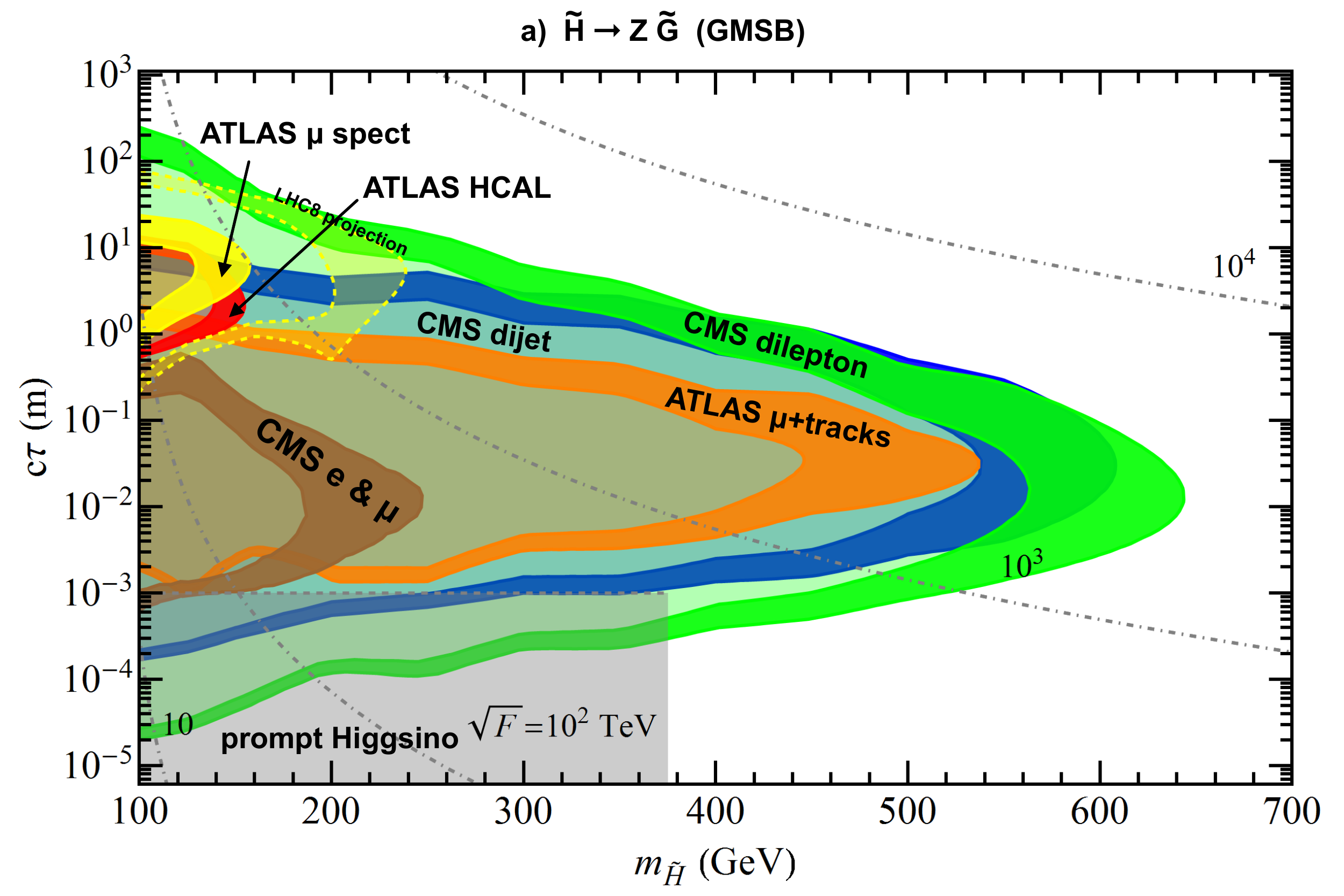}}
\subfigure{
\includegraphics[scale=0.4]{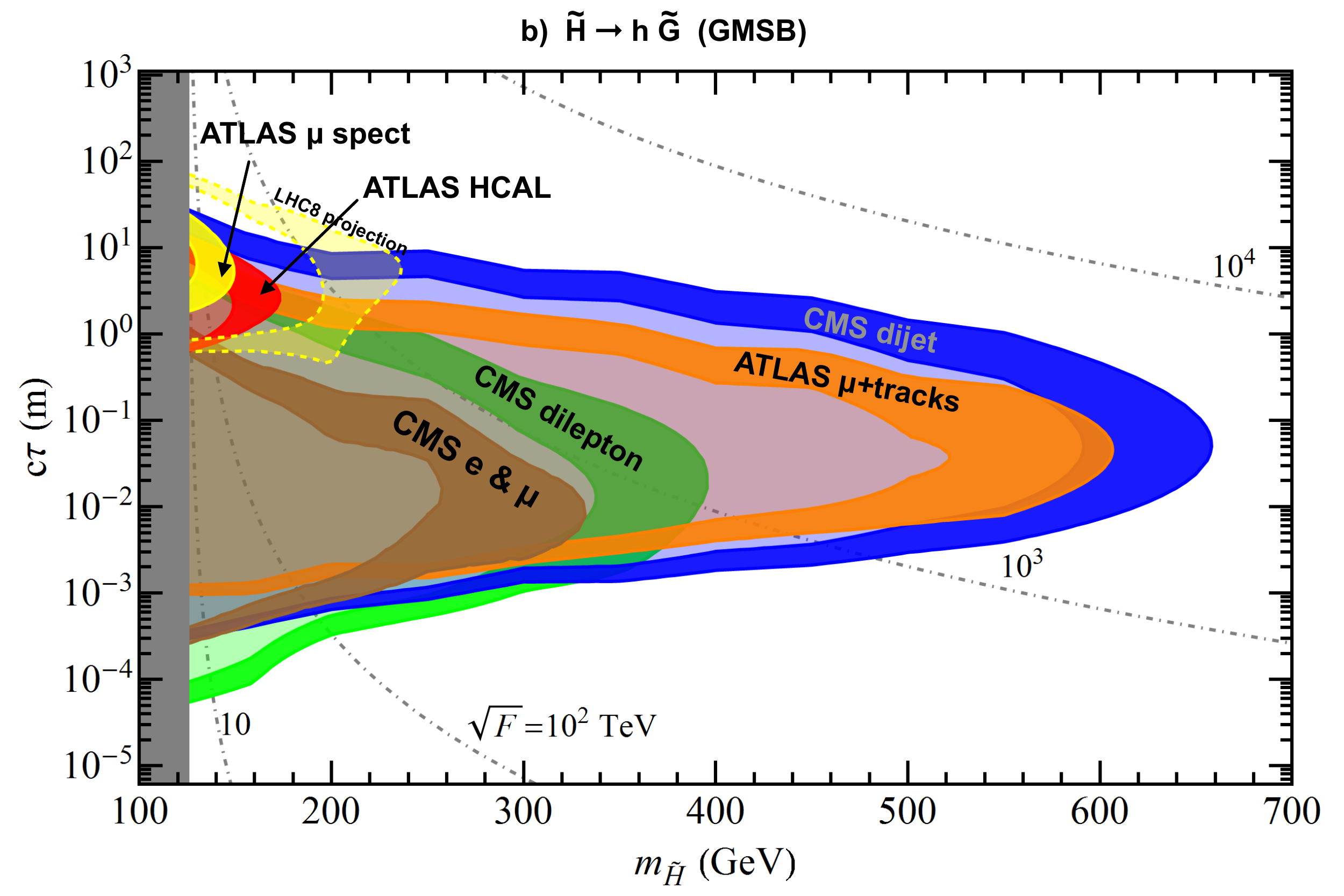}}
\subfigure{
\includegraphics[scale=0.4]{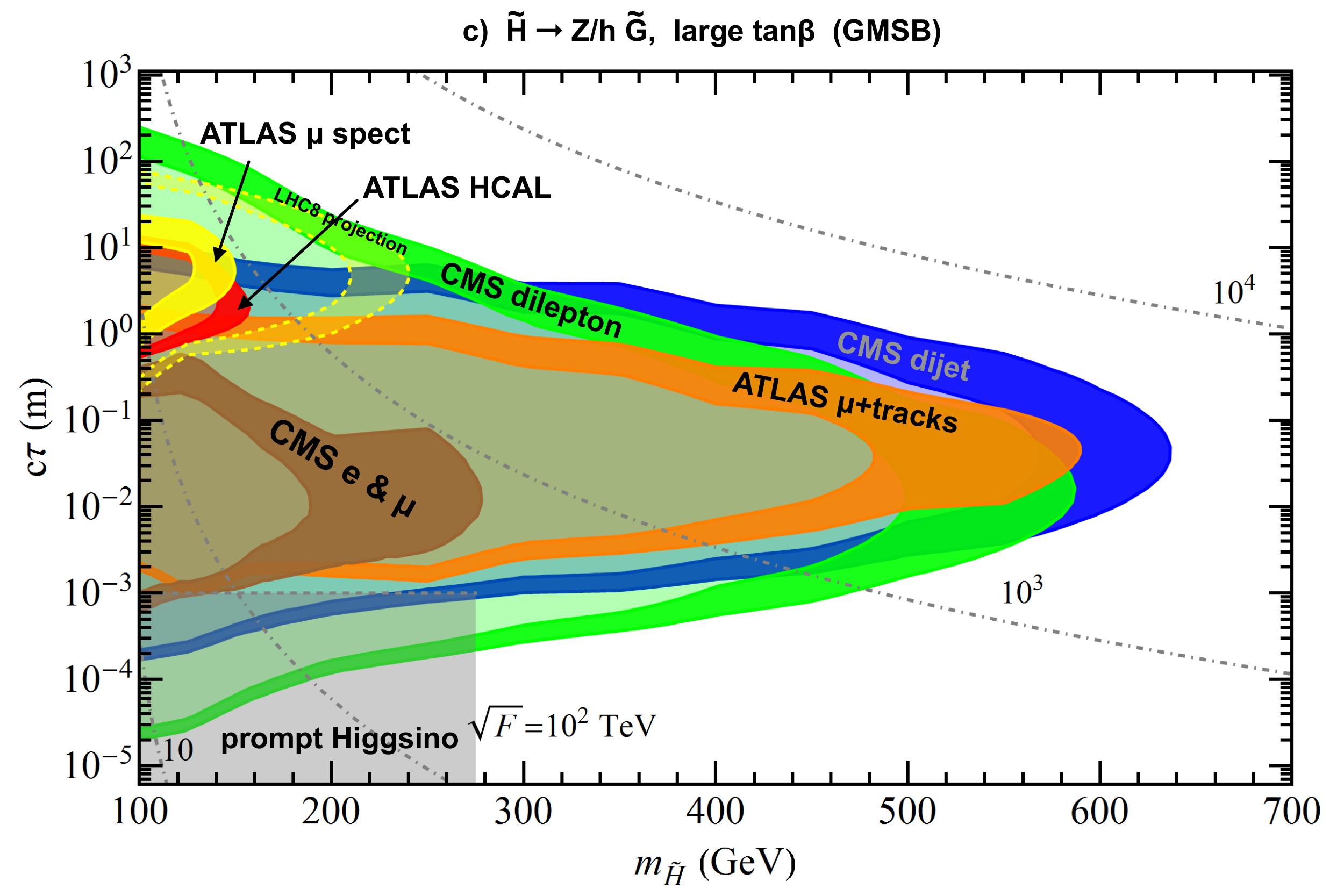}}
\caption[]{Recast constraints on displaced $\tilde H^0$ decays via GMSB: a) pure $\tilde H^0 \to Z\tilde G$, b) pure $\tilde H^0 \to h\tilde G$, c) large-$\tan\beta$.  Colored bands indicate acceptance variations up/down by 1.5.  The dot-dashed lines indicate contours of the SUSY-breaking scale $\sqrt{F}$.  Prompt limits (gray) are derived from~\cite{Khachatryan:2014mma}.  They are conservatively cut off at 1~mm.}
\label{fig:GMSBhiggsino}
\end{figure}

Finally for GMSB, we consider Higgsino multiplet co-NLSPs.  As in the RPV case above, we assume that all Higgsino states are nearby to one another (split by no more than $O$(10~GeV)), with heavier states decaying promptly.  The lightest Higgsino will preferentially decay to some mixture of $Z\tilde G$ and $h\tilde G$, with $\gamma\tilde G$ suppressed.  The lifetime and branching fractions of the lightest Higgsino exhibit simple behavior if mixings with the bino and wino are small, and the scalar Higgs sector is close to the ``decoupling limit.''  For instance, when $\tan\beta=1$, the lightest Higgsino coupling to either $Z\tilde G$ or $h\tilde G$ vanishes, depending on the relative signs of the $\mu$ parameter and $M_{1,2}$.  For $\tan\beta\gg 1$, the $h\tilde G$ and $Z\tilde G$ decay modes have similar partial widths if $m_{\tilde H} \gsim m_h$.

We present the limits in these three extreme cases in Fig.~\ref{fig:GMSBhiggsino}:  a) pure $\tilde H \to Z\tilde G$, b) pure $\tilde H \to h\tilde G$, and c) large-$\tan\beta$.  Though the HSCP searches again do not apply, the GMSB Higgsino brings into play all of our other displaced decay recasts, now including as well the CMS displaced dilepton search.  For decays that include direct $Z$ bosons, this last search can be seen to play a major role, competing significantly with and even beating the CMS dijet search.  This is especially obvious at very low and very high lifetimes, where the dilepton search benefits from both lower track impact parameter thresholds and looser $p_T$ reconstruction thresholds.  Similarly, leptonic decays play a major role in the prompt searches~\cite{Khachatryan:2014mma}, with maximal sensitivity for pure $\tilde H \to Z\tilde G$, capitalizing in part on 3- and 4-lepton channels.  Taking the large-$\tan\beta$ case as a baseline example, $m_{\tilde H} = 100$~GeV is now covered from zero lifetime up to $c\tau \sim 100$~m, and $m_{\tilde H} = 300$~GeV is covered up to $c\tau \sim $ few~m.  The highest mass reach is for $c\tau \sim 10$~cm, extending out to about 600~GeV.  For the pure $\tilde H \to h\tilde G$ case, our displaced search recasts represent the only extant limits, as was the case for the RPV Higgsinos.

We have touched upon most of the possible displaced NLSP decays in gauge mediation.  The last obvious remaining option, which we now briefly discuss for completeness, would be wino co-NLSPs.  In some ways, this should overlap significantly with our $\tilde H$ results, but there are some notable differences.  First, wino cross sections are about two times larger.  Second, the $\tilde W^0$ has a significant branching fraction into photons, hence subjecting it to the displaced/delayed photon searches.  Third, when the wino multiplet is somewhat separated in mass from the bino and Higgsino, the mass splitting between charged and neutral states is protected at leading order in the mass mixing by an accidental custodial symmetry, with the first nontrivial mixing contributions often comparable to or smaller than the electroweak radiative mass splitting of $\approx$~170~MeV.  This famously leads to the long-lived decays $\tilde W^\pm \to \pi^\pm\tilde W^0$, with $c\tau \simeq 5$~cm, searched for in~\cite{Aad:2013yna,CMS:2014gxa}.  In such a case, there can be nontrivial competition between the above decay and $\tilde W^\pm \to W^\pm\tilde G$.  There can also be peculiar cases with an initial stage decay $\tilde W^\pm \to \pi^\pm\tilde W^0$, leaving a disappearing track, followed by a secondary displaced decay $\tilde W^0 \to (\gamma/Z)\tilde G$.  Finally, there are also some corners of parameter space with $m(\tilde W^\pm) < m(\tilde W^0)$ due to chargino and neutralino mixings~\cite{Kribs:2008hq}, causing every event to contain two displaced $W$s.  The signatures would be much more similar to those of $\tilde H \to Z\tilde G$, though missing the displaced dileptons and, if $c\tau > O$(cm), containing a track or track stub pointing to the displaced decay (similar to the slepton NLSPs).  All together, the potentially rich displaced phenomenology of wino co-NLSPs in gauge mediation clearly deserves a more detailed investigation, and would bring together a surprisingly varied set of displaced search results.

\subsection{Mini-Split SUSY}

The last model framework that we consider is mini-split supersymmetry, where the scalars of the MSSM (excepting the SM-like Higgs boson) are all raised to the 1000~TeV scale~\cite{ArkaniHamed:2004fb,Arvanitaki:2012ps,ArkaniHamed:2012gw}.  This scale could represent a sweet-spot between masses that are high enough to avoid flavor constraints with arbitrary sfermion mass matrix structure, but low enough to provide a 125~GeV Higgs from the stop loop corrections.  The separation between MSSM scalars and fermions can arise automatically in several SUSY-breaking mediation scenarios (surveyed in~\cite{Arvanitaki:2012ps} and discussed on general terms in~\cite{ArkaniHamed:2012gw}).  However, while some of the important virtues of SUSY such as gauge coupling unification and dark matter can be preserved, the original motivation of naturalness is partially abandoned.  The apparently finely-tuned Higgs mass might nonetheless be viewed as a byproduct of anthropic selection bias in the multiverse, in some ways similar to the unnaturally small cosmological constant~\cite{ArkaniHamed:2004fb}, or as a compromise against much larger ``tunings'' within the available ``model space'' of broken SUSY theories~\cite{ArkaniHamed:2012gw}.

While gauginos need not be present at any particular mass scale, the WIMP miracle is suggestive of TeV-scale masses, potentially within reach of the LHC.  One of the most interesting targets is the gluino, since the flow of R-parity in its decay must pass through the heavy squarks, leading to suppressed matrix elements and extended lifetimes.  Two types of LHC searches so far have directly targeted this signal:  searches for R-hadrons stopped in the calorimeters and decaying out-of-time with respect to collisions~\cite{Aad:2013gva,Khachatryan:2015jha}, and an ATLAS reinterpretation of its prompt jets+\met\ searches using models with displaced decays~\cite{ATLAS-2014-037}.  Neither of these are optimally sensitive, though the former strategy has the added benefit of permitting a lifetime measurement if a positive signal is observed, and the latter strategy can be carried out quickly with no changes to the event reconstruction and selection software.  In addition, searches for anomalous tracks from collider-stable R-hadrons, which we have discussed above in the contexts of both RPV and GMSB, continue to apply.

\begin{figure}[tp!]
\centering
\subfigure{
\includegraphics[scale=0.5]{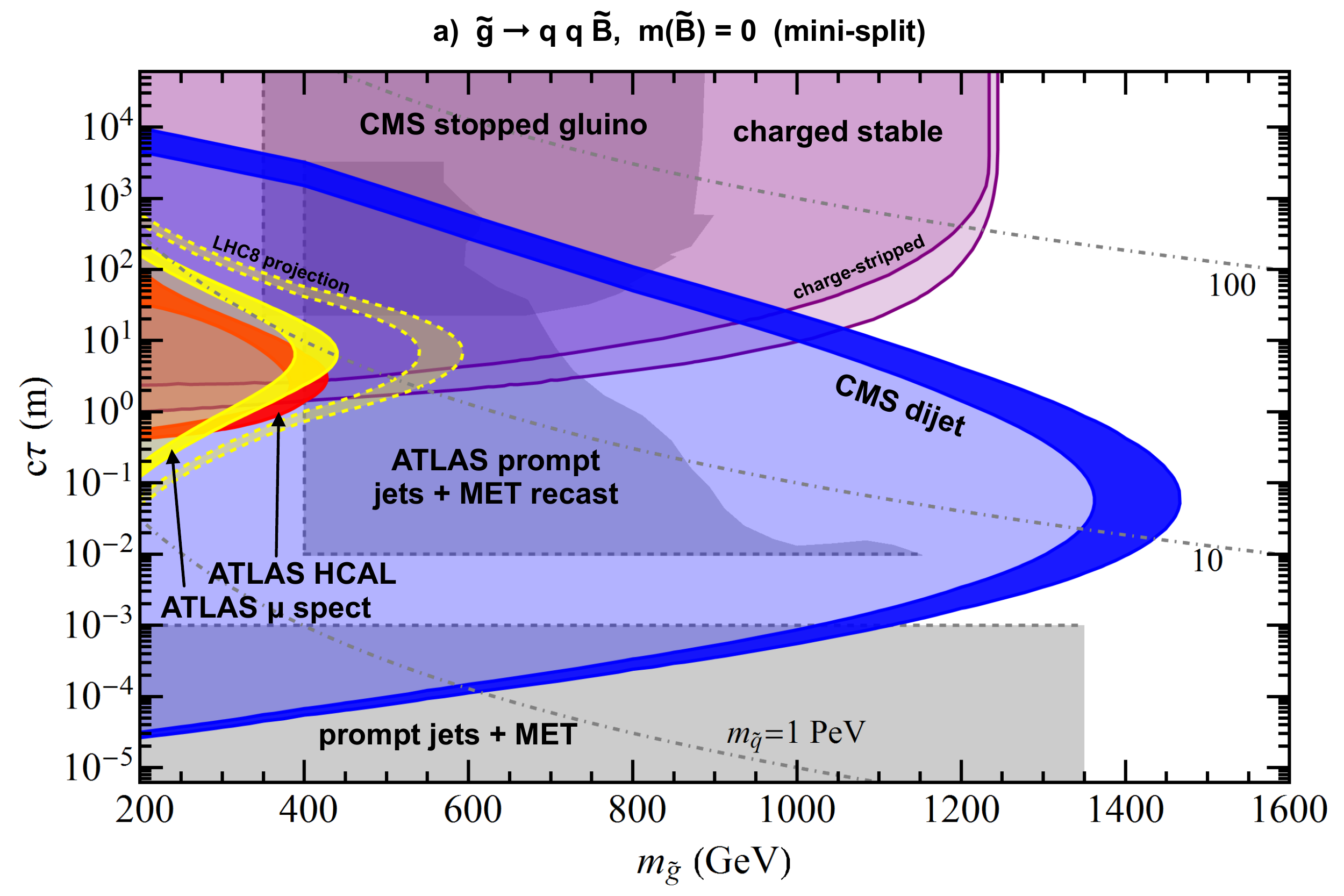}}
\subfigure{
\includegraphics[scale=0.5]{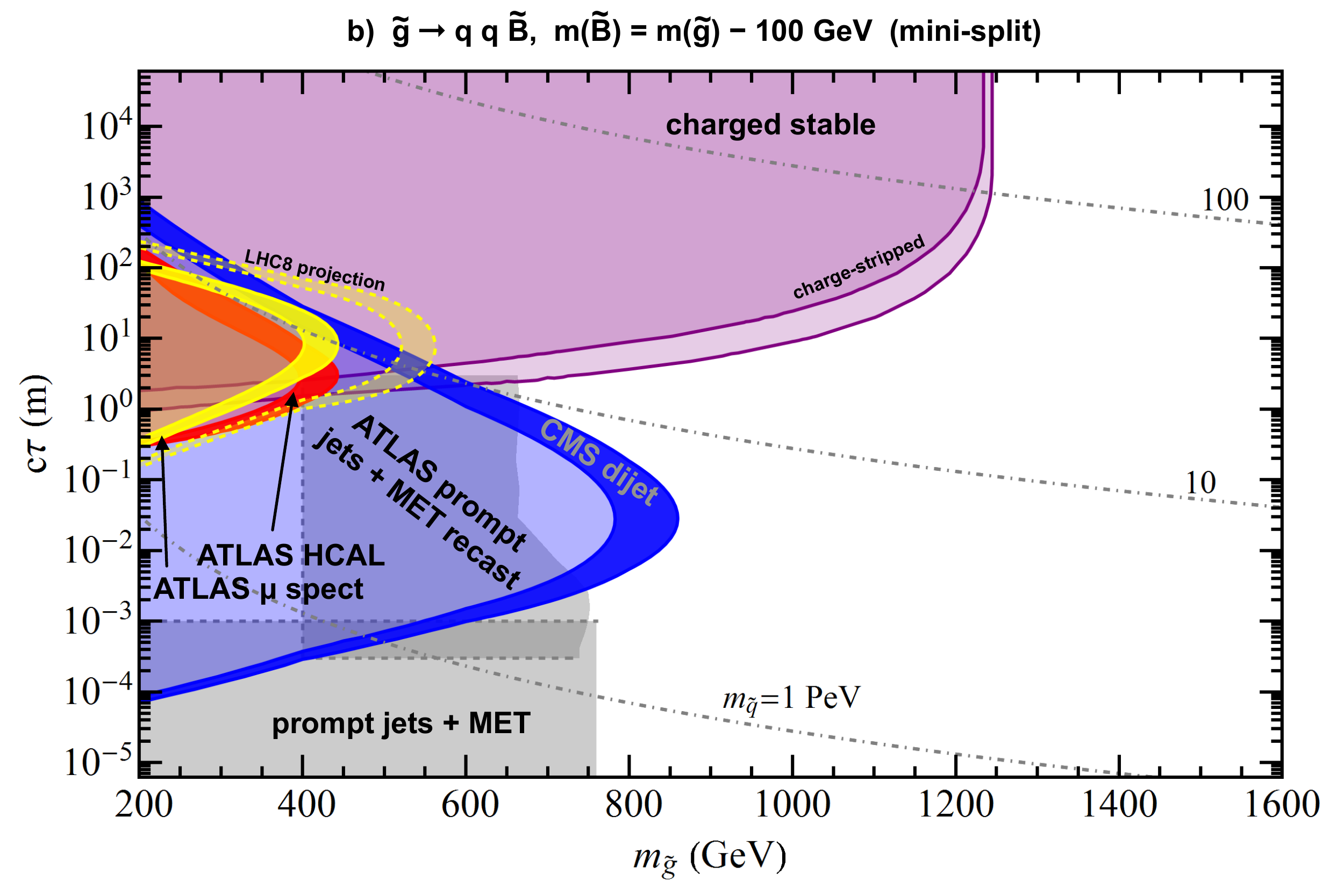}}
\caption[]{Recast constraints on displaced $\tilde g \to q\bar q \tilde B$ in mini-split SUSY: a) $m_{\tilde B} = 0$, b) $m_{\tilde B} = m_{\tilde g} - 100$~GeV.  Colored bands indicate acceptance variations up/down by 1.5.  The dot-dashed lines indicate the intermediate squark mass, assuming that either $d_R$ or $s_R$ dominates the decay.  Prompt limits (gray) are derived from~\cite{CMS-PAS-SUS-13-019,Aad:2014wea}.  They are conservatively cut off at 1~mm.  Additional displaced limits come from stopped R-hadron searches~\cite{Khachatryan:2015jha} and ATLAS's recast prompt limits~\cite{ATLAS-2014-037}.}
\label{fig:MiniSplitgluino}
\end{figure}

Here we put all of these long-lived gluino searches into context for a couple of specific assumptions for the decay kinematics.  In almost full generality, the gluino may decay into any flavor-combination of quark-antiquark pairs plus a $\tilde B$, $\tilde W$, or $\tilde H$.  The exact admixture of these decays is highly model-dependent.  Since the decay rate through any given off-shell squark channel scales like $1/m_{\tilde q}^4$, the lightest squark eigenstate would dominate if there is a somewhat spread-out scalar mass spectrum.  Yukawa effects in the renormalization group may favor light third generation squarks, suggesting dominant decays $\tilde g \to t\bar t\tilde\chi^0$ or $\tilde g \to t\bar b\tilde\chi^-$+c.c.~\cite{ArkaniHamed:2012gw}, though flavor-anarchic soft masses could instead lead to lighter first- or second-generation squarks.  Loop-induced decays $\tilde g \to g\tilde \chi^0$ might also become important~\cite{Gambino:2005eh}, though again depending in detail on the squark mass spectrum, as well as on the gaugino mass spectrum.  For our initial study here, we simply assume 100\% 3-body branching fraction $\tilde g \to q\bar q \tilde B$ for $q=udcs$.  The only free parameters are then the gluino and neutralino masses, as well as the gluino lifetime set by the squark mass scale~\cite{ArkaniHamed:2012gw},
\beq
c\tau \:\approx\: 10^{-5}~{\rm m} \, \left(\frac {m_{\bar q}} {\rm PeV}\right)^4 \left(\frac {\rm TeV} {m_{\bar g}}\right)^5.
\eeq
We reserve a more general survey of displaced mini-split phenomenology for future work.

Fig.~\ref{fig:MiniSplitgluino} shows our results, choosing either $m_{\tilde B} = 0$ or $m_{\tilde B} = m_{\tilde g} - 100$~GeV.  We find once again that, amongst explicit displaced decay searches, CMS displaced dijets offers superior sensitivity.  For the light neutralino case, exclusions extend out as far as 1400~GeV, and for the heavy neutralino out to 800~GeV.  The dedicated mini-split gluino searches, based on stopped R-hadron decays~\cite{Aad:2013gva,Khachatryan:2015jha} and recasts of prompt searches~\cite{ATLAS-2014-037}, do not tend to be competitive with this search combined with the HSCP searches.  Interestingly, the ATLAS muon chamber and low-EM jets searches, which have tended to give universally weaker coverage, potentially offer more stable sensitivity as the visible activity gets squeezed out when $m_{\tilde B} \simeq m_{\tilde g}$.  This owes largely to their focus on lower-mass signatures, which for compressed SUSY spectra becomes a major virtue.  However, the overlap with HSCP coverage remains substantial.

\section{Discussion and Conclusions}
\label{sec:conclusions}

The initial 7~and 8~TeV runs of the LHC have launched an impressive exploration of the vast parameter space of the MSSM and its extensions, yielding the most sensitive searches to date and in many cases already probing up to TeV mass scales.  In this paper, we have sought to initiate a systematic extension of this program into the similarly vast realm of SUSY with non-prompt final-stage decays, as occur in a number of common and well-motivated scenarios within the MSSM such as R-parity violation, gauge mediation, and mini-split spectra.  This has been accomplished by recasting seven existing searches for stable charged particles and highly displaced decays, and combining these with prompt searches.  Our present survey has focused mainly on signals containing a sizable fraction of hadronic decays, including ``natural'' spectra with light stops and Higgsinos.  The overarching conclusion is that, while very few long-lived particle searches are explicitly designed to be sensitive to these signals, the extremely low backgrounds and reasonably high acceptances of those searches nonetheless allow us to place tight limits.  Indeed, we have typically found large patches of parameter space where multiple distinct search channels overlap.  That said, we have identified several places where significant improvements might still be possible.

We first list here some conclusions of our investigations regarding the performance of the searches:
\begin{itemize}
\item  In the long-lifetime limit, several searches have been carried out for stable charged R-hadrons.  They are usually still sensitive down to $c\tau \sim$~meter, catching rare late-decaying particles, and providing substantial overlap with explicit displaced decay searches.  Limits on squarks and gluinos extend up to about 1~TeV.  However, there is of course no sensitivity to long-lived {\it neutral} particles, such as (N)LSP Higgsinos.
\item  The CMS displaced dijet search is extremely effective for essentially any decay topology involving any number of energetic quarks and gluons (including one or three, as well as many decays with leptons~\cite{CMS:2014wda}).  It is almost universally the most powerful displaced decay search when hadronic decays dominate, including decays involving weak bosons.  For $c\tau \sim 10$~cm, (stop) squark pairs are probed up to almost 1~TeV, gluinos typically up to above 1~TeV, and Higgsinos up to 600--800~GeV.  Though the trigger $p_T$ requirements are relatively harsh, good sensitivity is maintained for superparticles with strong or electroweak production cross sections down to 100~GeV mass, by picking up events on the high-$p_T$ tails.  It seems likely that similar search strategies, perhaps capitalizing on different sets of triggers and displaced vertex criteria, could expand the model reach.  In particular, within our own set of models the original dijet requirement reduces by $O$(10--100) the acceptance for ``monojet'' decay topologies such as from $\tilde q \to q\tilde G$ and $\tilde g \to g\tilde G$ in GMSB.  A standard jets+$\met$ style trigger could much more efficiently pick up this signal, and a single displaced-jet requirement would likely eliminate the remaining background.  It would also be very useful to investigate this style of search for more traditionally motivated signals such as $\tilde\tau\to\tau\tilde G$, which would require accepting low-multiplicity/low-mass vertices.  For this signature and many others, it is important to understand what happens when the displaced particle is charged and can leave its own track segment.
\item  Existing ATLAS searches for displaced decays inside the calorimeters and muon chambers should in principle supplement the region $c\tau \sim$~few~meters, where both the stable charged track and displaced dijet searches are becoming weaker.  However, these searches were highly optimized for low-mass pseudoscalar signals, and achieve relatively limited sensitivity for displaced SUSY models.  It seems likely that these searches could be modified to better capitalize on the energetic signatures of superparticles with 100's of GeV mass, where they are anyway most needed to extend the global search reach beyond $c\tau\sim$~1~m, especially for long-lived neutral particles where stable charged track searches are unusable.  This could possibly be done using existing data from the specialized ATLAS HCAL and muon triggers.  Such modified searches would still need to achieve very high efficiency in order to become competitive with the other searches, possibly benefiting from a single-candidate mode rather than their standard double-candidate.
\item  We have also considered three searches that capitalize on relatively clean leptonic signatures:  a displaced muon in association with a displaced tracker vertex (not necessarily geometrically overlapping), a displaced dilepton pair, and generic $e\mu+X$ signatures where the electron and muon are each displaced.  Interestingly, the $\mu$+tracks search shows nontrivial sensitivity even to fully hadronic decays, provided that they contain energetic bottom quarks.  However, within the scope of models studied here, all of these searches truly become relevant for signals involving weak bosons, such as stop and Higgsino decays in GMSB.  The displaced dilepton search in particular offers improved sensitivity relative to displaced dijets when $Z$ bosons are available, since the former can be constructed with both lower impact parameter thresholds and lower $p_T$ thresholds.  The $\mu$+tracks is also highly competitive, though somewhat weaker in lifetime coverage due to geometric restrictions on the vertexing.\footnote{It could be quite interesting to recast this search (as well as the displaced dijets) for the $\tilde t \to l^+ b$ leptonic RPV model, which was searched for in the $e\mu+X$ channel in~\cite{Khachatryan:2014mea}.  There are likely other models we have not touched upon for which this search could be uniquely sensitive.}
\item  A variety of prompt searches have been directly applied to the decay topologies that we consider here.  It is, however, mostly unclear how effective these searches are when the decays become appreciably displaced, especially for searches involving leptons or traditional $b$- and $\tau$-jets.  A first analysis in this direction was performed by ATLAS~\cite{ATLAS-2014-037} for gluinos in mini-split SUSY, indicating an approximately logarithmic degradation of mass reach with increasing lifetime (presumably stemming from the onset of a linear falloff in displaced decay acceptance when $c\tau > O$(m)).  For searches at CMS, which rely much more on tracking and vertexing in the construction and validation of particle-flow jets, more significant degradations might be expected.  While we have shown that CMS's dedicated displaced dijets search can much more efficiently pick up the gluino signal in the lifetime range where ATLAS shows results, we expect that there would be additional benefits to exploring other searches involving ``many $b$-jets'' and/or ``many $\tau$-jets'' (possibly plus $\met$), ideally with some allowances for uncharacteristically high displacements.  Such searches would (or perhaps already do) bridge the possible weakening in fully hadronic coverage around $O$(mm) lifetime stemming from CMS's 500~$\mu$m displaced dijet impact parameter cut.  As noted, the benefit of searching down to smaller nonzero displacements was already made clear in our GMSB Higgsino results, where displaced dileptons was able to push the lifetime reach down by as much as an order of magnitude.
\end{itemize}

We next discuss some of the physics implications of our findings:
\begin{itemize}
\item  Natural supersymmetry with light, promptly-decaying stops has been coming under progressively more pressure from a series of LHC searches.  Moving to scenarios with displaced stop decays into $jj$ via baryonic RPV or $t^{(*)}\tilde G$ in GMSB, we are apparently forced into to even more unnatural regions of model space.  This is particularly true in baryonic RPV, where prompt decay limits are currently very weak.  However, very small $\lambda''$ and displaced stop decays are actually favored by cosmological arguments, as even $\lambda'' \gsim 10^{-6}$ with sub-TeV stops would efficiently wipe out the baryon asymmetry of the universe in a standard thermal history with baryogenesis no lower than the weak scale~\cite{Barry:2013nva}.  It is therefore becoming very difficult to simultaneously accommodate naturalness, baryogenesis, and LHC direct searches in such a scenario.  In GMSB, where an NLSP stop is rather nongeneric but has been an interesting logical possibility for some time~\cite{Chou:1999zb}, the region with $m_{\tilde t} < m_t$ is almost guaranteed to yield displaced decays (provided $\sqrt{F} \gsim 10$~TeV).  The $m_{\tilde t} < m_t$ possibility is now fully closed, as is essentially all model space below 500~GeV.
\item  Natural supersymmetry with light Higgsinos is traditionally very challenging to probe via direct electroweak Higgsino production, though some limits are becoming available in general GMSB models where decays into $Z$ bosons are appreciable~\cite{Khachatryan:2014mma}.  In the presence of baryonic RPV, the multijet decays of the lightest Higgsino yield a very striking and highly constrained displaced signature, and furnish the only extant LHC direct production limits in that topology.  The cosmological implications are more model-dependent, but again tend to disfavor the as-yet unprobed prompt decays~\cite{Barry:2013nva}.  RPV Higgsinos with $c\tau \gsim 10$~m would effectively act stable, and again become very difficult unless, as usual for neutralinos, they are produced in the decays of heavier colored superparticles and appear as $\met$.  In the GMSB case, as noted above, almost all of the searches that we have recast become sensitive, the only exception being stable charged particles.  The mass/lifetime coverage is qualitatively similar to the RPV case, though the smaller fraction of visible energy, the typically smaller number of hard partons, and the smaller branching fractions into individual final states all contribute to slightly lower mass reach.  Assuming that a natural Higgsino mass must be roughly below 400~GeV (naively corresponding to less than 5\% fine-tuning of the Higgs boson mass), the displaced GMSB decay searches can probe a large fraction of the available space with $\sqrt{F}$ between 10~TeV and a several 1000~TeV.  In particular, for $m_{\tilde H} \simeq$ few~100~GeV, there is now a fairly firm constraint $\sqrt{F} \gsim$~1000~TeV, unless the decay is dominated by $h\tilde G$.
\item  Mini-split SUSY with sub-TeV gluinos is close to being fully ruled-out for any squark mass scale.  An immediate escape hatch is to compress the spectrum to $m_{\tilde g} - m_{\tilde \chi} \lsim 100$~GeV.  We have also not studied in detail the limits on decays involving top quarks, though many other searches then open up, and it would be surprising if the limits become appreciably weaker.  Non-minimal decay topologies involving electroweak-ino cascades could also be interesting to study, but would likely only yield signals that are even more visible to displaced decay searches.
\item   Generic colored superparticles as (N)LSPs in either RPV or GMSB are to large degree ruled out for any lifetime if the mass is below about 1~TeV, again with the gluino limits tending to be several 100~GeV stronger than squark limits.  Light RPV squarks with prompt decays would also face direct search difficulties similar to stops, but partially compensated by the higher multiplicity of flavor/chirality states.
\end{itemize}

What else remains to be done?  Within the context of RPV (both baryonic and leptonic), a more thorough survey of the current status of different LSPs and flavor structures along the lines of~\cite{Evans:2012bf} and~\cite{Graham:2012th,Graham:2014vya} seems warranted.  Leptonic RPV in particular has a quite large set of possible couplings.  Spectra with ``electroweak'' LSPs besides Higgsinos, namely sleptons or gauginos, also deserve further attention, as they can become much more visible than they would be if their decays were prompt.  For general gauge mediation, we have emphasized in Section~\ref{sec:GMSB} that the full set of possible NLSPs is (rather remarkably) almost fully covered.  The major exceptions are again sleptons and winos, with the latter offering an interestingly varied array of different signatures.  We again expect that the existing set of displaced and prompt searches have much to say about all of the above models, though in many cases coverage may still be entirely lacking, unnecessarily weak, or ambiguous given the current limitations of making public the general analysis acceptances.

With the upcoming Run~2 of the LHC, the mass reach for the models that we have explicitly studied might be expected to roughly double, assuming that similar analyses will be undertaken.  We encourage the experiments to continue their displaced decay search programs at an even greater level of breadth so that interesting signals are not left behind.  We also hope that future recasts are better facilitated by more explicit discussions of analysis acceptances, less tied to one or two specific fully simulated models in limited kinematic ranges.  Endeavors like ours should ideally not require as much from-scratch calibration, extrapolation, and guesswork, as detailed here in Section~\ref{sec:searches} and in the Appendices below.  Of course, the need to facilitate more general model interpretations becomes even more pressing if a discovery is made.  Works along the lines of CMS's stable charged particle efficiency maps~\cite{CMS-EXO-13-006} are a step in the right direction.  But we emphasize that even coarse parametrizations such as the ones that we have developed can prove invaluable, especially if directly compared against full internal simulations by the collaborations.  We hope that our work, which has further clarified the extreme power and broad model reach of these searches, spurs further activities in these directions, and we look forward to the next round of LHC displaced search results.


\acknowledgments{We thank Joshua Hardenbrook, Ben Hooberman, David Shih, and Wells Wulsin for useful discussions. ZL and BT were supported by DoE grant No. DE-FG02-95ER40896 and by PITT PACC. ZL was also supported in part by the Andrew Mellon Predoctoral Fellowship and a PITT PACC Predoctoral Fellowship from Dietrich School of Art and Science, University of Pittsburgh, and in part by the Fermilab Graduate Student Research Program in Theoretical Physics. 


\appendix

\section*{Appendix: Detector Simulations and Calibrations}
\label{sec:detector}
\renewcommand{\thesubsection}{\Alph{subsection}}

The set of recastings performed in this paper are all based on a small set of custom detector simulation codes.  These are meant to capture the main features relevant to each analysis while bypassing the many highly complex details of the real detector response.  Since the displaced SUSY models studied here have never appeared in public collaboration results, calibration of our simulations must rely on the specific physics models that appear in the experimental literature.  Our aim has been to build simulations that reproduce the known experimental acceptances for new physics at the $O(1)$ level, which we are often able to accomplish even with very minimal treatment of detector issues.  In most cases this level of agreement is adequate to define reasonable estimates of the true sensitivity contours in the mass-lifetime plane, as the sensitivity can be an extremely steep function in both variables.  However, typically we can achieve even $O(10\%)$ agreement with the experimental results, either ``out of the box'' or by adjusting ad hoc efficiency factors.  A single tuning of the latter type is often able to reproduce the results for a broad range of models.

Below, we provide complete descriptions of our detector simulations and their calibrations.

\subsection{CMS Heavy Stable Charged Particles}
\label{sec:CMSHSCP_calibration}

The simulation required for the CMS heavy stable charged particle search~\cite{Chatrchyan:2013oca} can be quite minimal, since CMS searches for the same physics signal that we do: long-lived stop and gluino R-hadrons.  The only novelty that we introduce is the finite lifetime, which effectively adds an additional factor to the overall R-hadron acceptance.  Decays must occur completely outside of the muon system, with no visible particles from the decay pointing back.  We define the outer edge of this active volume simply as a cylinder with 8~m radius and 10~m half-length.\footnote{Reference~\cite{CMS-EXO-13-006} recommends using a slightly smaller radius of 7~m and a slightly longer half-length of 11~m.}

Calibration is trivial, as we define our analysis in Section~\ref{sec:CMSHSCP_analysis} to reproduce the results of~\cite{Chatrchyan:2013oca} for long lifetimes.

\subsection{CMS Displaced Dijets}
\label{sec:CMSdisplacedDijets_calibration}

The CMS displaced dijet search~\cite{CMS-EXO-12-038,CMS:2014wda} utilizes that detector's highly precise tracking capabilities.  Nonetheless, the tracker is not perfect, especially in the extreme cases of $O$(m) displacements.  At such large displacements, particles may not traverse enough detector layers to furnish a reconstructable string of hits, or may lead to hit patterns that not usable given a detector geometry and track-finding software that is highly optimized for tracks originating very close to the beampipe.  In particular, CMS observes a dramatic drop in tracking efficiency versus vertex radius in simulation~\cite{Chatrchyan:2014fea}, falling to zero beyond 60~cm.  The tracker also becomes highly inefficient for soft particles, especially for transverse momenta that are small enough for the particles to spiral-out.  The efficiency experiences a rapid turn-on near 1~GeV.

It is not possible to reproduce the tracking and vertex-finding performance in detail here.  Instead, we rely on simplistic parametrizations, and validate them against the results presented in the analysis note~\cite{CMS-EXO-12-038}.  As a ``zeroth-order'' approximation, we could consider the tracker to be perfect within some fiducial volume.  Our most naive version applies hard cutoffs at $r = 60$~cm and $p_T = 1$~GeV.  Based on the discussions in~\cite{CMS-EXO-12-038,CMS:2014hka}, and Fig.~3 in~\cite{CMS:2014hka}, we also include a corresponding longitudinal position cutoff at $z = 55$~cm, as well as a cutoff in transverse impact parameter at 30~cm.  However, we have found by comparing to~\cite{CMS-EXO-12-038} that this zeroth-order treatment is far too idealized.  CMS's new physics models with $O$(cm) lifetimes are reproduced fairly well, with reconstruction rate estimates already typically within about 25\% of CMS's.  (This result by itself indicates the amazing performance of the CMS tracker system for charged particles originating near the beampipe.)  But the efficiencies for the longer $O$(m) lifetimes come out too large by up to a factor of 3.5, and distributions of variables such as the reconstructed vertex radius often significantly disagree.

We therefore use a slightly more aggressive parametrization for our nominal detector simulation.  We continue to apply a hard cutoff at $p_T = 1$~GeV.  For the geometric limitations, instead of hard cutoffs, we apply a nontrivial track-finding probability.  We construct this probability by starting off with a prompt tracking efficiency of 90\%, and multiplying it by the product of three linearly-falling probability factors, one each for $r$, $z$, and transverse impact parameter.  Each factor starts at unity, and falls to zero at the geometric cutoff.  Finally, in addition to these track-by-track efficiencies, we apply an efficiency for identifying each candidate vertex.  This efficiency is the square of a linear falloff function versus radius, intercepting zero at 60~cm.  The combination of these ad hoc efficiency factors significantly improves our predicted rates and distributions relative to CMS.\footnote{The additional ad hoc vertexing efficiency causes a more rapid degradation of reconstruction rates versus radius than what is reported by CMS in their supplementary online materials~\cite{CMS-EXO-12-038-web}.  However, an alternative approach of convolving the reported rates versus radius with our simulated number of decays versus radius similarly disagrees with the CMS analysis note~\cite{CMS-EXO-12-038}.  We do not attempt to resolve this apparent discrepancy, but haven chosen to parametrize our simulation to reproduce the integrated rates of the note, rather than the differential rates of the online material.  This is in any case the more conservative choice.}

For the rest of the detector, we use a very minimal treatment to capture the essential elements of the geometry.  The tracker is surrounded by imaginary surfaces corresponding the ECAL and HCAL faces.  The ECAL face corresponds to a closed cylinder of radius 1.3~m and half-length 3~m.  The HCAL face corresponds to a closed cylinder of radius 1.8~m and half-length 3.75~m.  Charged particles that fail tracking due to the above inefficiencies are analytically propagated through the 3.8~T magnetic field to the appropriate calorimeter face:  ECAL for electrons, HCAL for charged hadrons (failed muon tracks are discarded).  Photons and neutral hadrons are propagated to the appropriate calorimeter face along straight lines.  Particles are absorbed at the calorimeter face, and replaced by anonymous massless pseudoparticles carrying the original particle energy, but with momentum oriented to toward the impact point.\footnote{For photons and electrons produced from displaced vertices in between the ECAL and HCAL faces, the energy is deposited without propagation, and the momentum vector pointed towards the vertex.  Any particle produced within the HCAL body is also absorbed in this way.  For this purpose we assign an outer HCAL surface of radius 2.8~m and half-length 5.5~m.  Particles produced outside of this volume are ignored.}  For successfully reconstructed tracks, we use the momentum vector at their extrapolated point of closest transverse approach to the beamline (where the 2D impact parameter is defined).  Track and calorimeter energy/momentum measurements are treated as perfect, since the effects of energy smearing are subdominant to our other modeling uncertainties.  The tracks and calorimeter pseudoparticles are the inputs into our jet clustering, excluding any identified isolated electrons or muons as defined in the next subsection.

\begin{figure}[tp!]
\centering
\subfigure{
\includegraphics[scale=0.40]{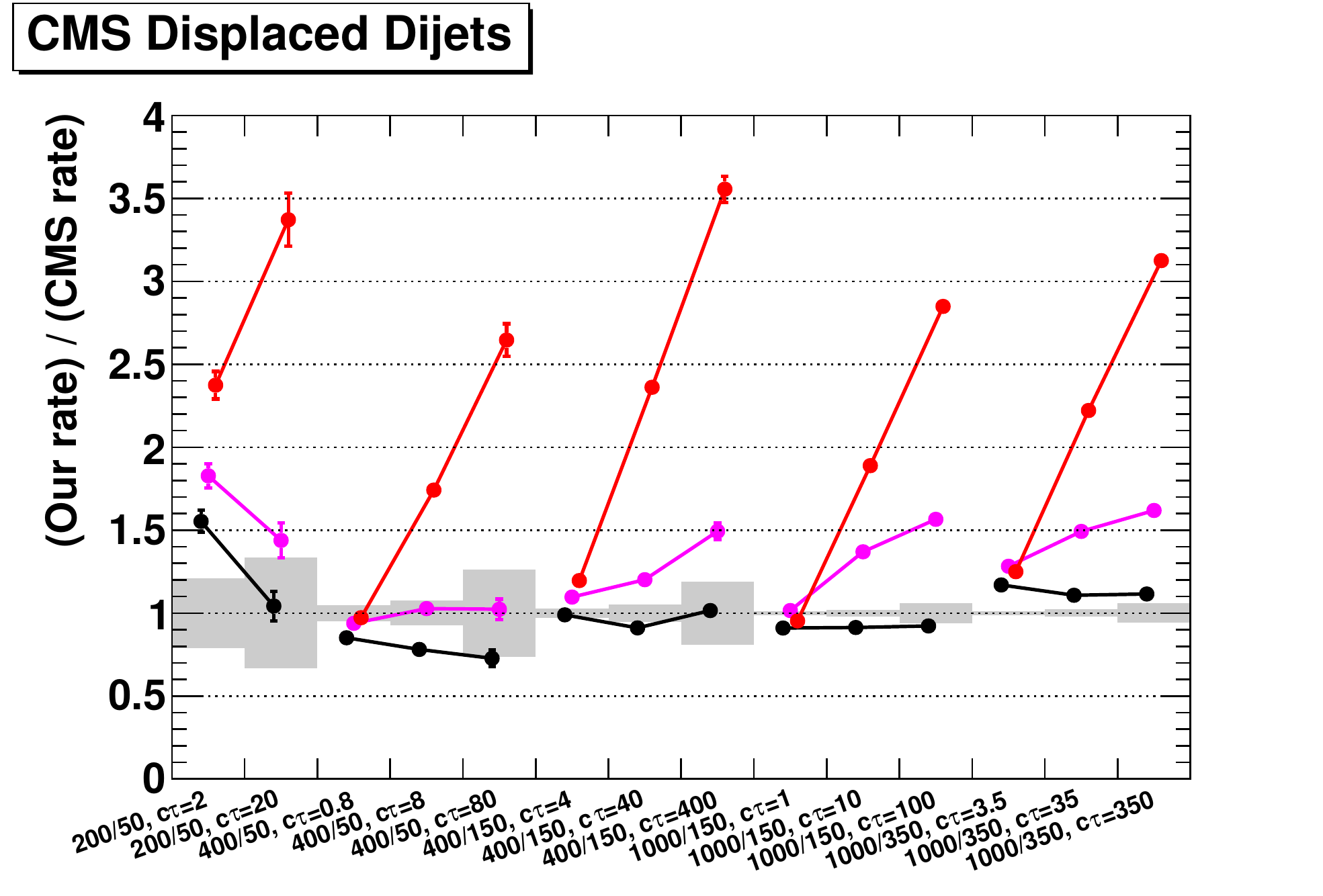}}
\subfigure{
\includegraphics[scale=0.40]{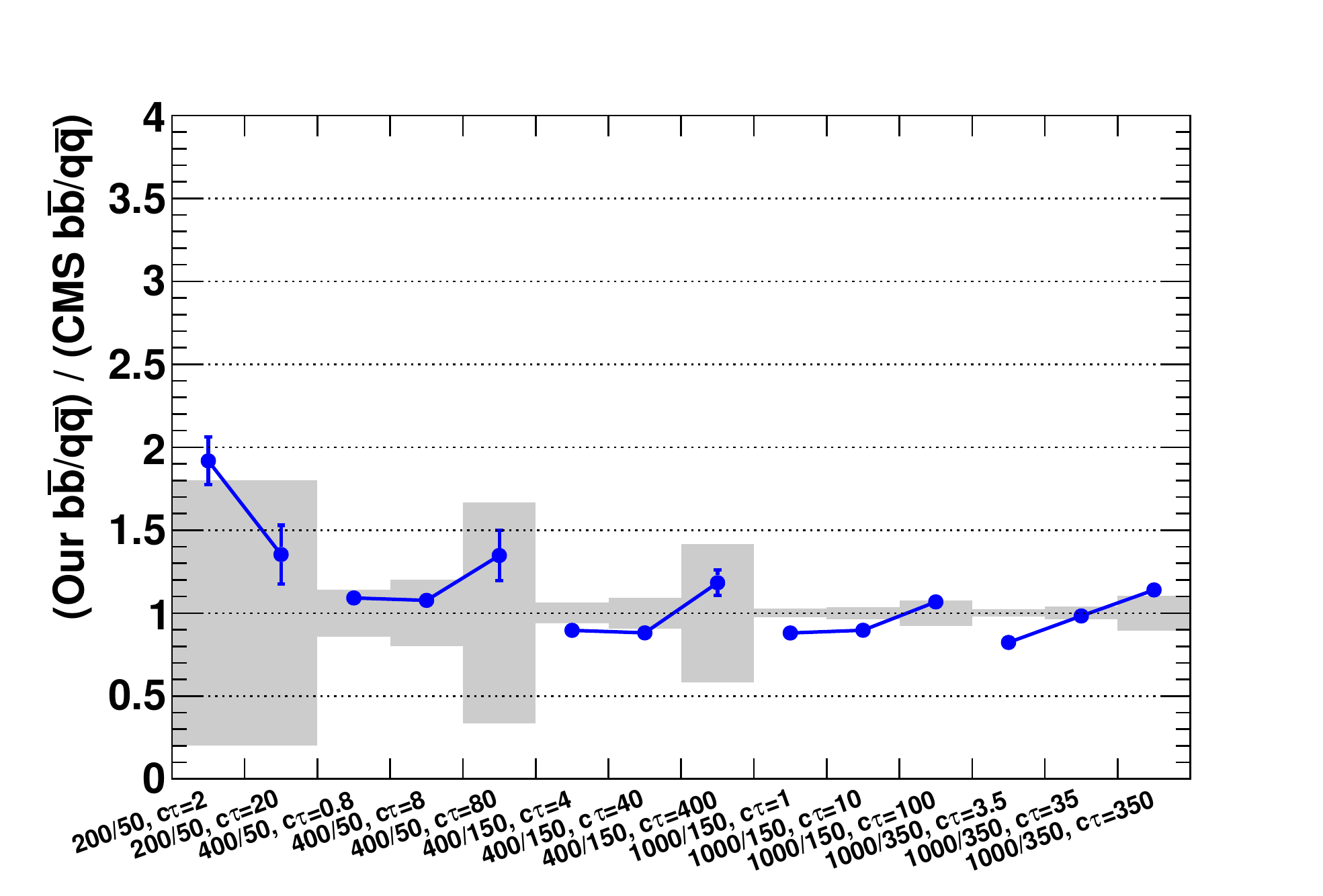}}
\caption{A comparison of our detector simulation to the CMS displaced dijet analysis, illustrating the ratio of individual displaced decay reconstruction rates to CMS for light quark decays ({\bf left}) and the ratio of ratios between $b$-quarks and light quarks ({\bf right}).  The 14 benchmark models are labeled by scalar~/ pseudoscalar masses in GeV and by $c\tau$ in cm.  On the left plot, we show our nominal simulation (black), our nominal simulation without the vertexing efficiency factor (pink), and our ``zeroth-order'' simulation with perfect tracking within the fiducial volume (red).  Error bars are monte carlo statistics from our simulations.  Grey bands indicate CMS's efficiency uncertainties.}
\label{fig:calibration_CMSdijet_AxEpsilon}
\end{figure}

We plot our calibration results of the per-decay reconstruction rate (``acceptance times efficiency'' in CMS's language) relative to CMS in Fig.~\ref{fig:calibration_CMSdijet_AxEpsilon}.  These comparisons are based on the set of Hidden Valley~\cite{Han:2007ae} inspired simplified models used by CMS, consisting of a heavy Higgs-like scalar, produced through gluon fusion, that decays into a pair of long-lived pseudoscalars.  The pseudoscalars then undergo displaced decays into dijets.  (We have not undertaken calibrations against the $\tilde \chi^0 \to \mu jj$ signatures that are studied in the more recent preprint~\cite{CMS:2014wda}.)   The steep slopes for our zeroth-order detector's reconstruction rates relative to CMS indicates that simulation's failure to correctly account for inefficiencies at high displacements.  A remnant of this slope remains after applying our tracking efficiencies but before applying our vertexing efficiencies.  The nominal detector is generally in agreement with CMS for all lifetimes to within 20\%, with the notable exception of the 200~GeV scalar model.  We do not consider this a serious issue, as this model has high sensitivity to initial-state radiation modeling.  The other models exhibit a good spread of overall masses, mass hierarchies, and lifetimes.  In particular, the 400/50 and 1000/150 models tend to produce dijets near the edge of the jet clustering radius, and we see that we tend to slightly underestimate their reconstruction rates relative to models with widely-separated jets.  Nonetheless, the highest dijet mass covered is 350~GeV, whereas some of our SUSY models go above 1~TeV.  We assume that there are no dramatic changes in efficiencies as we scale up in mass, though a broader set of simulated models from CMS would help to clarify the actual behavior.

Fig~\ref{fig:calibration_CMSdijet_AxEpsilon} also indicates our ability to reproduce reconstructions with heavy flavor, by showing a double-ratio of reconstruction rates.  The numerator is our estimated ratio of rates for $X \to b\bar b$ relative to $X$ decays to light flavors.  The denominator is CMS's estimate of the same ratio.  For this analysis, we pretend that the secondary displacements from the bottoms either cannot be resolved or are effectively ignored by the adaptive vertex finder.  The agreement is generally seen to be quite reasonable, with a handful of outliers disagreeing at more than 20\%.  The fact that we achieve such good agreement without separately displacing the bottoms strongly suggests that the dominant differences in light flavor and heavy flavor efficiencies stems from the different visible particle multiplicities and kinematics.  There remains a question of whether small primary displacements at the mm-level may have resolvable secondary displacements according to CMS's adaptive vertex fitter, such that only the ``best'' of the truth vertices from each decay actually contributes.  (E.g., $\tilde t \to \bar b \bar b$ could produce up to three separate vertices, one from the radiation before the $b$'s hadronize, and two more from the $b$-hadron decays.)  Presumably, this would degrade the efficiency, since fewer tracks would be usable from any given decay.  However, CMS does not provide enough information to infer exactly what happens for primary displacements below the cm-scale.  In the main results in our paper, we only consider the extremely conservative assumption that vertices with bottom/charm quarks and sub-cm displacements experience complete reconstruction failure.

\begin{figure}[tp!]
\centering
\subfigure{
\includegraphics[scale=0.40]{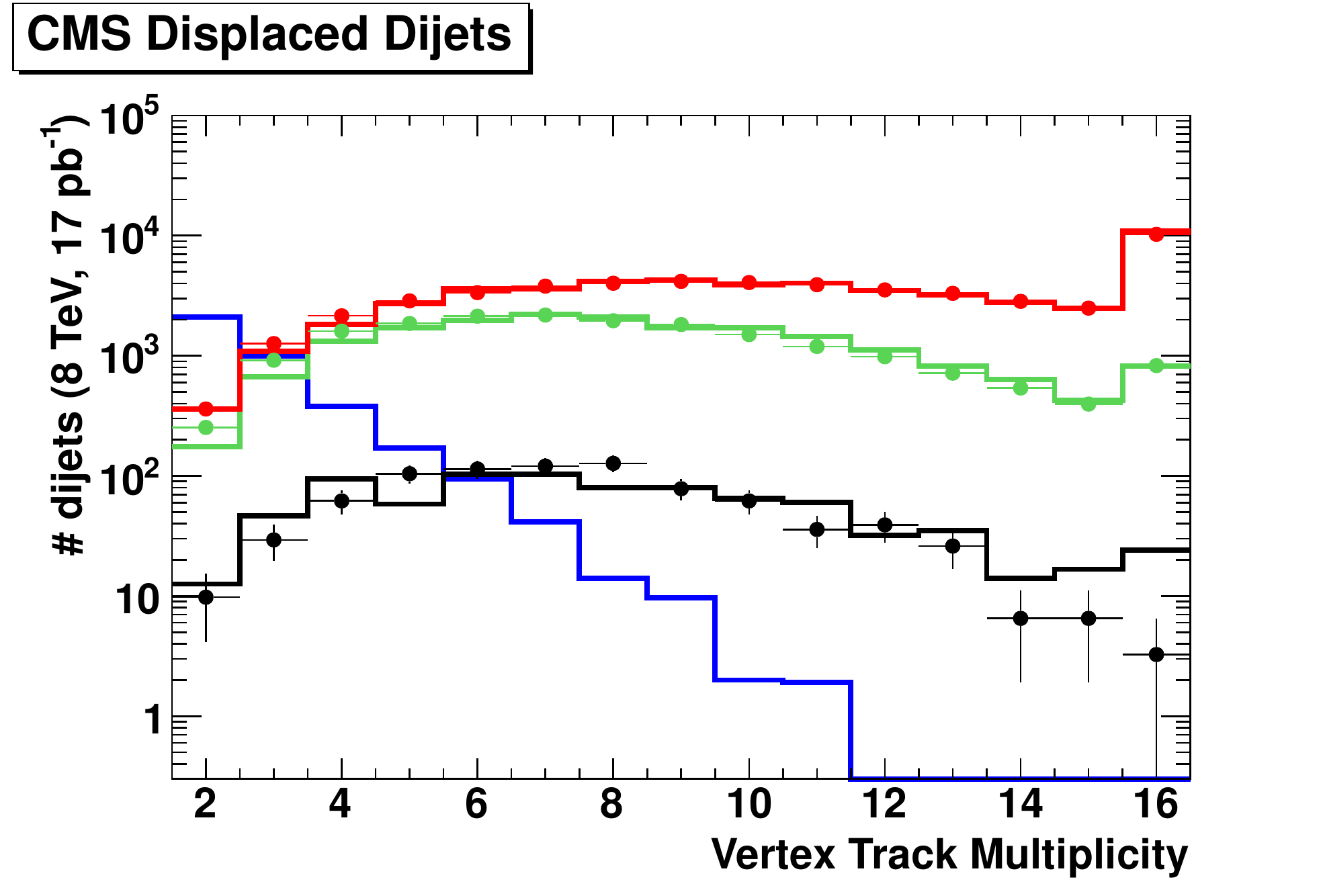}}
\subfigure{
\includegraphics[scale=0.40]{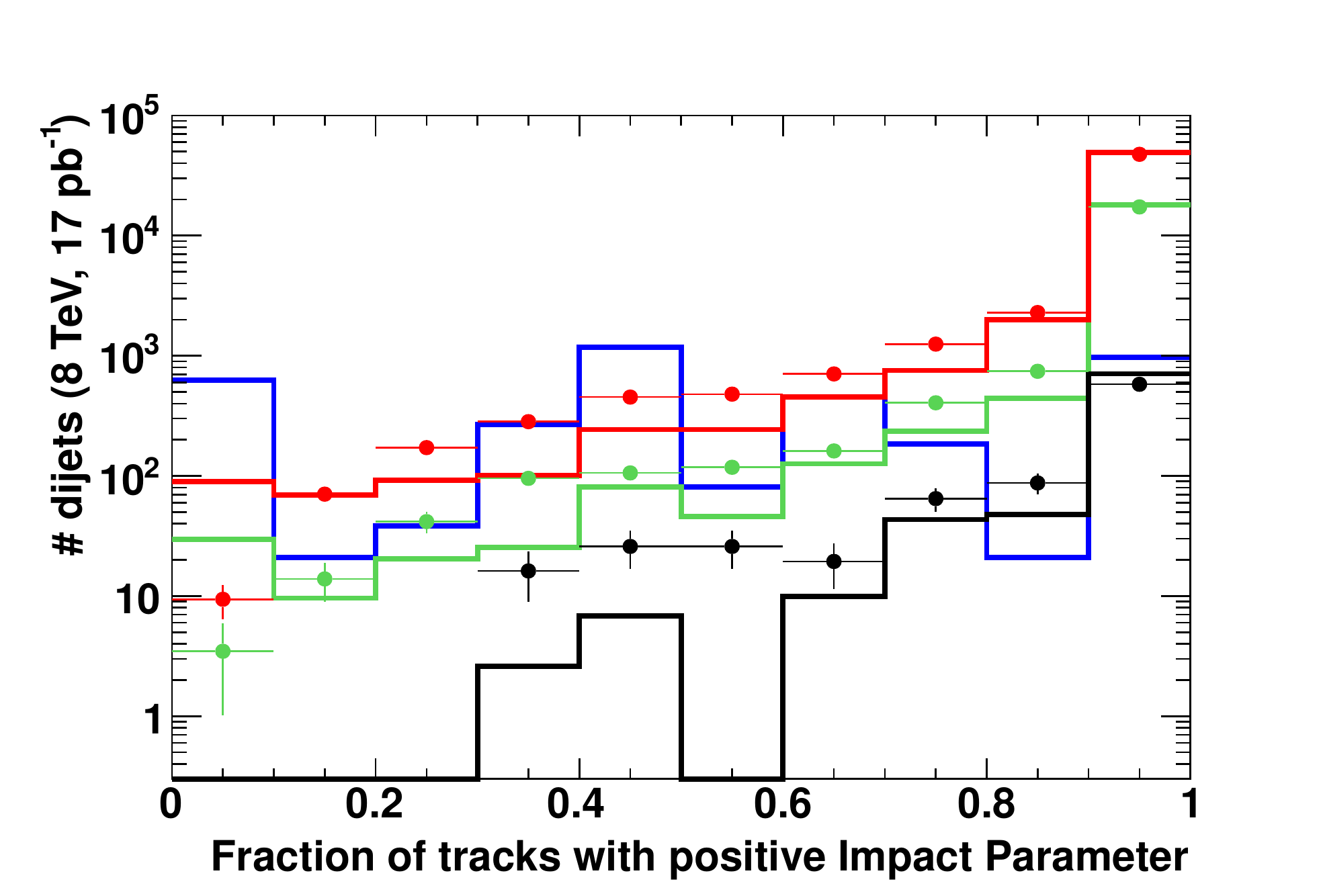}}
\subfigure{
\includegraphics[scale=0.40]{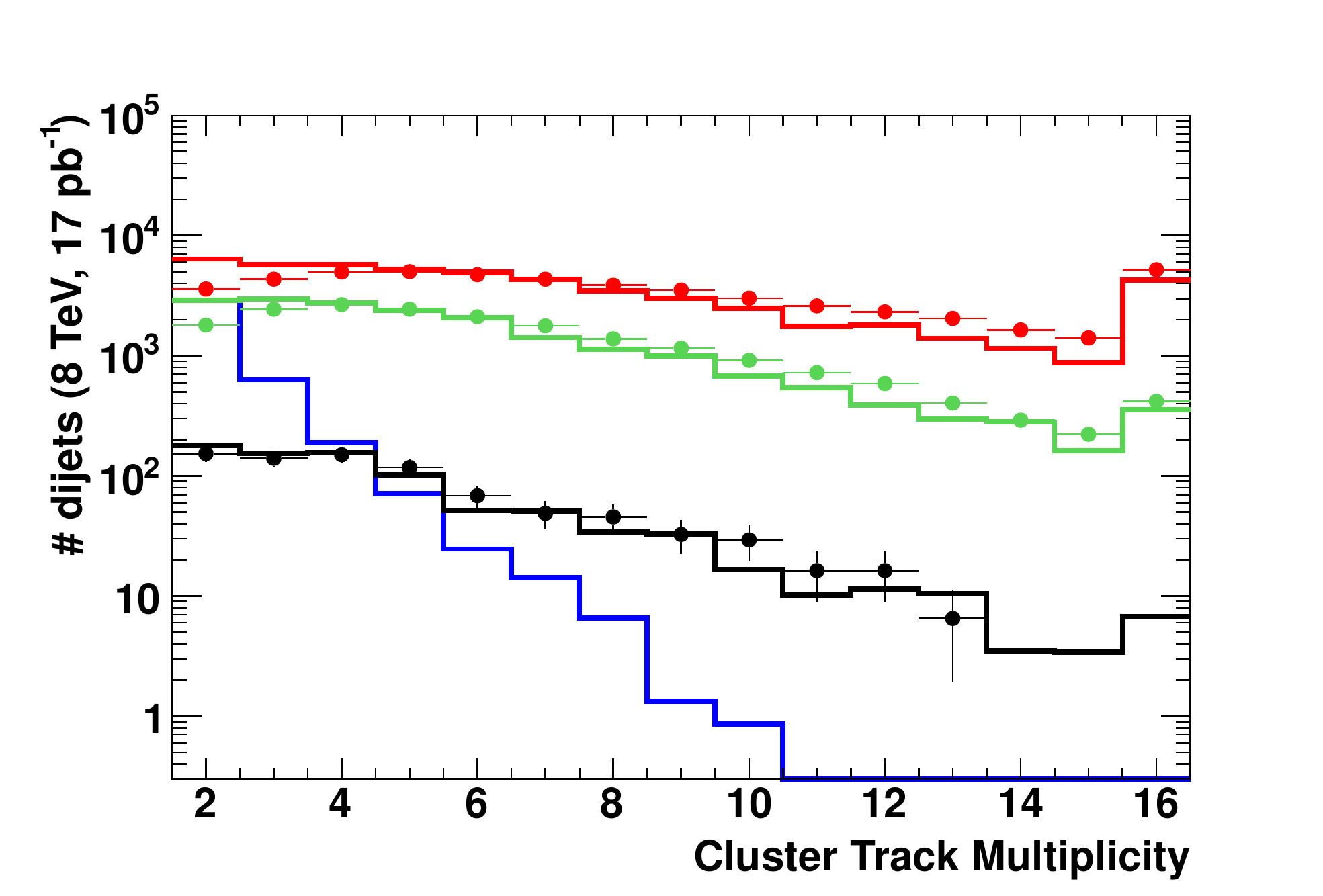}}
\subfigure{
\includegraphics[scale=0.40]{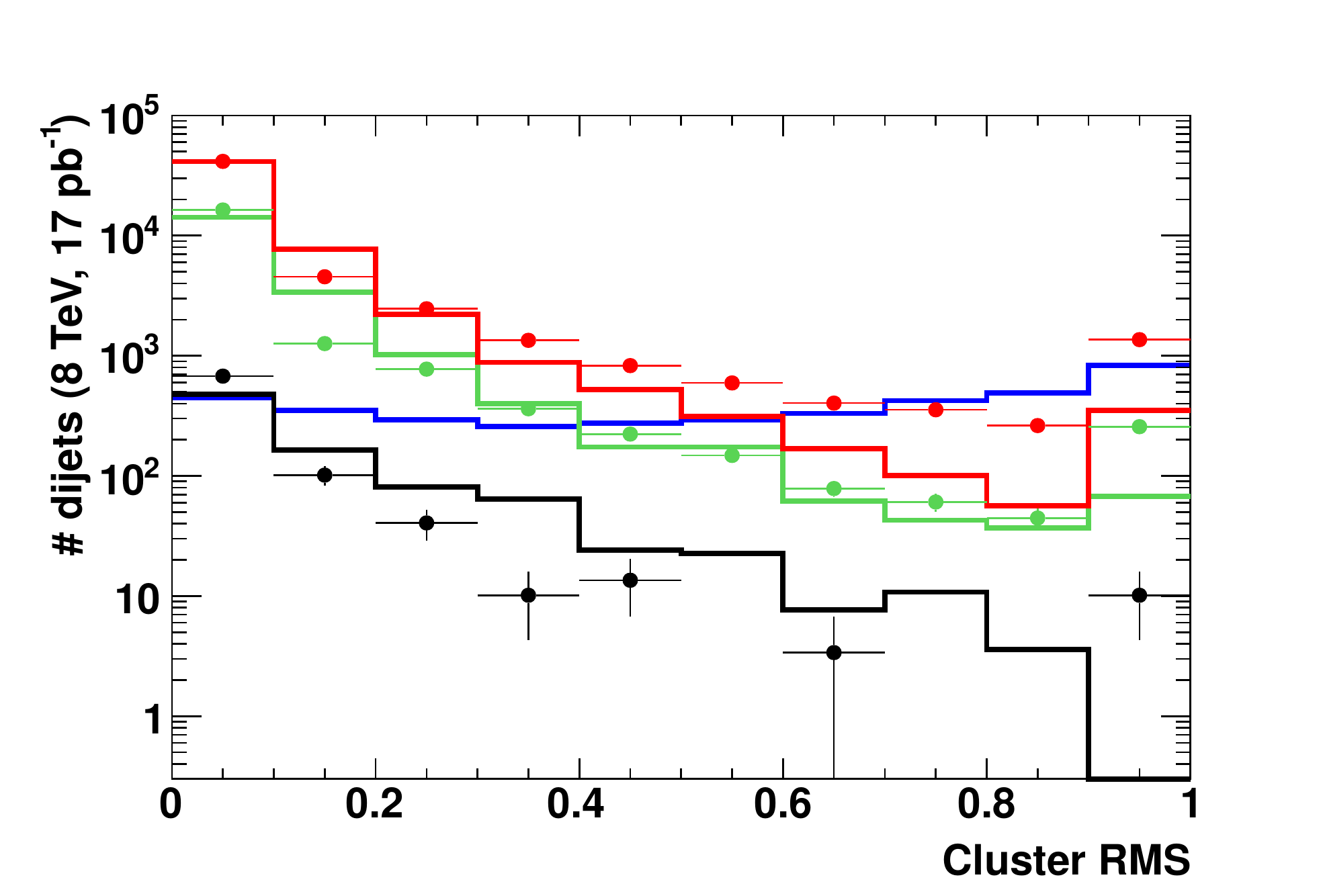}}
\subfigure{
\includegraphics[scale=0.40]{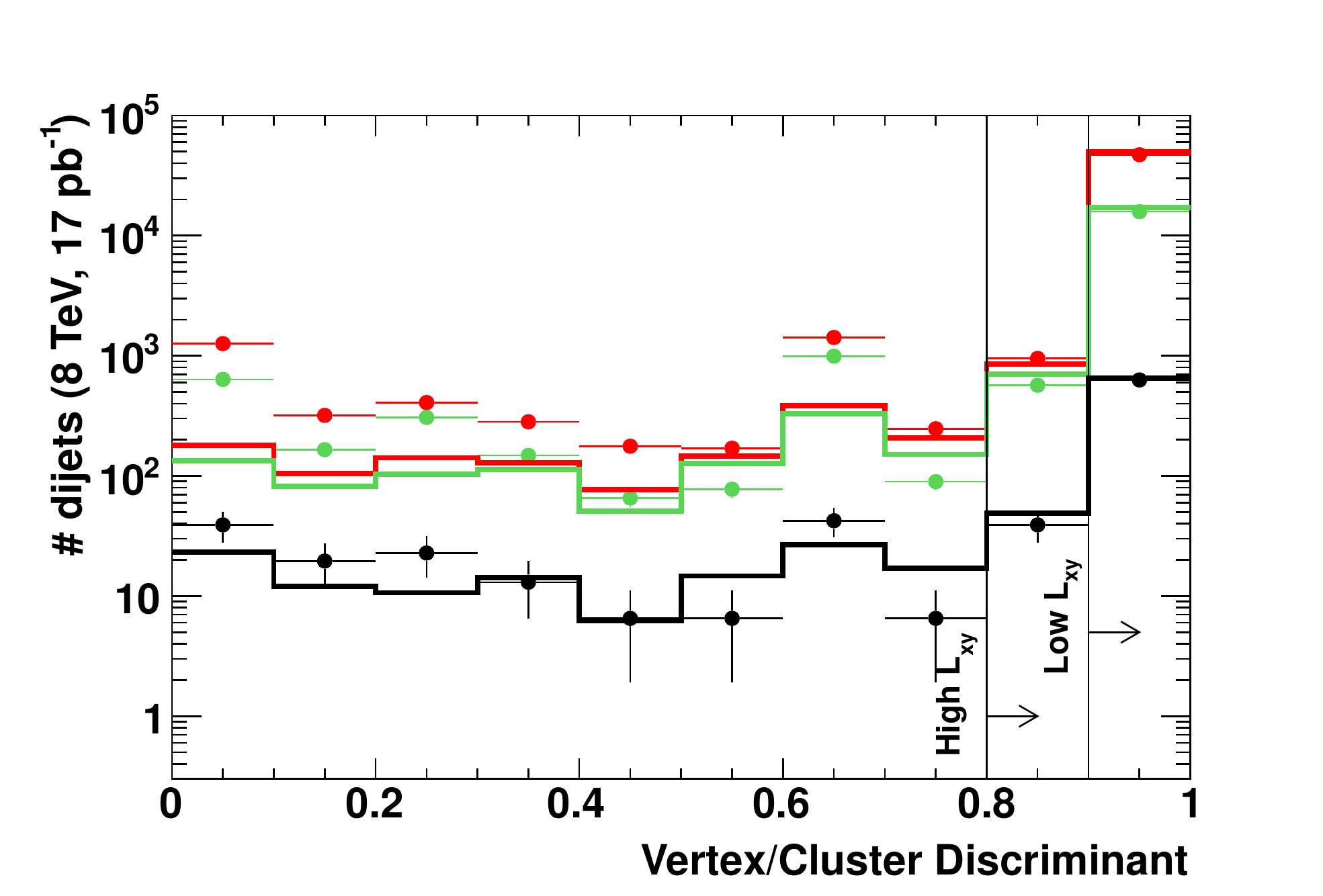}}
\caption{A comparison of our detector simulation to the CMS displaced dijet analysis, illustrating expected reconstructed dijet counts versus several discriminator variables (Fig.~1 of~\cite{CMS-EXO-12-038}).  Continuous histograms indicate CMS predictions for QCD background (blue) and different scalar~/ pseudoscalar masses: 1000~GeV~/ 350~GeV with $c\tau = 35$~cm (red), 400~GeV~/ 150~GeV with $c\tau = 40$~cm (green), and 200~GeV~/ 50~GeV with $c\tau = 20$~cm (black); all with $\sigma \to 10$~$\mu$b for visualization.  Data points with error bars are our simulation predictions, and are color-matched and normalized to the corresponding CMS model histogram.  The multivariate discriminant cuts for High-$\Lxy$ and Low-$\Lxy$ signal regions are also indicated.}
\label{fig:calibration_CMSdijet_variables}
\end{figure}

Finally, we comment on our modeling of the variables that go into CMS's multivariate vertex/cluster discriminant variable.  The discriminant is defined as a ratio of products of normalized p.d.f.'s over four variables:  vertex track multiplicity, vertex positive IP fraction, cluster track multiplicity, and cluster RMS.  The ``cluster'' is formed as described in Section~\ref{sec:CMSdisplacedDijets_analysis}.  We plot our predicted Hidden Valley signal distributions for the four variables and the multivariate discriminant in Fig.~\ref{fig:calibration_CMSdijet_variables}, along with the CMS predictions.  For both our own simulation and CMS's, we form the discriminant using CMS's p.d.f.'s.  (The signal discriminant distribution that would be predicted by CMS is generated by us through toy monte carlo from the individual variables' histograms.)  The agreements are mostly reasonable, though there can be differences in the tails.  The fraction of events passing the discriminant cuts are nonetheless nearly identical to CMS.  The signal's discriminant distributions at high values appear to be mostly driven by the multiplicity variables, which we model relatively well.

\subsection{CMS Displaced Dileptons}
\label{sec:CMSdisplacedDileptons_calibration}

For the CMS displaced dilepton search~\cite{CMS:2014hka}, we use the same detector simulation discussed in the previous subsection, though without the vertex reconstruction penalty as a function of radius.  We also apply the $p_T$-dependent lepton ID efficiencies provided in the appendix of~\cite{CMS-SUS-13-010}, though we divide these by 0.9 to approximately deconvolve the prompt track-finding efficiency which we have already accounted for.

Similar to the displaced dijet search, the benchmark models feature a Higgs-like scalar decaying to a pair of long-lived pseudoscalars, but with the latter decaying to $e^+e^-$ or $\mu^+\mu^-$.  Unlike the displaced dijet search, detailed tables of acceptances and efficiencies are not provided.  However, a handful of specific numbers are given, and the observed cross section limits can also be used to infer overall reconstruction rates.

For the 1000~GeV~/ 150~GeV models, CMS's individual pseudoscalar reconstruction rates for $c\tau = 1$~cm in the electron (muon) decay channel are given as 36\% (46\%).  Our predictions are in decent agreement, at 36\% (40\%).  For $c\tau = 20$~cm in the electron (muon) decay channel, CMS gives 14\% (20\%).  Our predictions here are 9\% (10\%), indicating too-low reconstruction rates by $O(1)$.  This is likely due to our ad hoc track-finding efficiency's linear falloff being too steep for this analysis.  Indeed the supplementary online material of~\cite{CMS-EXO-12-037} indicates fairly stable reconstruction efficiency for this model out to $r \simeq 50$~cm, where our simulation would predict nearly zero.  It seems quite likely that the low track multiplicity of the decay contributes to a higher rate of successful displaced track reconstructions, relative to that of the displaced dijets.  However, we conservatively continue to use our tracking efficiency factors derived for the latter analysis, especially as in some SUSY models the lepton pair may be produced in association with hadronic tracks from the same vertex.  This does not lead to a runaway loss of efficiency at high lifetime for simple dilepton decays, as we will see.

\begin{figure}[tp!]
\centering
\subfigure{
\includegraphics[scale=0.40]{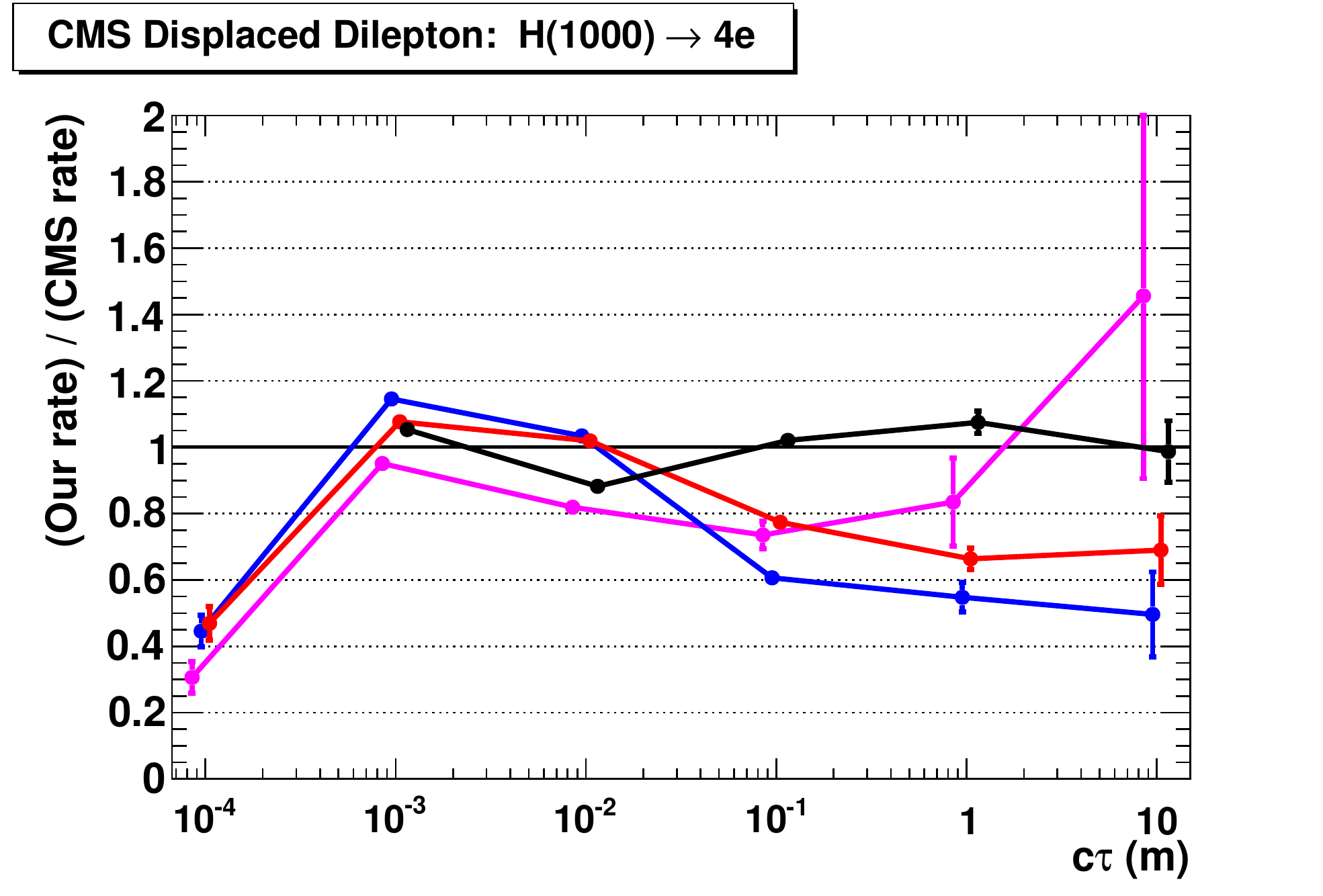}}
\subfigure{
\includegraphics[scale=0.40]{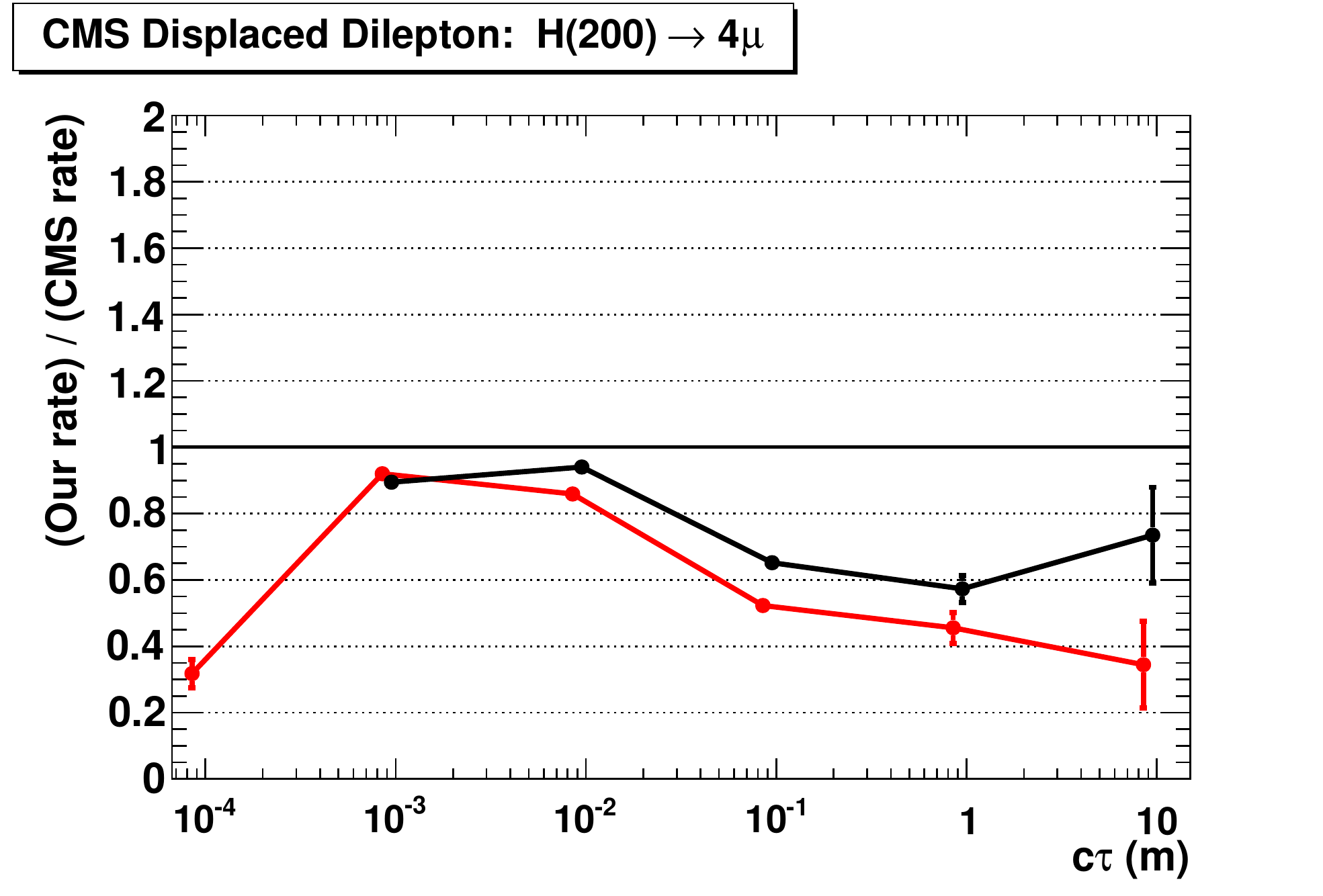}}
\caption{A comparison of our detector simulation to the CMS displaced dilepton analysis by inverting their reported limits, illustrating the approximate ratio of event reconstruction rates to CMS for 1000~GeV scalar cascade decays to electrons ({\bf left}) and 200~GeV scalar cascade decays to muons ({\bf right}) versus pseudoscalar lifetime (based on the limits presented in Figs.~4 and 5 of~\cite{CMS:2014hka}).  On the left plot, we show pseudoscalar masses of 20~GeV (pink), 50~GeV (blue), 150~GeV (red), and 350~GeV (black).  On the right plot, we show pseudoscalar masses of 20~GeV (red) and 50~GeV (black).  Error bars are monte carlo statistics from our simulations.  All models are evaluated at $c\tau$ in powers of 10, but are offset slightly horizontally for clarity.}
\label{fig:calibration_CMSdilepton}
\end{figure}

The calculation of limits from the signal reconstruction rates is in principle nontrivial, involving incorporation of backgrounds and various sources of systematic errors.  However, given that the analysis is ultimately both background-free (as inferred from a control region) and has zero observed events, and that the naive Poissonian 95\% C.L. limit for such an experiment is $\approx 3$~signal events, we can perform a back-of-the-envelope estimate of the reconstruction rates.  The explicit CMS per-candidate rates discussed above are reproduced at the 10\% level using this method, providing a good cross-check.  Fig.~\ref{fig:calibration_CMSdilepton} illustrates our estimated relative rates for a few example models.  The general behavior is that our simulation appears to be less efficient than reality, especially at very low and very high lifetimes.  The former is perhaps unsurprising given our coarse modeling of the impact parameter cut, and the latter could be due to our overzealous tracking efficiency falloff.  Nonetheless, some of the model points still appear to exceed the CMS rate, even at higher lifetimes, motivating us to keep these somewhat artificial effects to help prevent us from inferring too-strong limits on SUSY models.

\subsection{CMS Displaced Electron and Muon}
\label{sec:CMSemu_calibration}

For the CMS displaced electron and muon search~\cite{Khachatryan:2014mea}, we continue to use the detector simulation described in Appendix~\ref{sec:CMSdisplacedDijets_calibration}.  To better match the efficiencies reported by CMS in the present analysis, we apply a flat event-by-event weight of 0.80. 

\begin{figure}[tp!]
\centering
\includegraphics[scale=0.40]{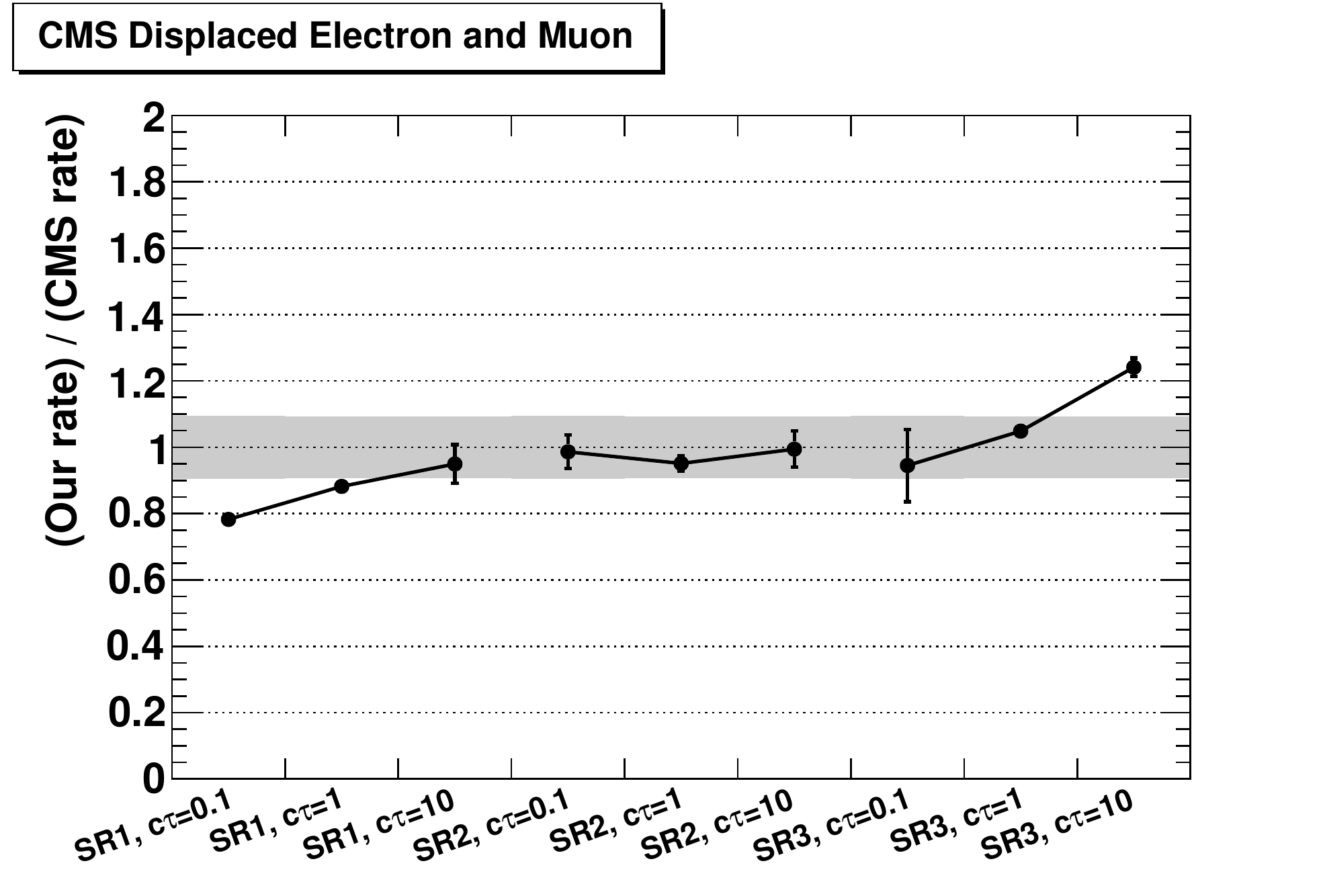}
\caption[]{A comparison of our detector simulation to the CMS displaced electron and muon analysis, illustrating the ratio of individual displaced decay reconstruction rates to CMS for the 500~GeV stop benchmark model.  The bins are labeled by analysis signal region (SR1 for lower impact parameters, through SR3 for higher impact parameters) and by the stop's $c\tau$ in cm.  Grey bands indicate CMS's efficiency uncertainties, with the stop cross section uncertainty quadrature-subtracted.}
\label{fig:calibration_CMSemu}
\end{figure}

The physics model studied by CMS is RPV stop pair production, but with each stop decaying into $bl^+$, with equal branching fractions into each of the three lepton generations.  It is not explicitly stated in this analysis whether the displaced leptons could suffer any loss of efficiency if produced from the decay of a charged stop-hadron, leading to an ``exploding track'' topology rather a displaced vertex with no string of hits tracing back to the primary vertex.  In any case, since the analysis focuses on impact parameters below 2~cm, it is mostly sensitive to decays that occur before reaching the pixels.  We treat charged and neutral stop-hadrons identically, assuming that this is not a major issue.  CMS gives explicit reconstruction rates for a 500~GeV stop at lifetimes of 0.1~cm, 1~cm, and 10~cm.  Our detector simulation does a good job of reproducing all of these numbers to 20\% accuracy, as indicated in Fig.~\ref{fig:calibration_CMSemu}.

\subsection{ATLAS Muon Spectrometer}
\label{sec:ATLASmuonChamber_calibration}

The ATLAS muon spectrometer search~\cite{ATLAS:2012av} is again very difficult to model without access to both a full ATLAS detector simulation and the exact reconstruction algorithms.  Here, we simply parametrize all of these with fixed efficiency factors.  Our simulation defines an active trigger volume within $r = [4.0,6.5]$~m and $|\eta| < 1.0$.  Displaced particles that decay in this region are given a 50\% chance to fire the muon RoI cluster trigger.  For events passing the trigger, two displaced decays must be reconstructed in either the muon barrel or muon endcap, respectively defined as the volumes $r = [4.0,7.5]$~m and $|\eta| < 1.0$, or $|\eta| = [1.0,2.5]$ and $z = [8,14]$~m.  Both of these decays must also occur within the data acquisition timing window, designed to follow particles moving through the detector at the speed of light.  We choose a maximum delay of 7~ns, which we have independently calibrated to the extremely timing-sensitive ATLAS 120~GeV/ 40~GeV model results.  Our calibrated choice of maximum delay indeed corresponds to the end of the efficiency plateau versus time for the trigger~\cite{ATLAS-2009-082}.  We assign each displaced vertex that survives this cut a reconstruction rate of 40\%, irrespective of which ones were capable of firing the RoI trigger.  Finally, in order to very approximately account for possible isolation failures when a displaced decay points back to the detector volume, we limit the amount of visible transverse energy flowing back into the HCAL to 15~GeV.  For this purpose, we define the outer surface of the HCAL as a cylinder of radius 4.25~m and half-length 6~m.  This cut tends to have only modest effect on ATLAS scalar models, but could become important for our SUSY models at higher mass.

\begin{figure}[tp!]
\centering
\includegraphics[scale=0.40]{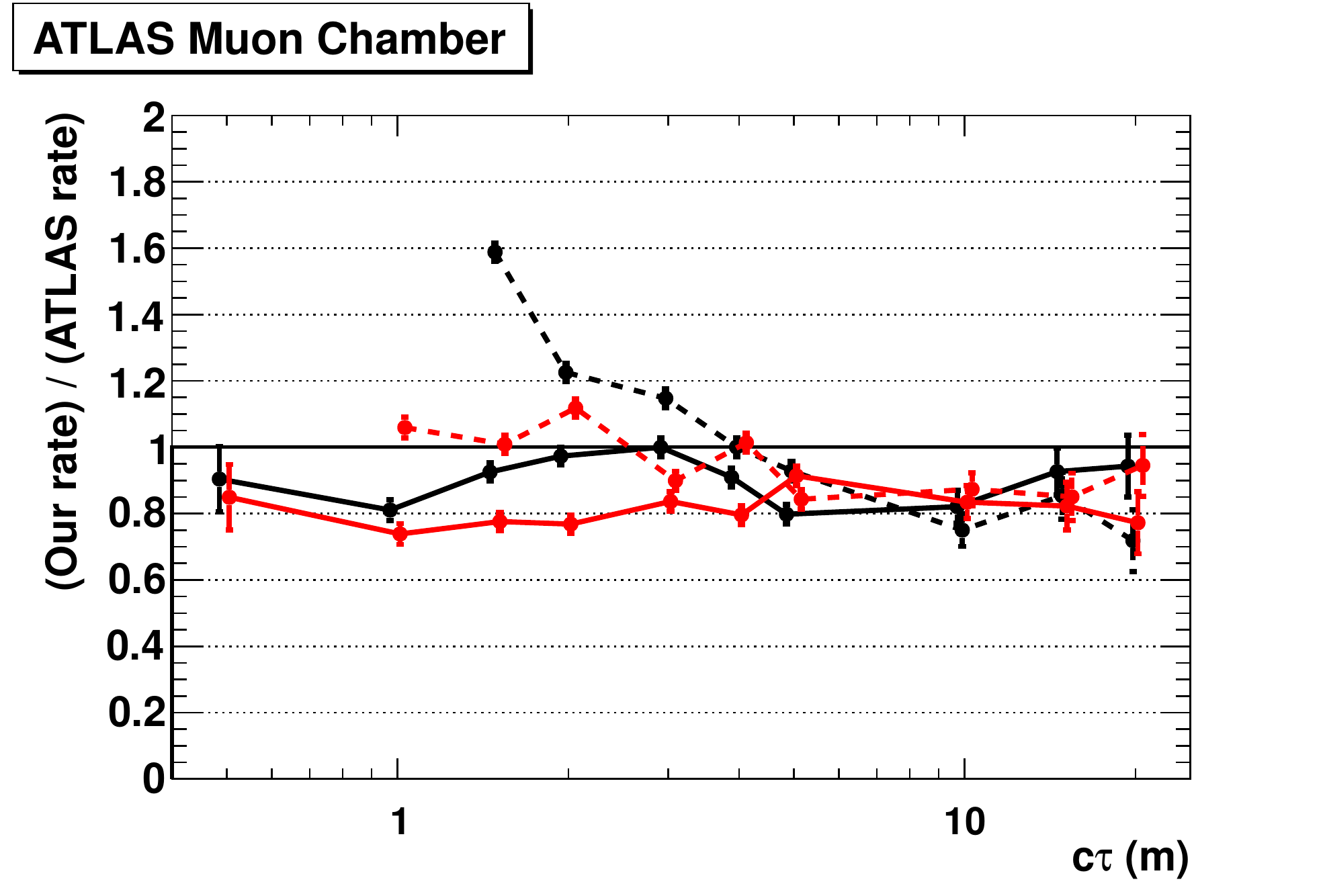}
\caption{A comparison of our detector simulation to the ATLAS muon chamber analysis using their reported limits, illustrating the ratio of event reconstruction rates to ATLAS versus pseudoscalar lifetime (based on the limits presented in Fig.~3 of~\cite{ATLAS:2012av}).  We show scalar~/ pseudoscalar masses of 120~GeV~/ 20~GeV (solid black), 120~GeV~/ 40~GeV (dashed black), 140~GeV~/ 20~GeV (solid red), and 140~GeV~/ 40~GeV (dashed red).  Error bars are monte carlo statistics from our simulations.  All models are evaluated at common $c\tau$ of (0.5,1,1.5,2,3,4,5,10,15,20)~m, but are offset slightly horizontally for clarity.}
\label{fig:calibration_ATLASmuonChamber}
\end{figure}

Similar to the CMS displaced dilepton analysis above, we calibrate against the ATLAS models by extracting their overall efficiencies from the reported limits.  Given that this analysis is background-free, the limits are assumed to correspond to $\approx3$ signal events.  The result of this comparison is shown in Fig.~\ref{fig:calibration_ATLASmuonChamber}.   We are able to reproduce all of the ATLAS results to within about 20\%.  (The one outlier point occurs on a very steep part of the efficiency curve.)

\subsection{ATLAS Low-EM Jets}
\label{sec:ATLASlowEM_calibration}

For our SUSY models to pass the ATLAS low-EM jets search~\cite{ATLAS-2014-041}, both displaced decays must occur within an effective HCAL volume:  a ``barrel'' with $|\eta| < 1.7$ and $r = [2.25,3.35]$~m; or an ``endcap'' with $|\eta| < 2.5$, $r < 2.0$~m, and $z = [4.25,5.0]$~m.  To approximate ATLAS's isolation cuts against activity in the ECAL and tracker, we apply a flat efficiency factor of 0.55 per displaced decay.\footnote{ATLAS's reported efficiencies are smaller than this.  However, those also fold in the efficiencies for passing the $p_T$ cuts within their studied physics models.}  (As usual, events with one or two charged displaced particles, which would leave high-$p_T$ tracks, are not considered.)

Our detector simulation assumes perfect and immediate absorption of the visible decay energy within our active HCAL volume.  However, we conservatively veto events where any visible final-state particle from either displaced decay points back towards the ECAL volume, which we take to be $r < 2.0$~m and $|z| < 4.0$~m.  The effect of this veto is modest for most of ATLAS's Hidden Valley models, but notably has a nearly 50\% effect on the acceptance for the highest mass models, actually improving our agreement (see below).  The cut is especially relevant for our SUSY models, which generally contain many more decay particles produced at large angles.  Practically, for us the cut serves as a proxy for any number of unknown requirements on the energy pattern and timing in the HCAL, in addition to the explicit limits on the nearby ECAL activity.  As such, it serves as our largest source of modeling uncertainty on this analysis.

\begin{figure}[tp!]
\centering
\includegraphics[scale=0.40]{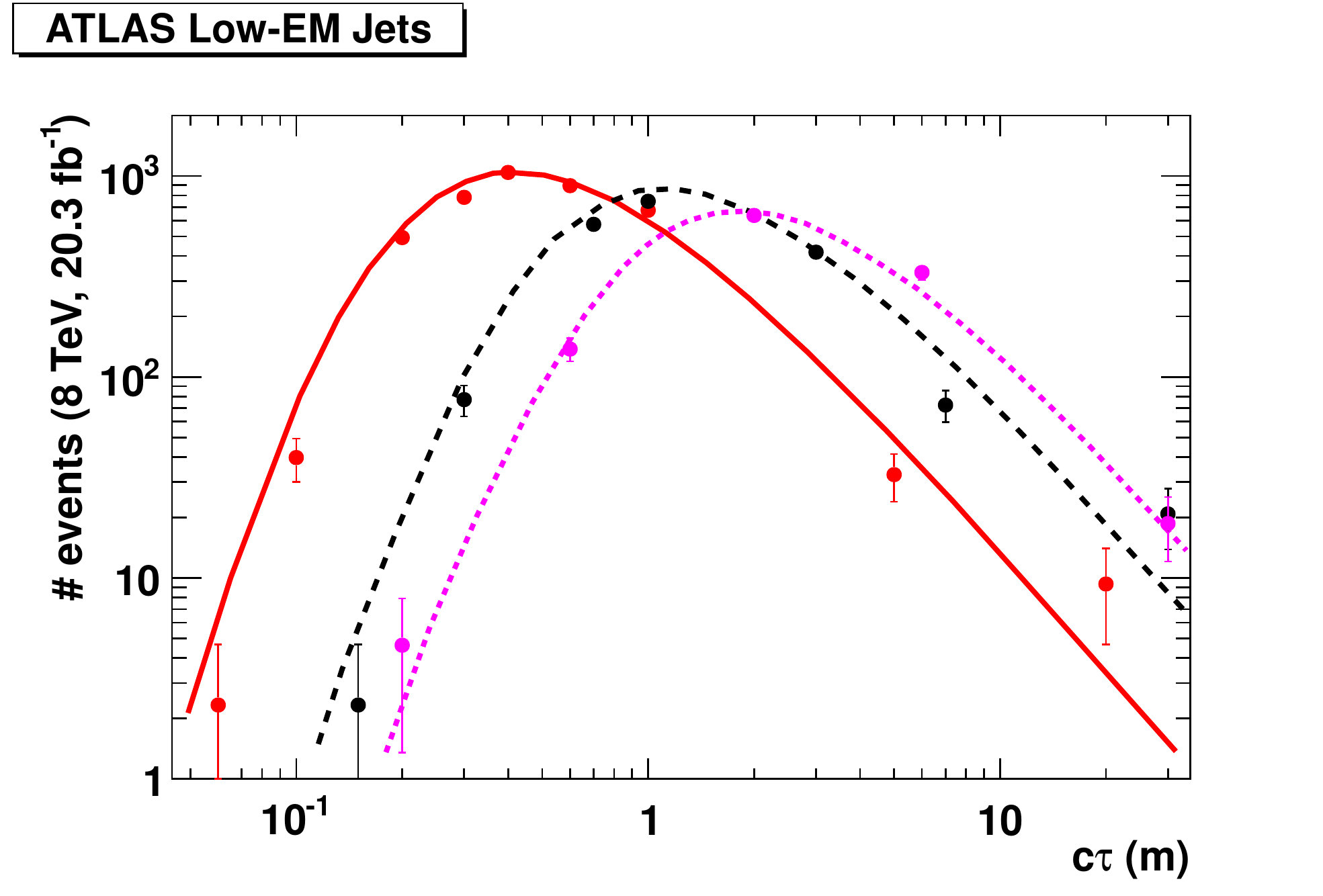}
\caption{A comparison of our detector simulation to the ATLAS low-EM jet analysis, illustrating expected event counts versus pseudoscalar lifetime (Fig.~4 of~\cite{ATLAS-2014-041}).  Curves indicate ATLAS predictions for different scalar~/ pseudoscalar masses: 126~GeV~/ 10~GeV (red solid), 126~GeV~/ 25~GeV (black dashed), and 140~GeV~/ 40~GeV (pink dotted).  Data points with error bars are our simulation predictions, and are color-matched to the corresponding ATLAS model curve.}
\label{fig:calibration_ATLASHCAL}
\end{figure}

To validate our detector simulation and calibrate the efficiency factor, we compare to ATLAS's physics models, which again feature singly-produced Higgs-like scalars that promptly decay into a pair of pseudoscalars with displaced decays.  The pseudoscalars dominantly decay into heavy quarks, with an $O(10\%)$ branching ratio to $\tau$-leptons, though we use a simpler model with decays only to light quarks.  (Our model therefore does not include energy losses to neutrinos, which we do not expect to be a major effect.)  We have found good agreement with the reconstruction efficiencies of the eight benchmark models, detailed in the appendix of~\cite{ATLAS-2014-041}, though for 100~GeV scalars our efficiencies come out $O(1)$ smaller.  In any case, these models barely pass the $E_T$ cuts, and are significantly more sensitive to the detailed turn-on of the efficiency with scalar mass.  We have also studied the efficiency versus pseudoscalar lifetime for three of the mass points.  The agreement is illustrated in Fig.~\ref{fig:calibration_ATLASHCAL}, and is typically at the 10\% level.

\subsection{ATLAS Displaced Muon Plus Tracks}
\label{sec:ATLASmuonTracks_calibration}

Similar to our simulation of the CMS displaced dijet analysis, for the ATLAS muon plus tracks search~\cite{ATLAS-2013-092} we assign each track a reconstruction probability.  We ignore all tracks from vertices with $r > 18$~cm or $z > 30$~cm, as well as cylindrical regions near the beampipe or the pixel layers: $r = [2.5,3.8]$~cm, $[4.5,6.0]$~cm, $[8.5,9.5]$~cm, and $[12,13]$~cm.  Within the active tracking volume, we use a tracking efficiency of $0.85 \times (1-r/(24\;{\rm cm})) \times (1-z/(30\;{\rm cm}))$.  For muon identification, which requires a very high-quality inner track, we apply an additional fixed efficiency factor of 70\%.  We apply another fixed factor of 70\% for successfully matching a muon to a displaced vertex.  These ad hoc factors reproduce the qualitative vertex reconstruction efficiency behavior versus $(r,z)$ in Fig.~3 of~\cite{ATLAS-2013-092}, as well as the ratio between event selection efficiencies before/after the muon-vertex matching requirement in Table~5 of that note.\footnote{The table suggests that the vertex reconstruction efficiencies, before track multiplicity and vertex mass cuts, are close to one.  However, matching between outer muons and inner tracks, as well as between muons and vertices, exhibits these nontrivial efficiencies.}  The latter is particularly relevant for the version of the analysis that we run for our recasts, which does not require the muon-vertex matching used in the nominal analysis.

\begin{figure}[tp!]
\centering
\includegraphics[scale=0.40]{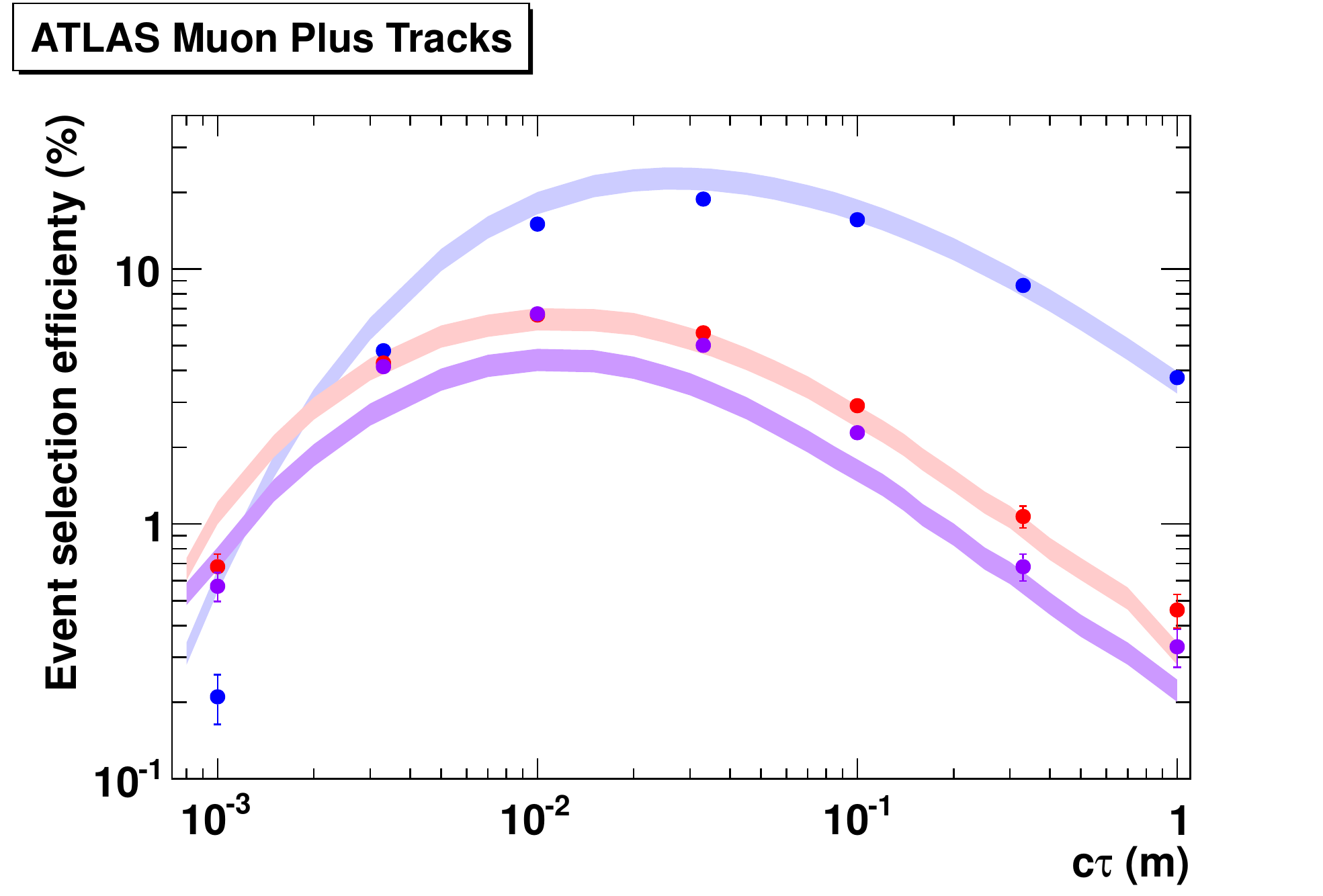}
\caption{A comparison of our detector simulation to the ATLAS muon plus tracks analysis, illustrating expected event reconstruction efficiencies versus neutralino lifetime (Fig.~4 of~\cite{ATLAS-2013-092}).  Colored bands indicate ATLAS predictions for different squark~/ neutralino masses: 700~GeV~/ 494~GeV (blue), 700~GeV~/ 108~GeV (red), and 1000~GeV~/ 108~GeV (purple).  Data points with error bars are our simulation predictions, and are color-matched to the corresponding ATLAS model curve.}
\label{fig:calibration_ATLASmuonTracks}
\end{figure}

ATLAS studies a set of SUSY models with leptonic RPV, where squark pairs promptly decay into jets plus long-lived LSP neutralinos, which then each undergo a displace decays into a muon plus two quarks.  Three mass points are considered: MH (700~GeV~/ 494~GeV), ML (700~GeV~/ 108~GeV), and HL (1000~GeV~/ 108~GeV).  Fig.~\ref{fig:calibration_ATLASmuonTracks} shows the event reconstruction efficiencies versus lifetime for these models after full selection cuts, as predicted by our simulation and by ATLAS's full simulation.  (The individual vertex reconstruction efficiencies are approximately half of the event reconstruction efficiencies, and show nearly identical behavior.)  The large gap in efficiencies between MH and ML/HL is reproduced, as is the shift in the peak versus lifetime and the falloffs at low and high lifetimes.  However, the $\approx 30\%$ gap between ML and HL is not reproduced, except at higher lifetimes, where the higher boost of HL tends to feel the tracking efficiency losses at large radii.  (The higher boost would also lead to smaller impact parameters at a given decay radius, but this is largely offset by the time-dilated decay length.)  The difference in modeling is possibly attributable to the fact that our simulation does not account for how displaced tracking and vertexing efficiencies change with track density, nor to possible issues in impact parameter reconstruction at small angles, all of which the lighter neutralino could be particularly sensitive to.  (To get a sense of this sensitivity, deleting one track from each vertex would cause the efficiency to fall by 20\% due to failures of the cut on the number of tracks.)  There could also be effects on the global muon reconstruction.  Nonetheless, the size of the mismodeling is below $O(1)$, and appears to be mainly relevant for relatively low-mass decays at relatively high boost.


\bibliography{lit}
\bibliographystyle{apsper}

\end{document}